\documentclass{book}

\usepackage{rbstyle}

\graphicspath{{./img/}} 

\begin{document}

	\frontmatter

		\loadgeometry{title}

		\begin{titlepage} \label{chap:sy_title}
\begin{center}

	\includegraphics[scale=0.5]{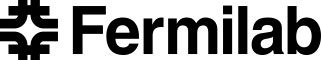}
	
	\vfill
	
	\textsc{\color{nal-blue} \LARGE Concepts Rookie Book}
	\title{Concepts Rookie Book}
	
	\vfill
	
	\textsc{Accelerator Division | Operations Department}
	\author{Accelerator Division | Operations Department}
	
	\today

\end{center}
\end{titlepage}
		
		\loadgeometry{base}
		
		\tableofcontents

		\listoffigures
		\listoftables
		
		\chapter*{Acknowledgments} \label{chap:con_ack}

	This book is a product of the FNAL Operations Department. Thanks to all 
	the former and current operators and experts who have spent their time 
	and energy collecting and organizing all of this information.
	
	\vfill
	
	\begin{center}
		Written, Compiled, and Edited by
		
		Dennis Barak, Beau Harrison, and Adam Watts
		
		Image updates by Jamal Johnson and Michael Wren
	\end{center}

	\mainmatter

		\loadgeometry{body}

		\chapter{Introduction} \label{chap:con_intro}

\Gls{fnal}\index{Fermilab} is a high-energy physics facility that studies the fundamental nature of matter and energy. The operator's role in this environment is to facilitate safe, efficient, and organized operation of the accelerators that deliver beam to experiments.

The purpose of a particle accelerator is to increase the kinetic energy of a concentrated group of charged particles, collectively known as \keyterm{\gls{beam}}. Collisions between the beam and a target material lead to the creation of new particles that were not present in the initial beam. \marginpar{\keyterm{\Gls{beam}} is a focused group of particles all traveling in the same overall direction.}For example, one of our experiments uses high-energy protons to strike a graphite target. The resulting pions and kaons decay into neutrino beam, allowing physicists to study the behavior of these fundamental particles.

The accelerators at Fermilab form a sequential chain of connected machines: The Proton Source (\index{Pre-Acc}Pre-Accelerator, \index{Linac}Linac, and \index{Booster}Booster), \index{Recycler}Recycler, Main Injector. These machines and the subsequent experiments are connected via non-accelerating ``beamlines.'' This rookie book is meant to give an overview of the machines and systems that provide beam to the experiments.

This book begins with basic information required to understand the later chapters. We then go over each accelerator in the chain to describe specific systems in more depth. Next, we give an introduction to fundamental accelerator physics concepts, as well as a description of specific beam instrumentation devices. We follow these chapters with a description of the utility systems that support the function of the accelerators. Finally, we discuss the role of operators in safety at the lab.

%Role of the Operator

%\keyterm{Red key terms} are the concepts that appear in the Introductory OJT and the Introductory OP2 test.

	\section{Essential Concepts}
	We begin by introducing some common concepts that are used in this text and in the Main Control Room. The ideas discussed here will be explained in greater detail at a later point, but a simple introduction is necessary here.
	
		\subsection{Particle Beam Composition, Intensity, and Energy}
		Particle accelerators primarily use the electromagnetic force to interact with beam. The number of particles in a beam is called the ``beam intensity,'' which we can measure either in terms of the total number of particles or in terms of the beam current\footnote{Current is defined as the flux of electric charge through a closed surface.}. Operators monitor beam intensity throughout the accelerator chain in order to know when and where beam is lost. 
				
		\marginpar{\keyterm{\Gls{acceleration}} refers to the increase of beam kinetic energy.}
		
		For describing particle energy, we use the unit ``\ev{},'' or electron-volt. One \ev{} is the amount of kinetic energy an electron gains by passing through a potential difference of one volt. The following prefixes describe the energy scales relevant to Fermilab: \kev{} (kilo-electron volt, \ev{1E3}), \Mev{} (Mega-electron volt, \ev{1E6}), \gev{} (Giga-electron volt, \ev{1E9}), \tev{} (Tera-electron volt, \ev{1E12}).

		The purpose of a particle accelerator is to increase the kinetic energy of the beam particles; we refer to this as \keyterm{acceleration} of the beam.\footnote{Acceleration is technically any change in velocity, so even slowing down or changing the trajectory of the beam would still be ``particle acceleration.'' At the lab, however, we use the term ``acceleration'' to refer to a \textit{positive} increase in the particle velocity.} 
		
		\subsection{Longitudinal and Transverse Directions}
		The three dimensions commonly used to describe particle motion are longitudinal, horizontal, and vertical. The \keyterm{\gls{longitudinal}} dimension is the direction in which beam travels. The \textit{horizontal} and \textit{vertical} dimensions are perpendicular to one another and to the longitudinal direction. The horizontal and vertical dimensions form the \keyterm{\gls{transverse}} plane. We refer to the longitudinal directions as ``upstream'' or ``downstream''\footnote{Just like the water in a river, beam travels from an ``upstream'' location to a ``downstream'' location.}; up, down, right, or left describe the transverse directions.
		
		\marginpar{
			\centering
			\includegraphics[width=5cm]{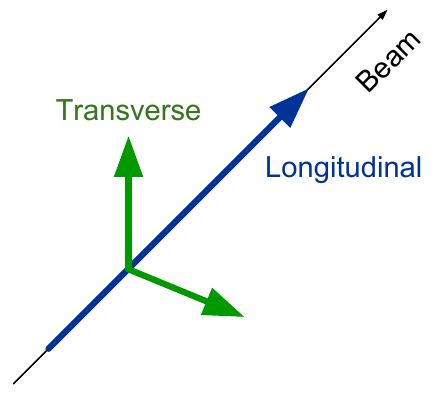}
			\captionof{figure}[The longitudinal direction and the transverse plane]{The \keyterm{\gls{longitudinal}} direction and the \keyterm{\gls{transverse}} plane}
		}
		
		\subsection{RF and Magnets}
		The major components that make up particle accelerators are magnets and RF cavities. Magnets control the beam's trajectory and size, and \keyterm{\Glspl{RFcavity}} accelerate the beam. The concepts behind these systems will be introduced in this section and then covered in detail later in the book. 

		\marginpar{\keyterm{\Glspl{RFcavity}} are hollow structures resonating with electromagnetic waves that accelerate the beam.}
		
			\subsubsection{RF}
			Radio Frequency (RF) systems accelerate charged particle beams. The portion of the RF system acting on the beam is the RF cavity, which is an electromagnetically resonant structure. When used inside accelerators, RF cavities generate a strong longitudinal electric field that accelerates beam to a desired energy level. RF systems will be covered in more detail in \textit{Chapter~\ref{chap:con_RF}}.
			
			\subsubsection{Magnets}
			
			Magnets change the direction of particles' trajectory. Accelerators either use permanent magnets or electromagnets.  Electromagnets are useful because we can change their fields by adjusting the amount of current we supply them. On the other hand, permanent magnets generate fixed-strength fields. The name and function of a magnet depends on the number of magnetic poles that it contains. For example, dipole magnets contain two poles, quadrupole magnets contain four poles, sextupole magnets contain six poles, and so on. 
			
			Dipole magnets change the trajectory of an entire beam of particles. This is commonly referred to as ``bending'' the beam\footnote{For this reason, dipoles are often called ``bend'' magnets and the dipole field is called the ``bend'' field.}. Dipoles bend beam either horizontally, vertically, or in both directions. 
			
			Quadrupole magnets, often called ``quads,'' focus the beam and keep it constrained to the machine. A single quadrupole provides beam focusing in one plane and beam defocusing in the other. ``Focusing'' (F) quadrupoles focus horizontally and defocus vertically. ``Defocusing'' (D) quadrupoles defocus horizontally and focus vertically. To obtain net focusing in both transverse directions, quadrupoles are arranged in an alternating pattern called a ``lattice''. One common lattice, with equal spacing between quads, is called ``FODO,'' where the ``O'' represents a drift space in which there are no main quadrupoles.
			
			The main magnets that form an accelerator are typically very large and are often grouped into common circuits. Smaller magnets, called ``trims'' or ``correctors,'' provide a field at a particular location that is independent of the main circuit. They can be used to correct for field imperfections in the larger main magnet systems and to intentionally change the trajectory of the beam at a particular location. 
			
			\textit{Chapter~\ref{chap:con_magnets}} covers the different magnet types and their effects in more detail.
			
			\subsubsection{Transverse and Longitudinal Orientation}
			
			% The following figure is for the physical dimensions subsubsection
			\marginpar{
				\centering
				(a) Dipole
				
				\includegraphics[width=4.5cm]{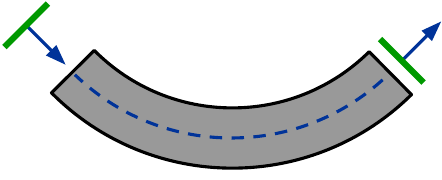}
				
				(b) An RF cavity or a quadrupole

				\includegraphics[width=3.5cm]{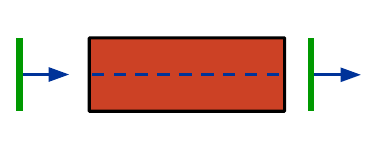}
				\captionof{figure}[Effect of a dipole]{The longitudinal direction (\textcolor{blue}{blue}) and the transverse plane (\textcolor{nal-green}{green}) as seen from above.}
				\label{fig:dimensions-2}			}
			
			The direction in which beam travels at a particular point defines the orientation of the transverse and longitudinal dimensions. Dipole magnets change the direction of the beam, and thus change the orientation of the transverse plane and longitudinal dimension. \textit{Figure~\ref{fig:dimensions-2}a} shows a dipole magnet that bends the beam by $90^\circ$, thus rotating the \textcolor{nal-green}{transverse plane} by $90^\circ$ as well. 
			
			Quadrupole magnets act on the beam transversely, and RF cavities act on the beam longitudinally, but neither change the orientation of either dimension. In other words, they do not alter the orientation of the transverse plane. \textit{Figure~\ref{fig:dimensions-2}b} shows that a quadrupole or an RF cavity does not affect the orientation of the longitudinal or transverse dimensions. 
			
		\subsection{Types of Accelerators}
		
		We now shift from discussing individual components to using a combination of components for a particular function.
		
			\subsubsection{Linear Accelerators}
			Linear accelerators, also called ``\index{Linac}linacs,'' are mainly composed of RF cavities placed in-line with one another to provide a large amount of energy gain per unit length. Magnets along the length of a \index{Linac}linac keep the beam contained inside the machine. By definition, beam makes only one pass in a linac because it is linear. 
			
			An important quality of a \index{Linac}linac is the energy gain it provides per unit length. The particles in a beam of like charges repel one another electrostatically, which causes a considerable increase in the beam's size\footnote{Ideally, the average transverse beam size does not increase from the beginning of an accelerator to the end.}. Additionally, this repulsive force\footnote{This force of repulsion is called the ``space charge force.''} is stronger at low energies than it is at high energies, and it is stronger for high-current beams than it is for low-current beams. Thus, the effects of this repulsion are minimized by accelerating the beam to higher energies quickly. 
	
			\begin{figure}[!htb]
				\centering
					\includegraphics[width=1.0\linewidth]{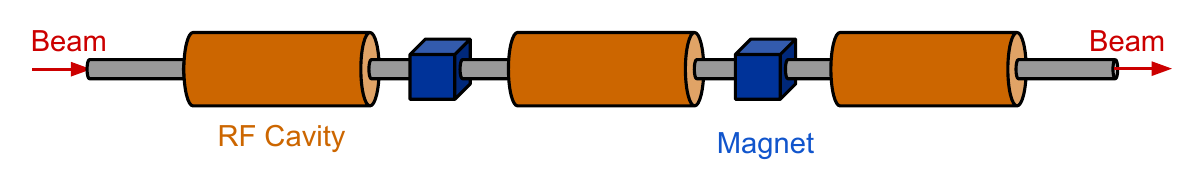}
					\caption{A simple linear accelerator}
					\label{fig:simplelinac}
			\end{figure}
			
			\subsubsection{Synchrotron Accelerators}
			
			\marginpar{A \keyterm{\gls{synchrotron}} is a type of circular accelerator in which the magnetic bend field and RF frequency increase together to accelerate beam.} 
			All of our circular accelerators are of the type known as a \keyterm{\gls{synchrotron}}. In a synchrotron, the magnet system and the RF system are ``synchronized'' as the kinetic energy of the beam increases; in other words, the magnet current and RF frequency must increase together in a controlled manner. This is necessary because the beam's path must remain constant even after receiving a kick\footnote{It is common to refer to the beam's energy increase from the cavity as a kick.} from an RF cavity's electric field, so the magnetic field increases to match any increase in the kinetic energy. We call this increase in magnetic field \keyterm{ramping} the magnets.
			
			\marginpar{\keyterm{Ramping} an electromagnet system refers to changing the magnetic field over a period of time.}
	
			\begin{figure}[!htb]
				\centering
					\includegraphics[width=0.7\linewidth]{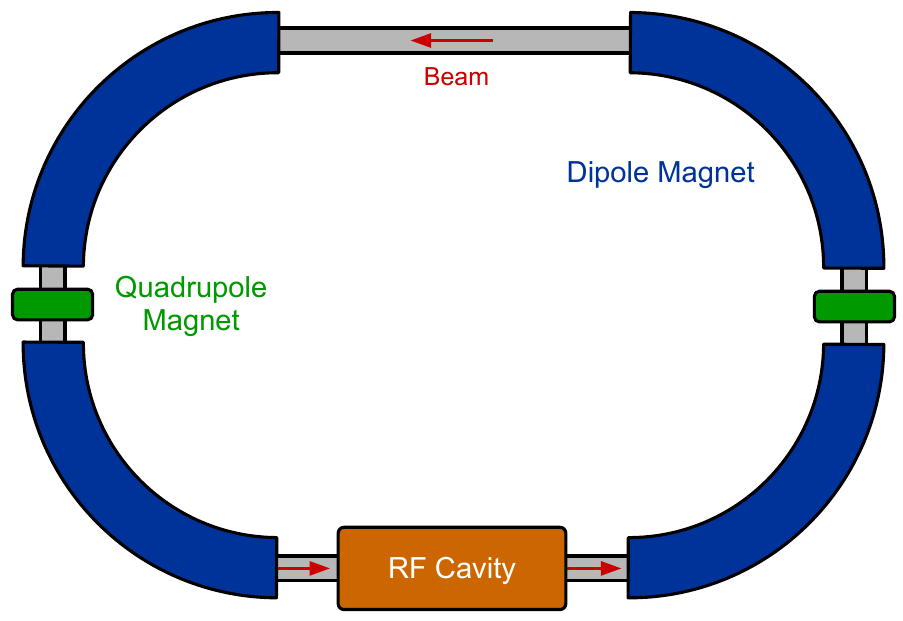}
					\caption{A simple synchrotron}
					\label{fig:simplesync}
			\end{figure}
			
			The circular beam path through the center of the magnets is known as the \keyterm{\gls{idealorbit}}. In a perfect machine, the beam would travel along the ideal orbit. However, imperfections such as aperture obstructions, field perturbations, and alignment errors require that the ideal orbit be adjusted to compensate. Furthermore, the adjusted beam path must end in the same place it started to ensure orbit stability. Thus a \keyterm{\gls{closedorbit}} is the beam trajectory that includes path corrections and returns to the same point on every revolution. The closed orbit is the reference against which we compare measured beam positions.
			
			\marginpar{\keyterm{Ideal Orbit} is the path of the beam through the center of the magnets. In a synchrotron, the synchronous particle travels on the ideal orbit.}
			
			\marginpar{\keyterm{Closed Orbit} is the beam path that returns to the same point on every revolution around a circular machine. It is the beam trajectory which includes intentional bumps and corrections.}
			
			%In a synchrotron, the longitudinal direction always points in the general direction of the beam. 
			%\textit{Figure~\ref{fig:dim-circ}}  shows how the longitudinal direction is always tangent to beam path in a synchrotron. 
			
			%\begin{figure}[!htb]
			%	\centering
			%	\includegraphics[width=0.9\textwidth]{dim-circ}
			%	\caption{The longitudinal and transverse directions in a synchrotron.}
			%	\label{fig:dim-circ}
			%\end{figure}
			
			\subsubsection{Beamlines}
			There are portions of an accelerator complex that transport beam from one accelerator to another or from an accelerator to a target. These are called ``beamlines,'' and they are mainly composed of magnetic elements, usually dipoles and quadrupoles. Because beamlines are fixed-energy, some use permanent magnets instead of electromagnets. 
			
			\begin{figure}[!htb]
				\centering
					\includegraphics[width=0.7\linewidth]{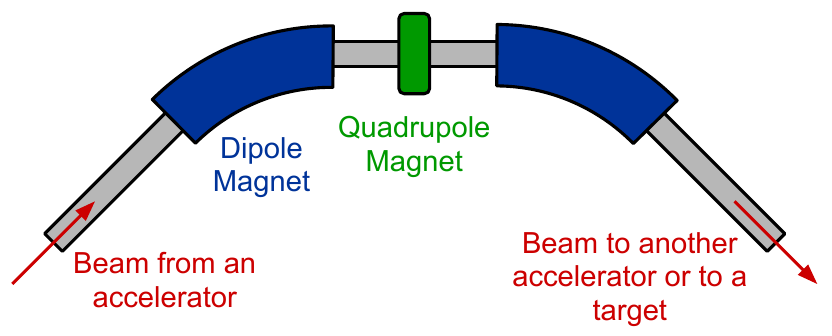}
					\caption{A simple beamline}
					\label{fig:simplebline}
			\end{figure}
			
			\subsubsection{Beam Transfers}
			\marginpar{\keyterm{\Gls{injection}} is the process of transferring beam into an accelerator.} \marginpar{\keyterm{\Gls{extraction}} is the process of removing beam from an accelerator.}
			The process of transferring beam into a machine is called \keyterm{\gls{injection}}, and the process of removing beam from a machine is called \keyterm{\gls{extraction}}. It is common to describe an accelerator by specifying the beam energy at injection and at extraction. Beamlines transport beam from one machine to another. The transfer of beam from one machine to another involves synchronizing the extraction of beam from one machine and the injection of beam into another. 
			
		\subsection{Accelerator Controls}
		Accelerator components are monitored and manipulated remotely through Fermilab's control system. The Fermilab controls system is very extensive, connecting a few hundred thousand devices to the \Gls{mcr} and remote consoles.
			
			\subsubsection{Device Parameters}
			Devices that are connected to the controls system have a name, called a ``parameter,'' that is up to fourteen (14) characters in length and in the form X:DEVICE901234. The prefix, X:, represents the machine in which the device is located. The final twelve characters, DEVICE901234, represent the individual device. Parameters can have up to four properties: readback, setting, control, and status. An example of each property is given below for a typical power supply.
			
			\begin{itemize}
				\item Analog Readback: Output voltage or current of a power supply
				\item Analog Setting: Requested output voltage or current of a power supply
				\item Digital Control: Ability to turn the power supply on or off
				\item Digital Status: Indicates whether the power supply is on or off
			\end{itemize}
			
			The digital signals are low voltages translated into a binary logic bit\footnote{A bit (a contraction of binary digit) can have only two values: either 1 or 0. The two values can also be interpreted as logical values (true/false, yes/no), algebraic signs ($+/-$), activation states (on/off), or any other two-valued attribute.} (0 or 1). In Operations we often call the proper state for a signal ``good'' or ``made up.'' 
			
			\subsubsection{Machine Cycles}
			Most of the devices at Fermilab are cycle driven, meaning that a defined sequence of tasks is performed at regular intervals. The cycle can be described in terms of its length or its rate. The cycle length is simply the length in time that a particular cycle takes. The cycle rate is the number of cycles that are completed per second. Typically, an entire machine cycles at the same time and rate. For example, the entire Proton Source operates at a fixed rate of 15 Hz. This means that all of the equipment performs a given task fifteen times a second. It is worth pointing out that not all machine cycles involve the acceleration of beam. The machines often complete cycles with no beam present. When a machine cycle does include beam, it is called a ``beam cycle.''

		\chapter{Magnets}\label{chap:con_magnets}
	Magnets steer the trajectory of the beam in particle accelerators. For linear machines like the \index{Linac}Linac and transfer beamlines, the magnets point the beam in the correct direction. In circular machines like the \index{Booster}Booster and Main Injector, the magnets keep the beam on a circular path by continuously bending the trajectory. To examine the force exerted on charged beam by a magnetic field, we refer to the \textit{Lorentz Force} expression given in \textit{Equation~\ref{eq:Lorentz2}}. The Lorentz force describes the force on a particle of charge $q$ and velocity $v$ due to electric field $E$ and magnetic field $B$.
	
	\begin{equation} \label{eq:Lorentz2}
		\vec{F}=q(\vec{E}+\vec{v} \times \vec{B})
	\end{equation}
	
		\marginpar{
			\centering
			\includegraphics[width=.9\linewidth]{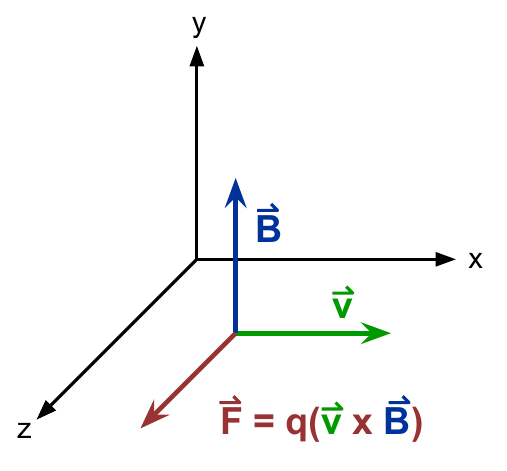}
			\captionof{figure}{Magnetic force on a charged particle}
			\label{fig:bforcevector}
		}
	
	We see that the magnetic contribution to the force $\vec{F} = q\vec{v} \times \vec{B}$ involves a vector cross-product; this means that the magnetic force will always be perpendicular to the beam velocity. \textit{Figure~\ref{fig:bforcevector}} shows the magnetic force $F$ on a particle with velocity $v$ due to a  magnetic field $B$. If the pictured vertical magnetic field is present and constant throughout the entire plane, the beam is contained in a circular orbit. This is how synchrotrons like Main Injector and Booster keep the beam on a circular trajectory.
	
	A vertically-oriented magnetic field exerts a horizontal inward force on the beam. If we can arrange a uniform vertical magnetic field spread out over a large area, as shown in \textit{Figure~\ref{fig:bendbeam}}, the magnetic force will continuously push inward and steer the beam in a circular path. This force is known as a ``centripetal force'' because it points radially inward. \textit{Figure~\ref{fig:bendbeam}} shows a uniform magnetic field pointing into the page and how it bends the path of a positively-charged particle in an arc.
	
	Instead of trying to create a uniform magnetic field over the entire accelerator area, we place magnets around the accelerator, This creates a magnetic field along parts of the beam trajectory. For linear accelerators and beam transfer lines,  we use magnets for small trajectory corrections. However, we need magnets along most of the beam path in a synchrotron to keep the beam on a circular path. 
		\marginpar{
			\centering
			\includegraphics[width=.9\linewidth]{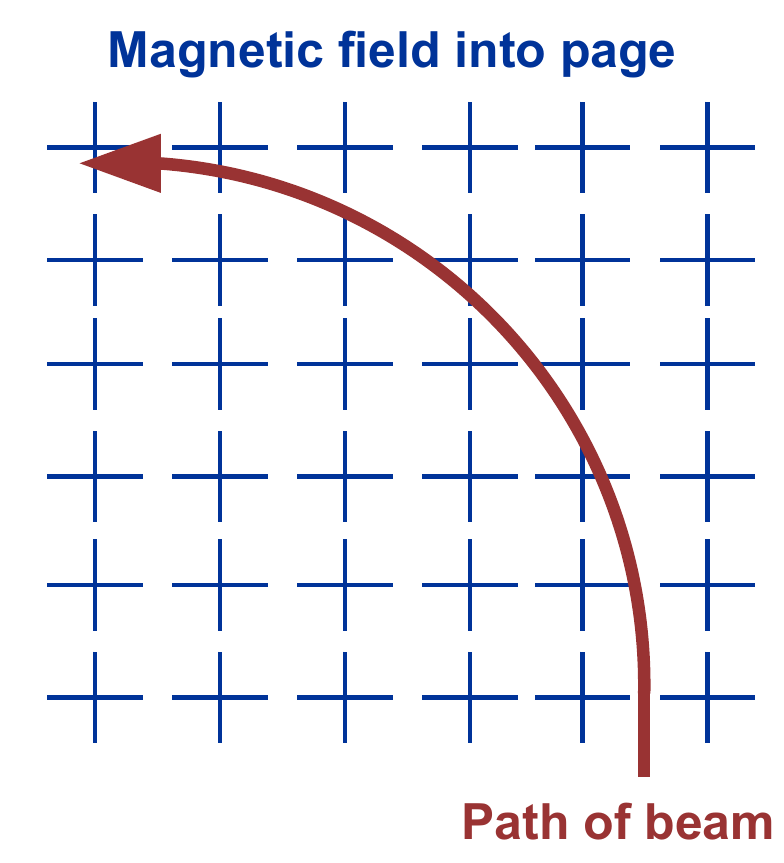}
			\captionof{figure}{Circular beam path due to a uniform magnetic field}
			\label{fig:bendbeam}
		}
	
	We will now examine why we need the ability to change the magnetic field strength in a circular accelerator as the beam momentum increases. \textit{Equation~\ref{eq:radius}} gives the radius of the curved path from \textit{Figure~\ref{fig:bendbeam}} as a function of particle momentum $p$, charge $q$, and magnetic field $B$. This equation is fully derived in Appendix \ref{chap:con_apndx_a}. 
	
	\begin{equation} \label{eq:radius}
		r=\frac{p}{qB}
	\end{equation}
	
	We see that the beam trajectory radius $r$ will increase with the beam momentum $p$ in an accelerator. To keep beam inside the machine, we increase the magnetic field strength $B$ to maintain a constant radius. Increasing the magnetic field is known as \keyterm{\gls{ramping}} the magnets, which corresponds to increasing the power supply current.

	\marginpar{\keyterm{\Gls{ramping}} the magnets refers to increasing magnetic field to keep the bend radius of the beam path constant through acceleration.}
	
	We have two basic ways of creating a magnetic field: the permanent magnet and the electromagnet. \textit{Permanent magnets} use ferromagnetic metals to create an unchanging field that is analogous to that of a refrigerator magnet. While permanent magnets require no power to operate, their magnetic field cannot be changed. Therefore, we use permanent magnets in machines where the beam energy does not change, like the Recycler and MI-8 line. We use \textit{electromagnets} when changing beam energy necessitates the ability to change the magnetic field. These magnets require power supplies that flow electrical current in loops to induce the magnetic field; by changing the power supply current, we can directly change the field strength of the magnet. Electromagnets are necessary in synchrotrons like Booster and Main Injector, where the magnetic field must increase to track the increasing momentum of the beam.
	
\begin{figure}[!htb]
		\centering
			\includegraphics[width=.8\linewidth]{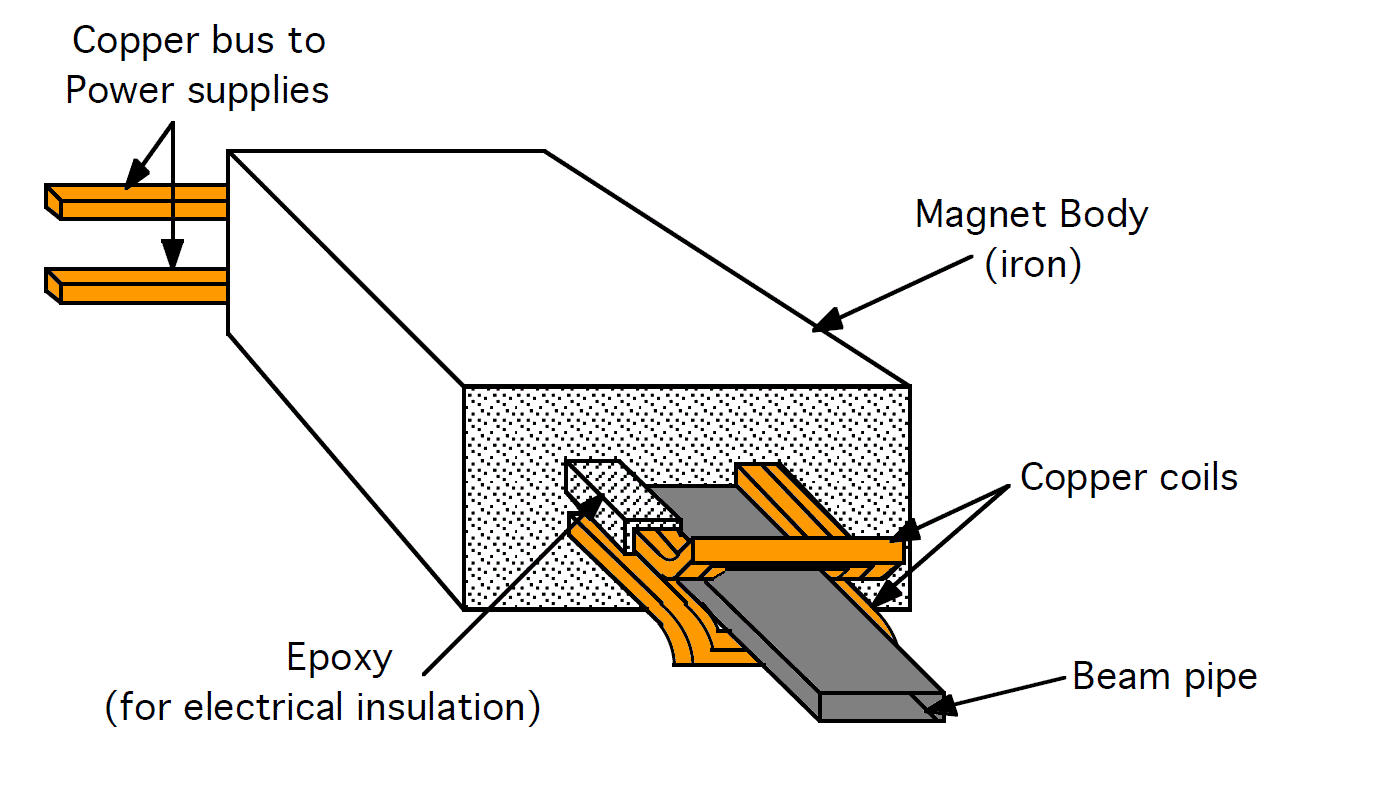}
			\caption{Model of a basic electromagnet.}
			\label{fig:dipolemagnet}
	\end{figure}

	\textit{Figure~\ref{fig:dipolemagnet}} shows an example of a typical electromagnet. Power supplies flow current through conducting loops called ``coils'' that induce a magnetic field inside the iron body. The iron magnet body shapes the field to permeate through the beam pipe. The copper bus is hollow in the center to allow for the flow of  low-conductivity water, or ``LCW,'' which draws heat away from the magnet without creating an electrical short. We explain LCW systems in more detail in \textit{Chapter~\ref{chap:con_utils}} of this book.
	
	Thus far, we have been describing the simplest function of the magnet, which is bending the overall beam trajectory. We call this type of magnet a ``dipole magnet,'' because it has a uniform magnetic field generated from two magnetic poles. As we will see in the coming sections, more complicated field shapes can produce other useful effects on the beam such as transverse focusing.
	
	\section{Dipoles}
		\keyterm{\Gls{dipole} magnets} provide the horizontal or vertical ``bending'' that steers the beam and keeps its path circular in a synchrotron. The ideal dipole field is constant in magnitude throughout the entire magnet, so the resulting force is independent of transverse particle displacement from the center of the beam pipe. 
		
		\marginpar{A \keyterm{\gls{dipole}} magnet's field is constant in the transverse plane, and bends the beam trajectory in an arc.}
		
		\begin{figure}[!htb]
		\centering
			\includegraphics[width=.45\linewidth]{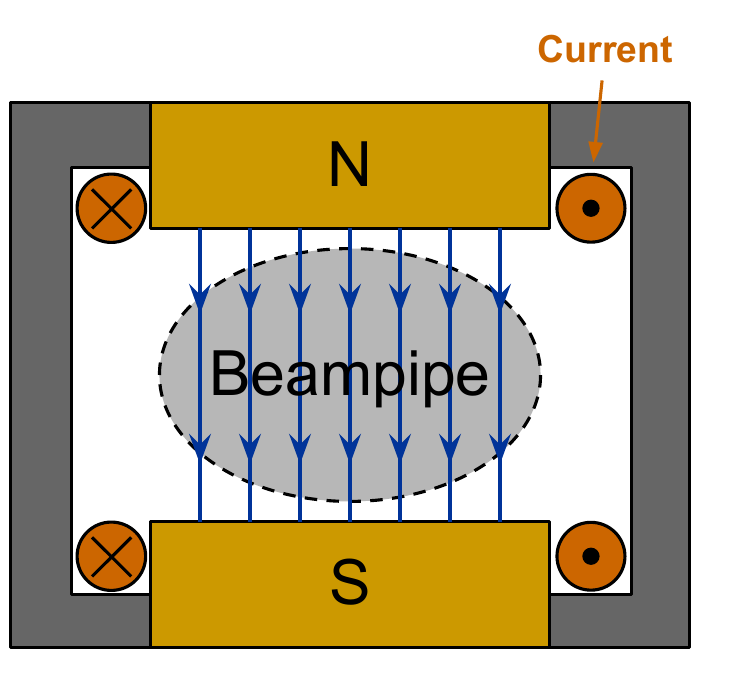}
			\includegraphics[width=.45\linewidth]{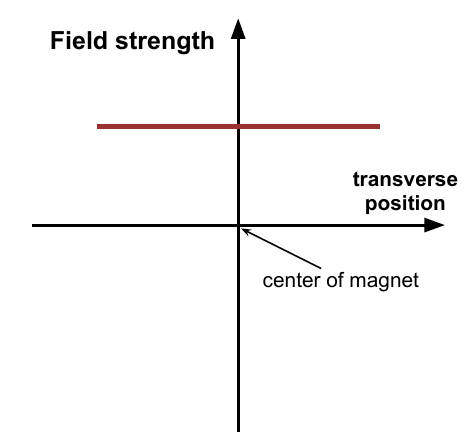}
			\caption{Dipole and the transverse dependence of its field.}
			\label{fig:dipole}
		\end{figure}
		
		\marginpar{
			\centering
			\includegraphics[width=.9\linewidth]{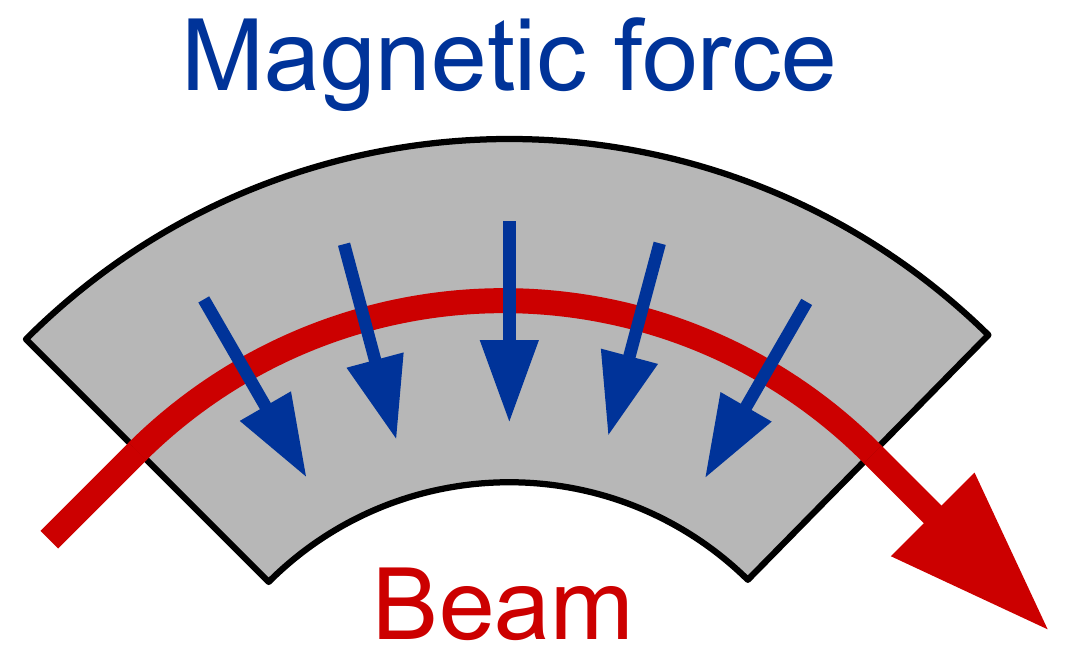}
			\captionof{figure}{Red beam path through a dipole with blue magnetic force}
			\label{fig:dipolebending}
		}
		
		\textit{Figure~\ref{fig:dipole}} shows the magnetic field lines due to the current in the copper coils of a dipole magnet. In the case of the dipole, the magnetic field lines point uniformly in one direction inside the beam pipe. This provides a bending force that does not depend on where the beam is in the magnet aperture.
		
\begin{figure}[!htb]
		\centering
			\includegraphics[width=.7\linewidth]{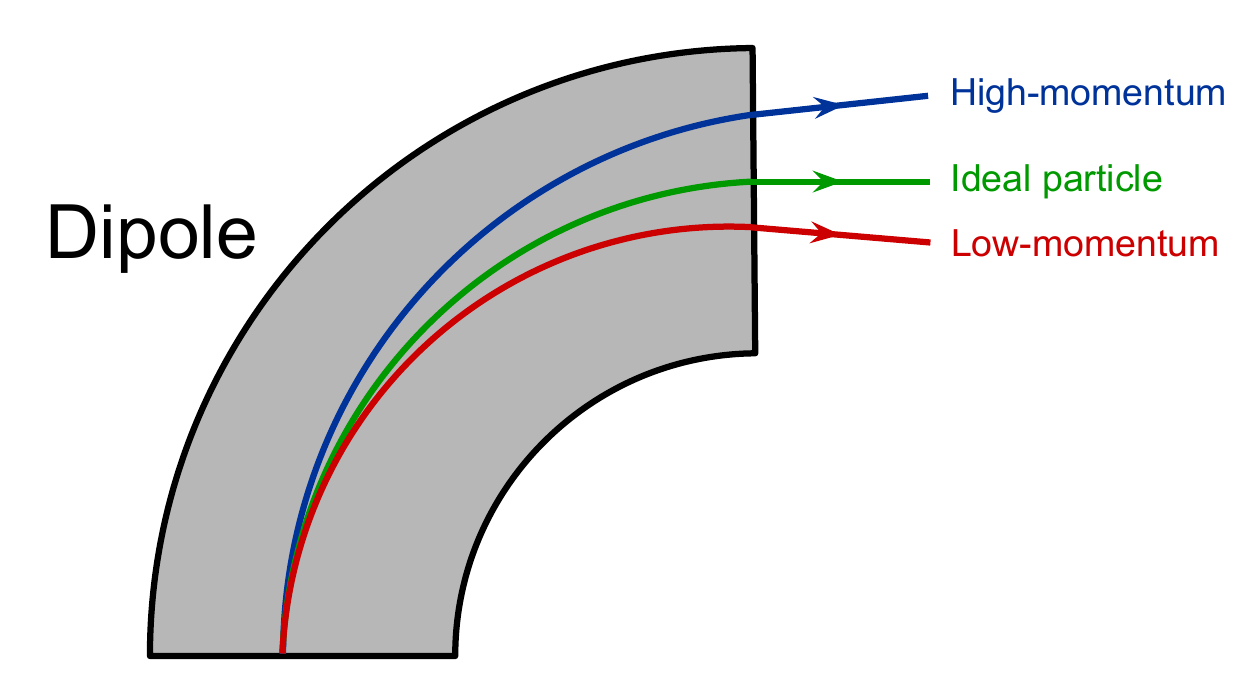}
			\caption{Dipole magnet dispersion.}
			\label{fig:dispersion}
		\end{figure}

		Since the force exerted on the particle due to the magnetic field is velocity-dependent, particles of different momentum will bend in slightly different arcs. This effect is known as \keyterm{\gls{dispersion}}, and it transversely separates the trajectories of particles with different momenta as shown in \textit{Figure~\ref{fig:dispersion}}. Think of this effect like a prism in optics: the dipole's dispersion separates particles of different momenta just as a prism separates light rays of different colors.

		\marginpar{\keyterm{\Gls{dispersion}} refers to the way dipoles bend beam along different trajectories depending on particle momentum.}
				
\begin{figure}[!htb]
		\centering
			\includegraphics[width=.7\linewidth]{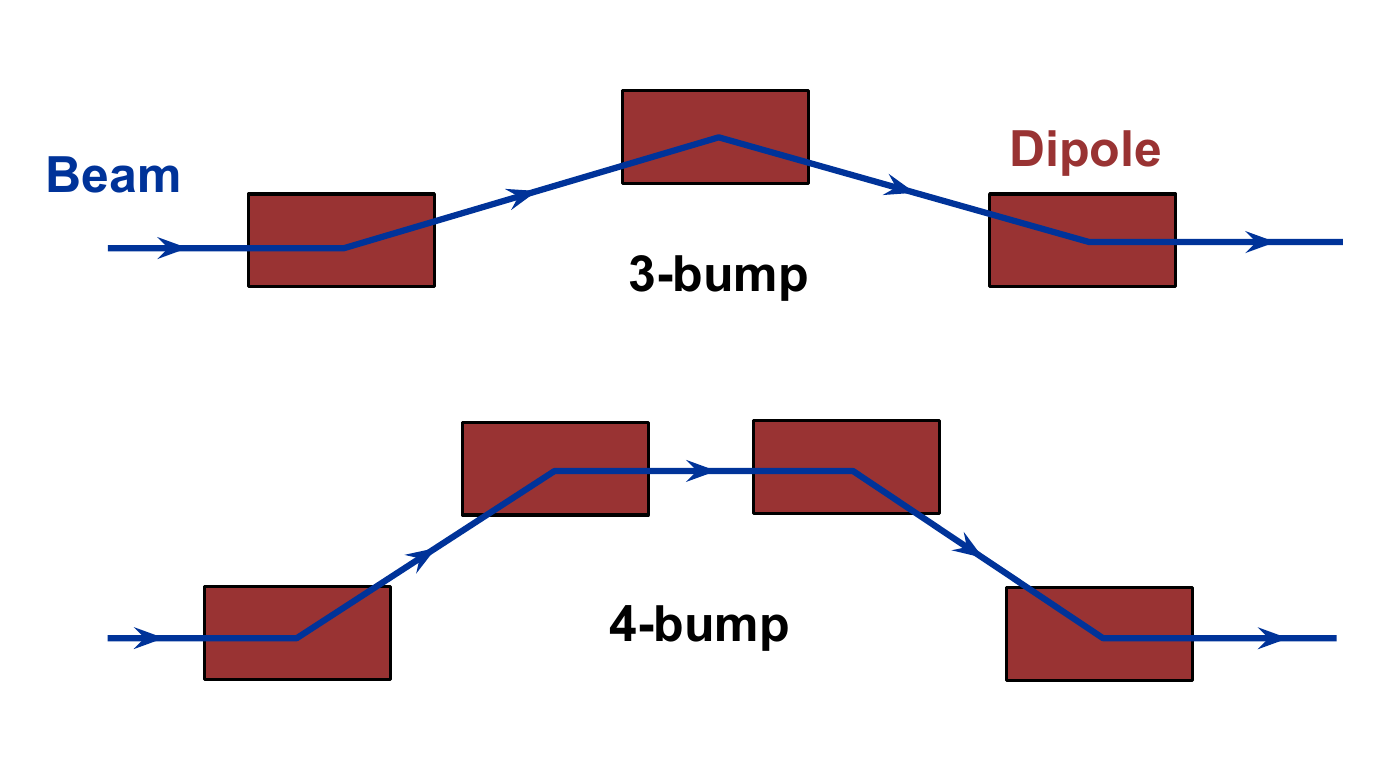}
			\caption{Dipole bumps.}
			\label{fig:bumps}
		\end{figure}

		Since beam particles travel so quickly, they do not spend much time in the aperture of the magnet. This means that the magnet only provides a small angle adjustment to the beam trajectory. In order to create localized position changes without affecting the overall orbit, we use three or more dipole corrector magnets to create what is known as a \keyterm{\gls{bump}}. Bumps allow us to change the position or angle of the beam in a specific location only; for instance, we may need to bump around something that is in the way of the beam. \textit{Figure~\ref{fig:bumps}} shows two position bumps using three and four magnets, colloquially known as a ``3-bump'' and ``4-bump.''
		
		\marginpar{\keyterm{\Glspl{bump}} are combinations of dipoles that cause localized position or angle adjustments in the beam without affecting the rest of its orbit.}
		
	\section{Quadrupoles}
		\keyterm{\Gls{quadrupole} magnets} provide transverse focusing of the beam, keeping it constrained within the beam pipe; they effectively operate like lenses. Inside a quad the field strength is zero at the center of the magnet aperture and increases linearly with transverse displacement. This means that the quadrupole exerts a linear restoring force very similar to that of a spring $F=-kx$, encouraging particles to move toward the center of the beam pipe. Just like a spring causing a mass to oscillate, the quadrupole's restoring force causes the beam to oscillate transversely. The number of transverse oscillations per revolution is called the ``tune,'' controlled by the quadrupole field strength.
		
		\marginpar{\keyterm{\Glspl{quadrupole}} act like lenses that focus the beam in one transverse plane but defocus in the other.} 
		
\begin{figure}[!htb]
		\centering
			\includegraphics[width=.45\linewidth]{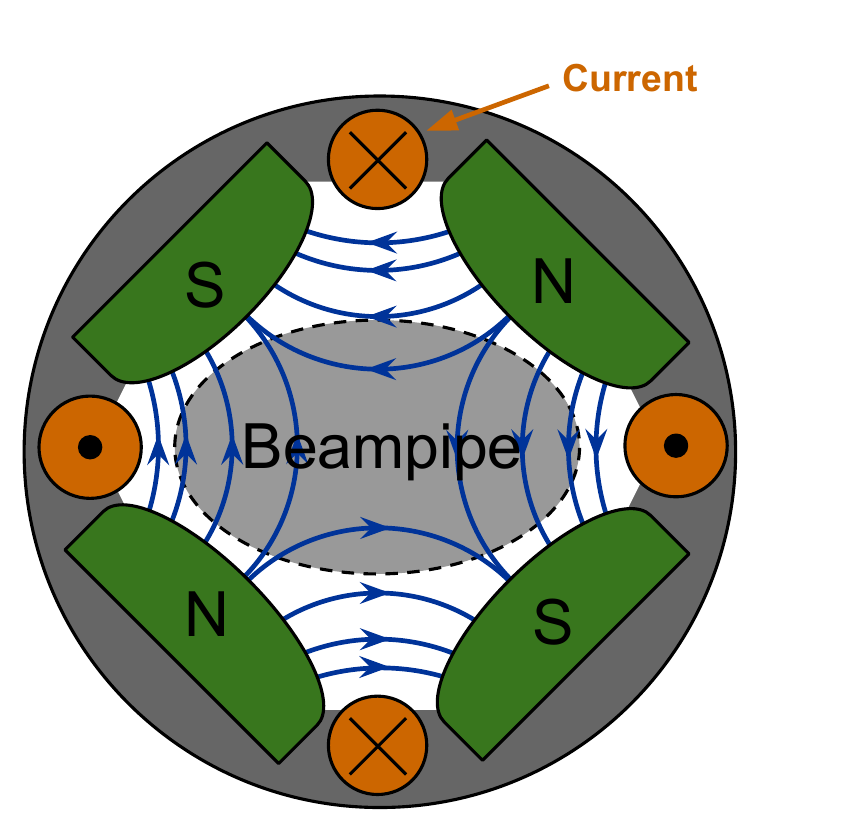}
			\includegraphics[width=.45\linewidth]{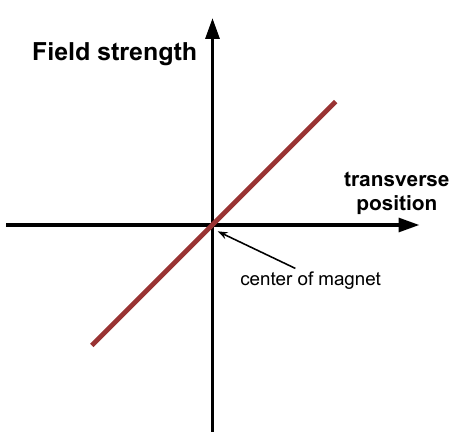}
			\caption{Focusing quadrupole and the transverse dependence of its field.}
			\label{fig:quadfields}
		\end{figure}

		\textit{Figure~\ref{fig:quadfields}} shows a quadrupole magnetic field that results from the four currents into and out of the page. The quadrupole field shape is useful for transverse beam focusing because the field provides a restoring force that is linearly proportional to the particle's displacement from the center of the beam pipe. Quadrupole magnets can be thought of as lenses that focus in one transverse plane while defocusing in the other; this is illustrated by \textit{Figure~\ref{fig:quadforce}}, which shows the force that results from the magnetic field of a quadrupole magnet. As it turns out, it is not possible for a magnetic field to focus in both transverse planes simultaneously\footnote{This is derived in \textit{Appendix~\ref{chap:con_apndx_a}}}.
		
		\marginpar{
			\centering
			\includegraphics[width=.9\linewidth]{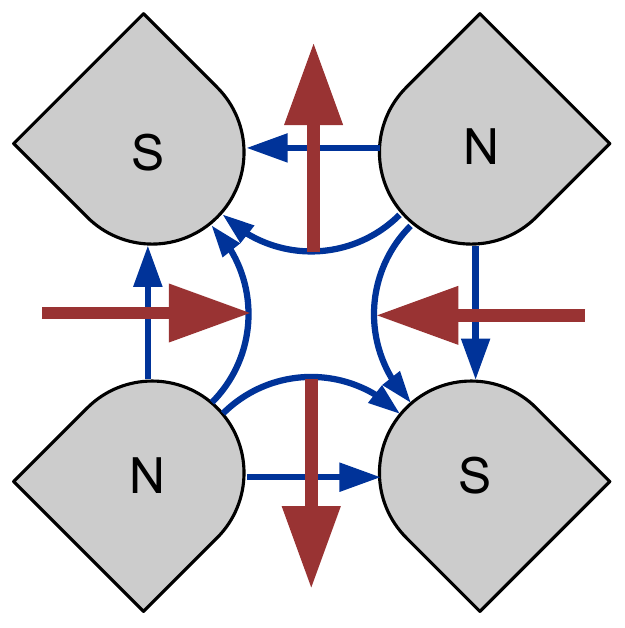}
			\captionof{figure}{Blue magnetic field and red force directions for a focusing quadrupole}
			\label{fig:quadforce}
			}
		
		By convention, we say that a ``focusing magnet'' focuses in the horizontal plane, and a ``defocusing magnet'' focuses in the vertical. To achieve overall beam focusing, we must arrange the magnets in an alternating repeating pattern known as a \keyterm{\gls{lattice}}. The smallest repeating pattern of quadrupoles is known as a \keyterm{\gls{cell}}, and is used to name the type of lattice. For example, the Main Injector has a ``FODO'' lattice: ``F'' is a focusing quad, ``O'' is a drift space with no quads, and ``D'' is a defocusing quad. In accelerator physics terminology, alternating quadrupoles to achieve net focusing is known as \textit{strong focusing}, or \textit{alternating-gradient focusing}.
		 
		 \marginpar{Quadrupoles are arranged in an alternating pattern known as a \keyterm{\gls{lattice}}, the smallest repeating piece of which is called a \keyterm{\gls{cell}}.}
		 
		 The strong-focusing process causes the beam to oscillate in the transverse plane about the center of the beam pipe. This motion is called \textit{betatron oscillation}, and the number of complete oscillations per revolution around a circular machine is called the \keyterm{\gls{tune}}. Betatron oscillation and tune are explained in more detail in \textit{Chapter~\ref{chap:con_physics}}.
		 
		  \marginpar{The number of betatron oscillations that a particle completes in one revolution is known as the \keyterm{\gls{tune}}.}
		
\begin{figure}[!htb]
		\centering
			\includegraphics[width=.7\linewidth]{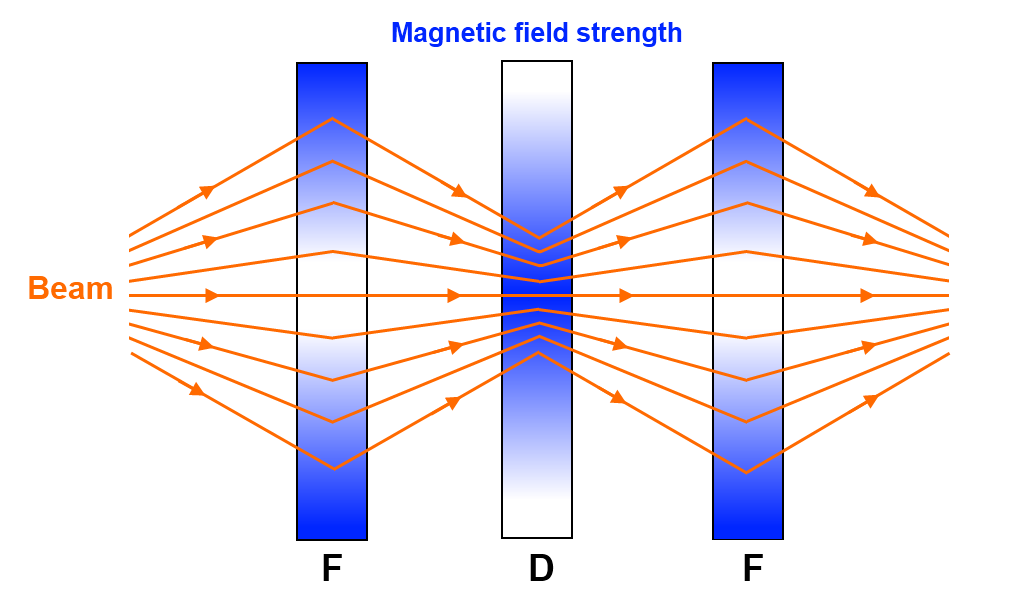}
			\caption{Strong focusing with alternating quadrupoles.}
			\label{fig:quad3}
		\end{figure}

		\textit{Figure~\ref{fig:quad3}} shows how the quadrupole magnetic field exerts a focusing force in one transverse plane while defocusing in the other. If the quadrupoles are arranged so the distance between magnets is less than or equal to twice the focal length, then we achieve overall focusing in both planes \cite{syphers}.
		
\begin{figure}[!htb]
		\centering
			\includegraphics[width=1.0\linewidth]{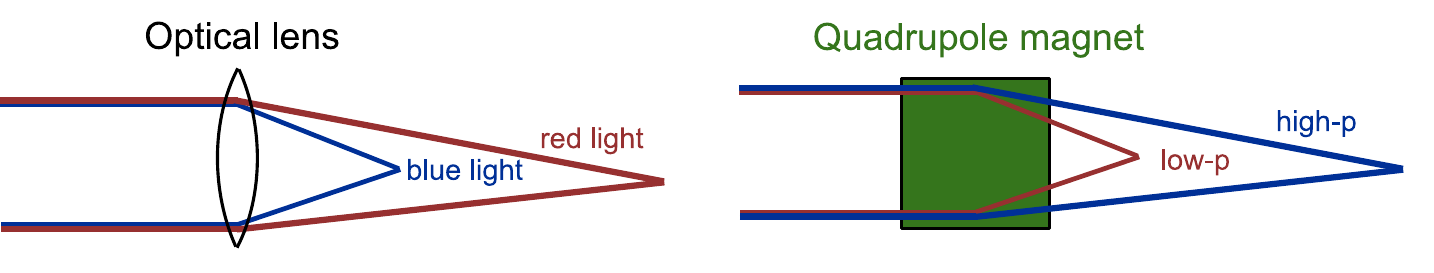}
			\caption{Chromatic aberration in optics and beam physics.}
			\label{fig:chromabb}
		\end{figure}

		Recall that the magnetic force on a charged particle depends on the particle's momentum, leading to the dispersion effect with dipoles; a dipole loses bending ability as the particle momentum increases\footnote{This effect is called ``magnetic rigidity'' and is expressed as $B\rho=\frac{p}{q}$ for particle momentum $p$ and charge $q$.}. Similarly, a quadrupole magnet loses focusing ability for higher-momentum particles. This effect is known as \textit{chromatic aberration}, which means that the focal length of a quadrupole magnet depends on the momentum of the beam particles. This effect also exists in optical physics, where the focal length of a lens depends on the color of the light passing through it. The analogy between beam and optical chromatic aberration is shown in \textit{Figure~\ref{fig:chromabb}}.
		
		The result of chromatic aberration is that the \textit{tune} of a given particle depends on its momentum. We can write an equation relating the tune spread of the beam to its momentum spread, shown in \textit{Equation~\ref{eq:chromaticity}}, and the proportionality constant is known as the \keyterm{\gls{chromaticity}} $\xi$.
		
		\begin{equation} \label{eq:chromaticity}
			\frac{\Delta\nu}{\nu}=\xi\frac{\Delta p}{p}
		\end{equation}
		
		\marginpar{\keyterm{\Gls{chromaticity}} is the dependence of the beam tune spread on the momentum spread.}
		
	\section{Sextupoles}
		
\begin{figure}[!htb]
		\centering
			\includegraphics[width=.45\linewidth]{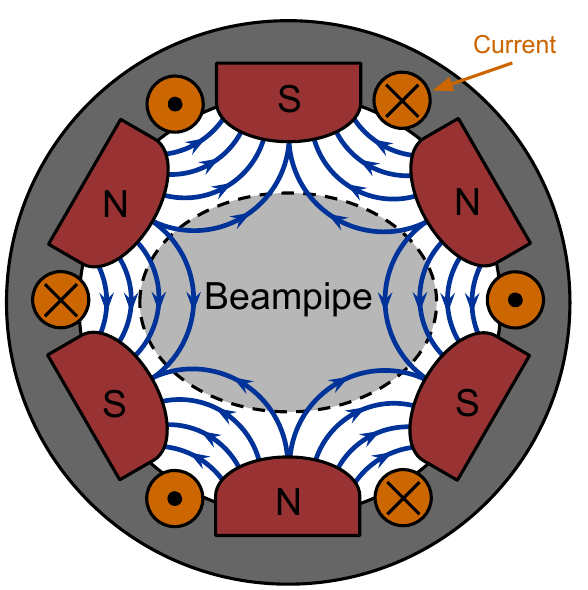}
			\includegraphics[width=.45\linewidth]{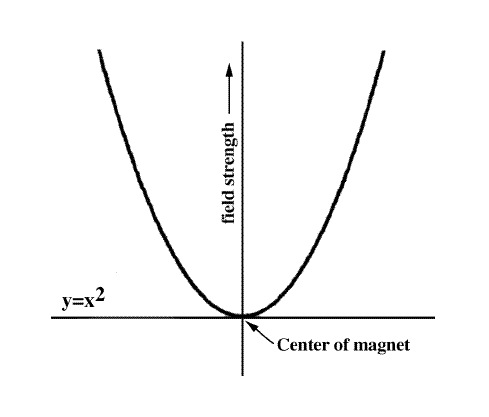}
			\caption{Sextupole and the transverse dependence of its field.}
			\label{fig:sext}
		\end{figure}

		\keyterm{\Gls{sextupole}} magnets provide a field whose strength varies with the square of the displacement from the center of the beam pipe, shown in \textit{Figure~\ref{fig:sext}}. This field dependence allows us to control the chromaticity effect by compensating for the loss of quadrupole focusing strength at higher beam momentum.
		
		\marginpar{\keyterm{\Glspl{sextupole}} are used to compensate for the chromaticity of the quadrupoles.}
		
\begin{figure}[!htb]
		\centering
			\includegraphics[width=.45\linewidth]{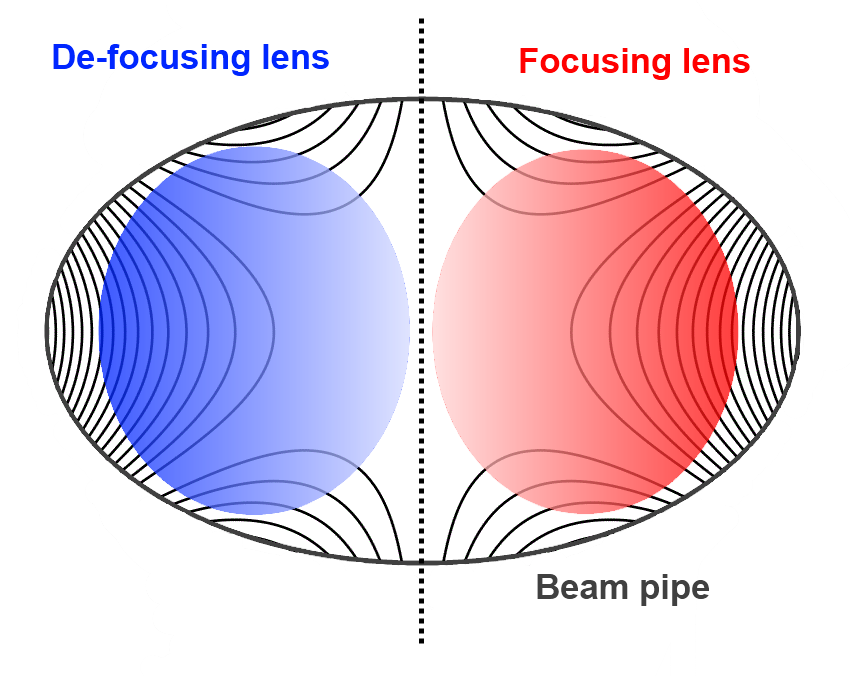}
			\includegraphics[width=.45\linewidth]{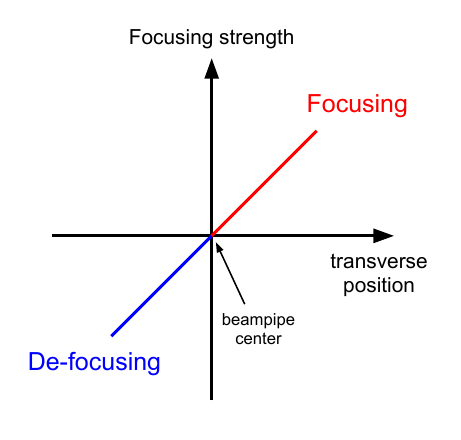}
			\caption{Sextupole as two lenses to control chromaticity.}
			\label{fig:sextfocusing}
		\end{figure}

		As it turns out, the \textit{focusing strength} of a sextupole magnet depends linearly on the transverse displacement from its center \cite{lee}, plotted on the right of \textit{Figure~\ref{fig:sextfocusing}}. Because of the dispersion of the dipole magnets, higher-energy particles will end up outside the center of the beam pipe, and lower energy particles will be on the inside. The sextupole thus focuses the higher-momentum particles more\footnote{This is the case below the transition energy. The situation is reversed after transition, so there is typically a deliberate chromaticity sign flip at transition in synchrotrons. Transition is explained in more detail in \textit{Chapter~\ref{chap:con_physics}} of this book.}, which compensates for the loss of quadrupole focusing strength. So if we place sextupole magnets in high-dispersion regions of the synchrotron, we can control the chromaticity by changing the current through their coils.
		
		\textit{Figure~\ref{fig:sextfocusing}} shows how the sextupole's focusing strength changes sign on either side of its center. Therefore we can think of a sextupole as two lenses on either side of the beam pipe center, one focusing and one defocusing. The sextupole focuses beam on one side, defocuses beam on the other, and the focusing strength depends on the transverse displacement. 
			
	\section{Octupoles}
\begin{figure}[!htb]
		\centering
			\includegraphics[width=.45\linewidth]{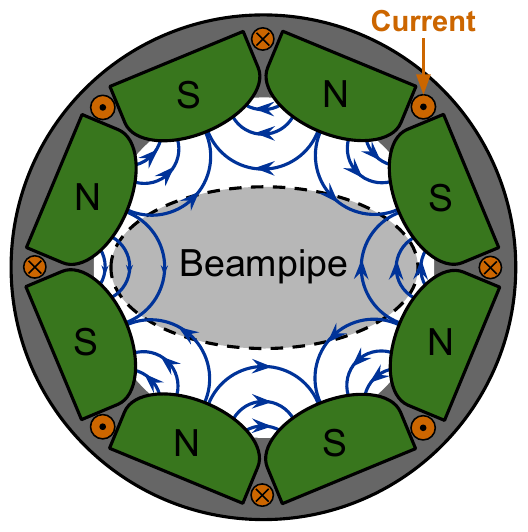}
			\includegraphics[width=.45\linewidth]{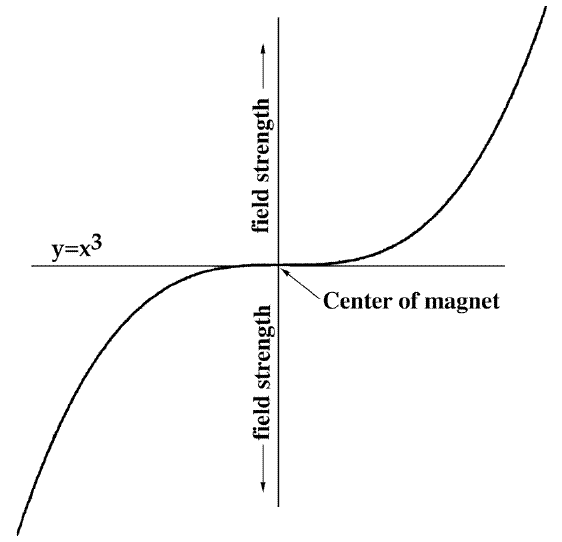}
			\caption{Octupole and the transverse dependence of its field.}
			\label{fig:octupole}
		\end{figure}

		The field strength of an \keyterm{\gls{octupole}} depends on the cube of the distance from the beam pipe center, illustrated in \textit{Figure~\ref{fig:octupole}}. Octupoles allow for control of how the tune spread depends on the amplitude of the betatron oscillations. We use octupole magnets in storage machines; rapid-cycling machines like the Booster do not typically need them.
		
		\marginpar{\keyterm{\Glspl{octupole}} control the way the frequency of betatron oscillations depend on their amplitude.}
		
	\section{Multipole Expansion}
		Real magnets do not actually fit into any of the above categories; their magnetic fields are combinations of many different field shapes. However, we can mathematically decompose any complicated magnetic field into a sum of the ``ideal'' magnetic fields mentioned above. This mathematical trick is called a ``multipole expansion,'' and is demonstrated for a general one-dimensional magnetic field $H(x)$ in \textit{Equation~\ref{eq:multipole}}.\footnote{This is a specific application of a \textit{Taylor power series} expansion, wherein any arbitrary function can be expressed as an infinite polynomial sum. For simplicity, \textit{Equation~\ref{eq:multipole}} is applied at the center of the beam pipe where $x = 0$.}
		
		\begin{equation} \label{eq:multipole}
			H(x)=\sum_{n=0}^{\infty}a_nx^n = a_0+a_1x+a_2x^2+...
		\end{equation}
		
		Each $a_n$ coefficient shows the strength of each ``ideal'' field component that makes up $H$: $a_0$ is the dipole strength, $a_1$ quadrupole, $a_2$ sextupole, $a_3$ octupole, $a_4$ decapole, and so on \textit{ad infinitum}. This is why sextupoles and octupoles are referred to as ``higher-order'' magnets, because their order $n$ exceeds that of the basic dipole and quadrupole. A magnet is typically named for the most dominant multipole field term, even though other $a_n$ coefficients may also be present.
	
	\section{Specialized Devices}
		We now introduce several specialized magnets that do not fall under the categories mentioned above. We need these magnets to accomplish specific tasks like beam injection and extraction, as well as fulfilling the role of multiple magnet types simultaneously. 
	
		\subsection{Combined Function Magnets}
\begin{figure}[!htb]
			\centering
				\includegraphics[width=.8\linewidth]{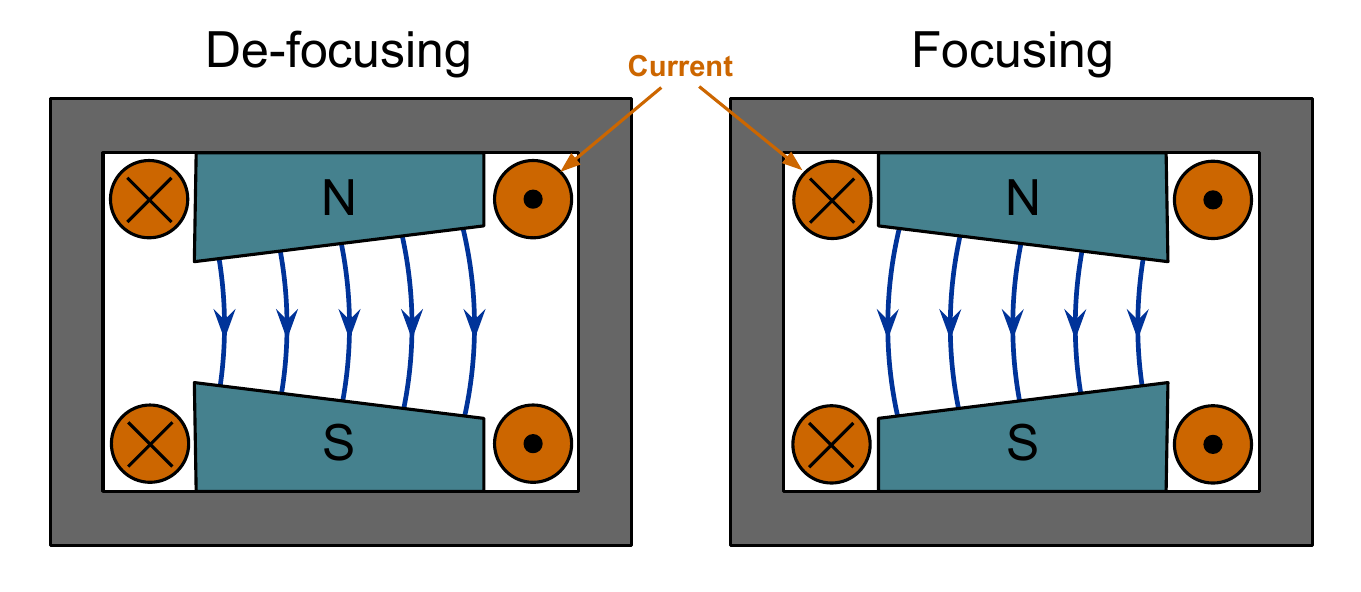}
				\caption{Combined-function magnet.}
				\label{fig:gradientmag}
			\end{figure}

			Not all accelerators use separate magnets for bending and focusing beam. The main magnets in Booster and Recycler are known as \keyterm{\glspl{combinedfunctionmagnet}}, or ``gradient magnets.'' These simultaneously bend and focus the beam, removing the need for separate dipoles and quadrupoles throughout the accelerator. \textit{Figure~\ref{fig:gradientmag}} shows a cross-section of both a focusing and defocusing combined-function magnet.
	
			\marginpar{\keyterm{\Glspl{combinedfunctionmagnet}} perform the roles of focusing and bending beam simultaneously.}
		
		\subsection{Lambertson and Septum Magnets}
			A septum magnet has two distinct apertures, one without any magnetic field and one with a dipole field. These magnets\footnote{We usually refer to these as ``septa,'' which is the plural of the word ``septum.''} provide the bending force for beam injection and extraction. The field-free aperture allows beam to pass through without being deflected, while the field aperture directs beam out of the magnet at an angle.
			
			\marginpar{
			\centering
			\includegraphics[width=1.0\linewidth]{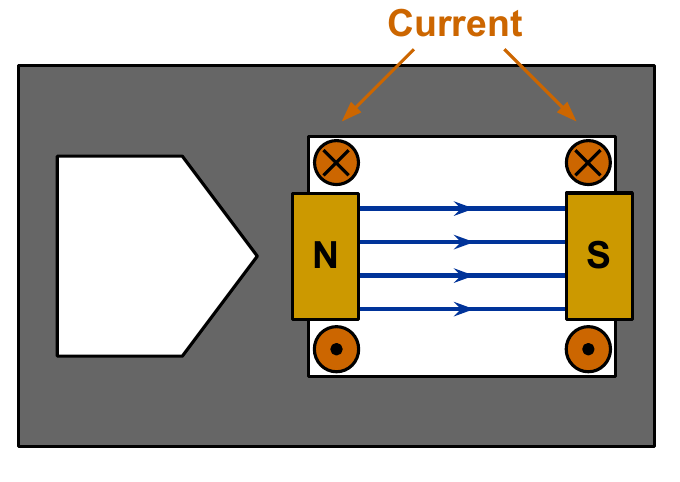}
			\captionof{figure}[Lambertson magnet]{The \keyterm{\Gls{Lambertson}} magnet has an extra field-free aperture to facilitate beam extraction}
			\label{fig:lambertson}
			}
			
			\keyterm{\Glspl{Lambertson}} are a type of septum magnet whose dipole field is oriented to bend beam perpendicular to the direction between apertures. If the field region of the Lambertson is not powered, however, no bend occurs and beam is not deflected by the Lambertson. \textit{Figure~\ref{fig:lambertson}} shows a simple model of a Lambertson magnet.
			
		\subsection{Kicker Magnets}
			\keyterm{\Gls{kicker} magnets} are dipoles that can turn on and off very quickly. The field strength is not nearly as high as main dipole magnets, so kickers only provide a small transverse deflection to the beam for injection or extraction.
			
			\marginpar{\keyterm{\Gls{kicker} magnets} are fast-acting dipoles that induce a small transverse oscillation.}
			
			The kicker system uses a power supply to charge a pulse-forming network, or \textit{PFN}. The PFN is a capacitor bank or large spool of cable that stores energy to deliver a quick high-voltage pulse when the kicker is triggered. When it is time for the kicker to fire, a high-voltage switch known as a ``thyratron'' tube discharges the PFN's voltage to the magnet. Because the kicker's magnetic field is only present for a short time when it is triggered, kickers are known as ``pulsed'' devices.
			
			These magnets cause a small transverse oscillation in the beam to jump the gap from the field-free to the field region of the Lambertson or septum magnet for beam extraction. They also cancel the beam oscillation after injection.
		
		\subsection{Permanent Magnets}
			Permanent magnets do not require power to provide a magnetic field, nor do they require water-cooling systems like the electromagnets. Composed of ferrite bricks stacked into particular shapes, these magnets can fulfill the role of any magnet we've talked about so far. Permanent magnets are used in the MI-8 line and the Recycler, where the beam energy (and thus the magnetic field) does not need to change.
	
		\subsection{Electrostatic Septa}
			\index{Electrostatic Septum}The \keyterm{\gls{electrostaticseptum}} uses electric fields to split incoming beam into multiple streams. Grounded wires in the center of the septum provide a potential difference with respect to the high voltage outer walls, creating a transverse electric field that pushes beam to either side of the beam pipe center. This is useful in Switchyard to split beam to multiple experiments using septa and Lambertson combinations; by adjusting the position of the entire septum device, we can change the amount of beam split on either side of the wires, which provides fine control over beam intensity delivered to each experiment. 
			
			\marginpar{The \keyterm{\gls{electrostaticseptum}} is a device that uses electric fields to split beam into multiple beams.}
			
			We also use a modified electrostatic septum in the Main Injector for resonant extraction out to Switchyard. This septum only generates an electric field on one side of the wires, and is field-free on the other side. The septum splits small amounts of circulating beam with the field aperture to slowly ``peel'' particles off and extract them, while leaving the rest of the beam to circulate through the other aperture.
			
			\begin{figure}[!htb]
			\centering
				\includegraphics[width=.8\linewidth]{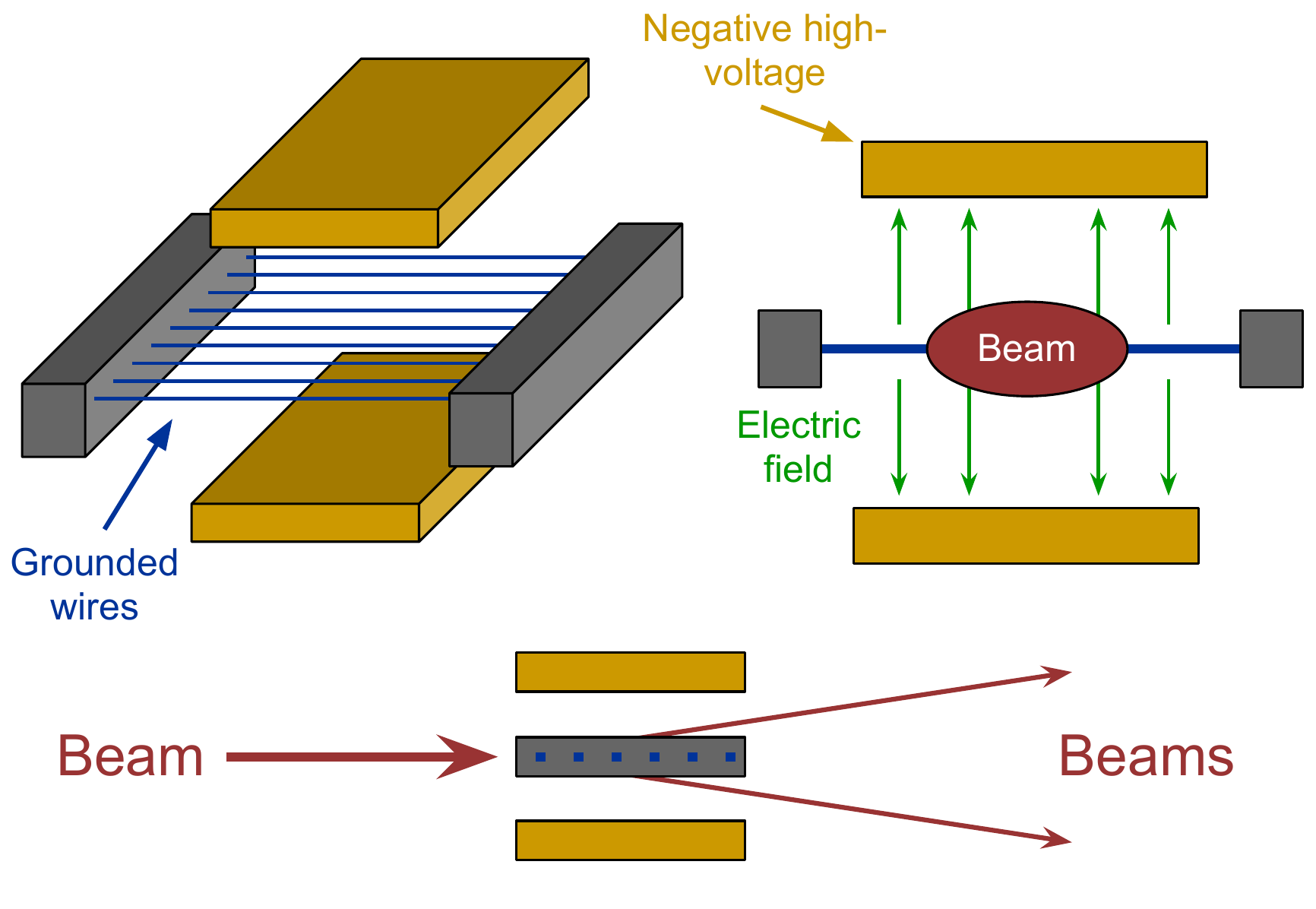}
				\caption{Electrostatic septum and beam splitting}
				\label{fig:esep}
			\end{figure}
			
	\section{Magnetic Hysteresis}
	
		The ferromagnetic material in the magnets retains a residual field when we change the power supply current. This means that a given current will not always provide the same magnetic field; the recent magnetic history is also a factor, and this effect is known as \textit{magnetic hysteresis}. \textit{Figure~\ref{fig:hysteresis}} shows the magnetic behavior of a ferromagnetic material as a function of the supplied current. This graph is colloquially known as a ``hysteresis curve.'' Notice that as the applied current increases, the magnetic field of the ferromagnetic material $B$ increases until it reaches saturation. If we then decrease the current, $B$ predictably decreases as well, but it does not come back to where we started. The new $B$ value represents the residual magnetism left in the ferromagnetic material. This effect is not good for accelerators, because we want a given current to produce the same magnetic field every time for consistency. 
		\marginpar{
			\centering
			\includegraphics[width=1.0\linewidth]{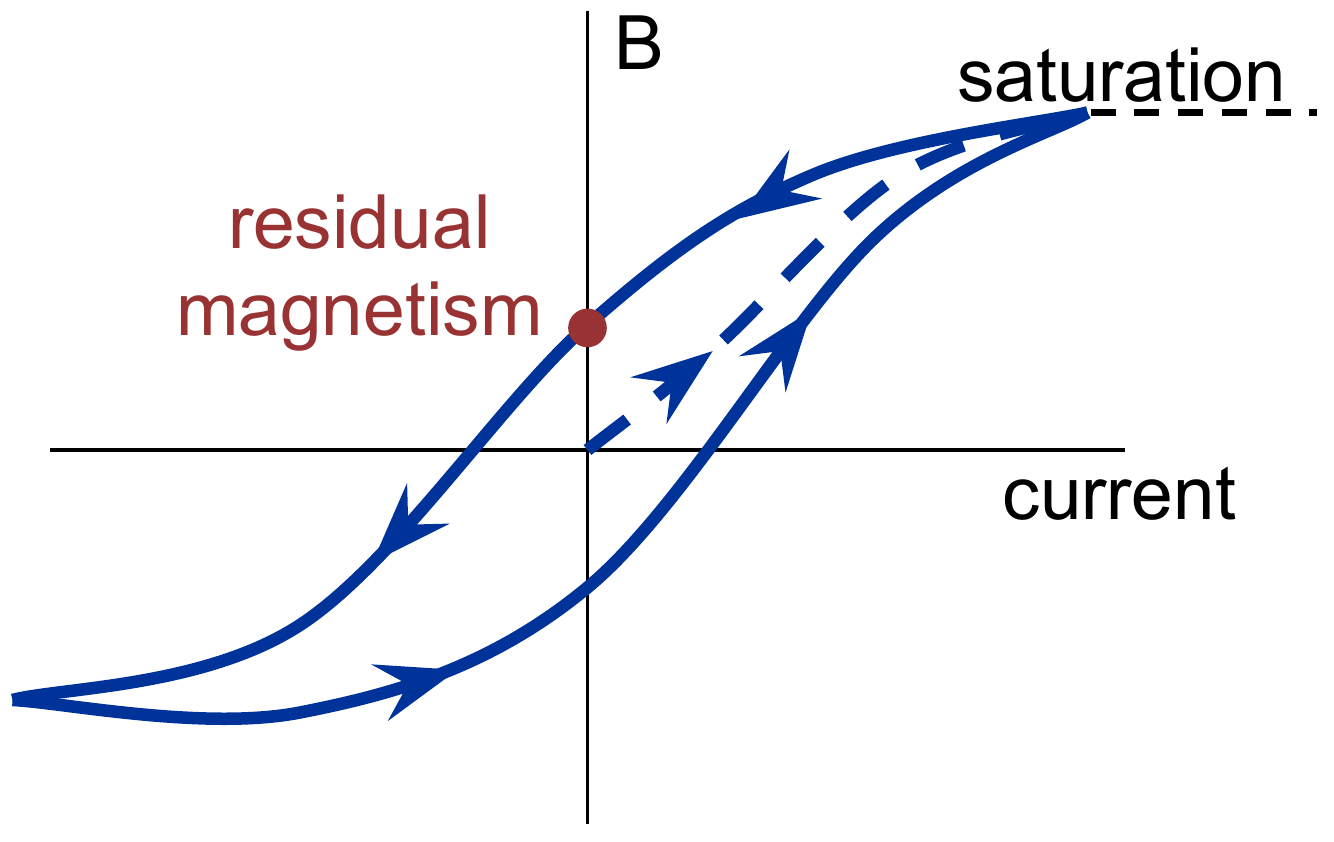}
			\captionof{figure}{Hysteresis of the magnetic field of a ferromagnetic material}
			\label{fig:hysteresis}
			}
		What this means for accelerator operations is that we sometimes need to cycle the fields in our magnets to reset the hysteresis of the iron. For example, we run the powered quads in the \Mev{400} and \gev{8} lines through a cycle when we turn them back on from an access; this involves changing the currents up and down several times, then returning the currents to their nominal values. This process ensures that the nominal magnet currents will produce the same field as they did before the machine was turned off.
		\chapter{Radio Frequency Systems} \label{chap:con_RF}
	\Gls{rf} refers to the high-frequency electromagnetic fields that accelerate and manipulate charged particle beams. These fields resonate as standing waves in enclosed devices called ``cavities,'' and oscillate quickly to push beam as it passes. The frequency of the field oscillation is usually in the radio portion of the electromagnetic spectrum, hence the term ``Radio Frequency.''

	This chapter begins with the motivation for using high-frequency electromagnetic fields, instead of magnetic or electrostatic fields, to accelerate the beam. We will then describe the cavities which resonate with the standing waves, and how the beam is accelerated by the contained fields. Next, we will show how the RF fields maintain longitudinal beam stability for stable acceleration. Finally, we will describe the systems that provide the RF power to the cavities, and how these systems work together for consistent and stable accelerating fields. 

	\section{Electric Fields and Kinetic Energy}
		Recall the \textit{Lorentz force equation}, which shows the force on a particle of charge $q$ with velocity $\vec{v}$ due to an electric $\vec{E}$ or magnetic $\vec{B}$ field. The electric field, velocity, and magnetic fields are all vector functions, as indicated by the arrows above the variables.
		
		\begin{equation} \label{eq:Lorentz}
			\vec{F}=q(\vec{E}+\vec{v} \times \vec{B})
		\end{equation}
		
		\textit{Equation~\ref{eq:Lorentz}} means that the force due to a \textit{magnetic} field must be perpendicular to the particle's velocity (i.e., a \textit{transverse} force)\footnote{In fact, the magnetic force must be perpendicular to both the velocity and the magnetic field; this is a consequence of the Lorentz force term for the magnetic field $q(\vec{v} \times \vec{B})$. For any two vectors $\vec{a}$ and $\vec{b}$ the cross product $\vec{a} \times \vec{b}$ is a vector that is perpendicular to both $\vec{a}$ and $\vec{b}$. In other words, $\vec{a} \times \vec{b}$ is orthogonal to the plane shared by $\vec{a}$ and $\vec{b}$}.
		
		However, there is no explicit velocity dependence for the \textit{electric} force. This means that an electric field can exert a force parallel to the particle velocity (i.e. a \textit{longitudinal} force). 
		
		A static magnetic field cannot increase the kinetic energy of the beam, because its force is always perpendicular to the particle velocity. This is a consequence of the \textit{Work-Energy Theorem}, which states that a system's kinetic energy change is equal to the work done on that system. However, only forces parallel to system motion can do work.\footnote{The work on a system is defined as $W=\int \vec{F}\cdot d\vec{l}$ for a force vector $\vec{F}$ and system displacement $\vec{l}$. The vector dot product shows that only force components parallel to $\vec{l}$ contribute to the work.} Thus, since a static magnetic force must be perpendicular to particle velocity, it cannot do any work on the particle. If the magnetic force cannot do any work, then it cannot change the kinetic energy of the particle. This is why we use \textit{electric} fields to increase particle kinetic energy in an accelerator.\footnote{Due to \textit{Faraday's Law}, it is actually possible to increase particle kinetic energy with a \textit{changing} magnetic field. In other words, a time-changing magnetic flux through a given surface is equivalent to a static electric flux through the boundary of that surface: $\oint \vec{E}\cdot d\vec{l}=-\frac{d}{dt}\iint \vec{B}\cdot d\vec{S}$. For example, the Betatron particle accelerator uses a magnetic field that oscillates in time to accelerate beam.} 
		
\begin{figure}[!htb]
		\centering
			\includegraphics[width=0.8\linewidth]{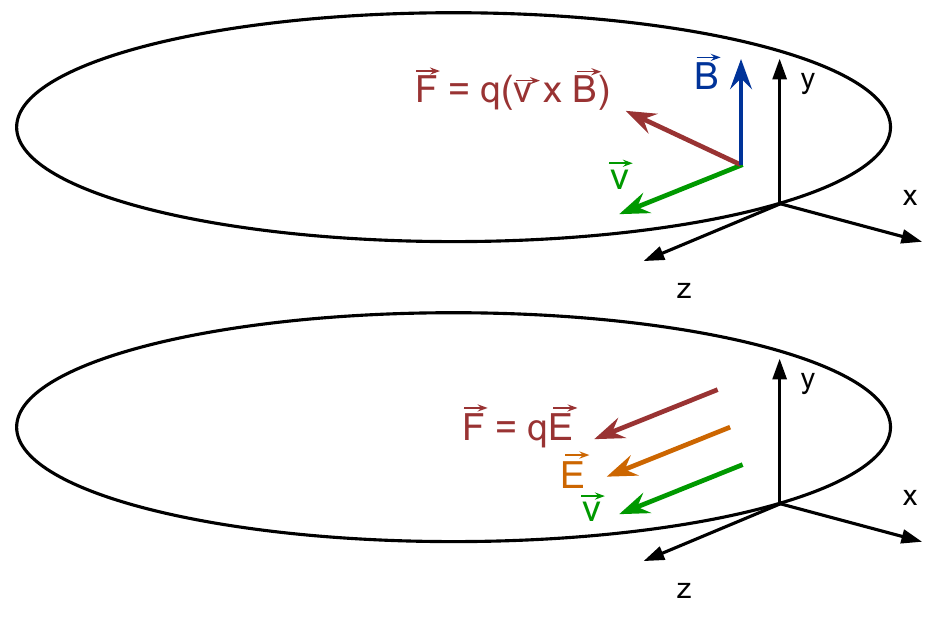}
			\caption{Electric vs. magnetic force on a charged particle.}
			\label{fig:EBforce}
		\end{figure}

		\textit{Figure~\ref{fig:EBforce}} illustrates the electric and magnetic forces on a moving charged particle as defined by \textit{Figure~\ref{eq:Lorentz}}. While the \textcolor{nal-red}{magnetic force} must be perpendicular to the \textcolor{nal-blue}{magnetic field} and the \textcolor{nal-green}{particle velocity}, the electric force has no such limitation. Thus we can choose an electric field direction such that the \textcolor{nal-orange}{electric force} points parallel to the particle velocity.
		
	\section{RF Cavities}
\begin{figure}[!htb]
		\centering
			\includegraphics[width=0.9\linewidth]{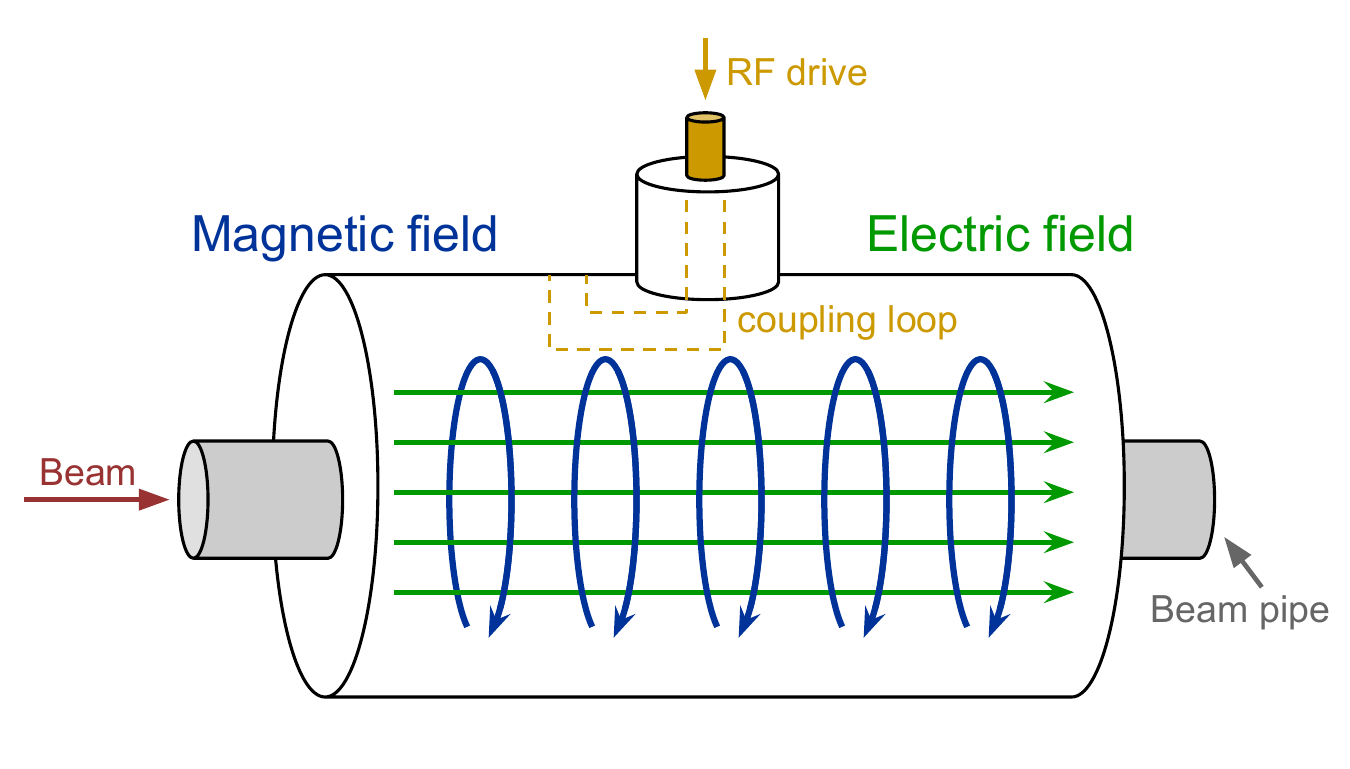}
			\caption{Simple ``pillbox'' RF Cavity.}
			\label{fig:RFcav}
		\end{figure}

		It is infeasible to use static electric fields in high-energy accelerators, because the necessary high voltages would cause electrostatic breakdown and arcing.\footnote{For example, to statically accelerate particles to the energy achieved by the Main Injector, it would require a voltage potential of \e{120e9} volts. The dielectric strength of air is about \e{3e6} volts per meter, which is the potential density required to initiate electrical arcing in air. Thus a ``static'' Main Injector accelerator would need to have a separation of \textit{40 kilometers} between electrodes to avoid arcing.}
		Instead, most modern high-energy accelerators use hollow conducting structures known as \keyterm{\glspl{RFcavity}} that contain electromagnetic waves. These cavities are electromagnetically resonant, in that they efficiently store oscillating energy at particular frequencies. Fermilab's RF cavities use standing waves generated from the interference between traveling and reflected waves inside the cavity. A standing wave doesn't appear to move through the cavity, because only the amplitude of the wave oscillates. \textit{Figure~\ref{fig:RFcav}} shows a model of a simple RF cavity, known as a ``pillbox cavity.'' The \textcolor{nal-blue}{magnetic} and \textcolor{nal-green}{electric} fields are shown at a particular point in the cycle, but remember that the field amplitudes actually oscillate in time at high frequency.
		
		\marginpar{\keyterm{\glspl{RFcavity}} resonate with electromagnetic waves that longitudinally manipulate and accelerate the beam.}
		
\begin{figure}[!htb]
			\centering
			\includegraphics[width=0.9\linewidth]{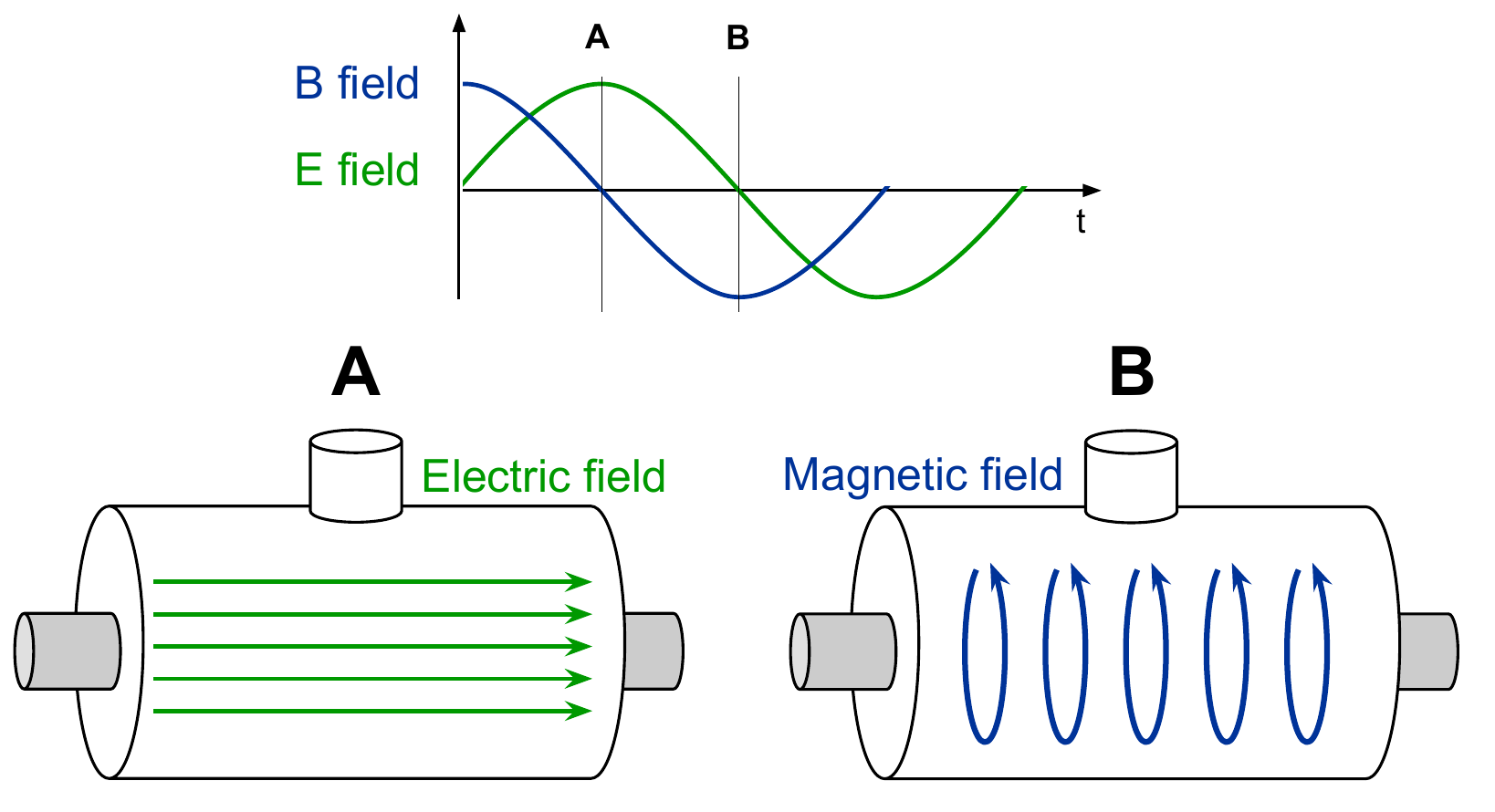}
			\caption{Timing difference between magnetic and electric fields.}
			\label{fig:cavfields}
		\end{figure}

		For a pillbox cavity, we choose the geometry and RF frequency so the electric field points longitudinally, and the magnetic field has a minimal transverse effect on the beam. Inside the resonant cavity, energy moves back-and-forth between the electric and magnetic fields; in other words, the magnetic field inside the cavity is zero when the electric field is at its maximum. \textit{Figure~\ref{fig:cavfields}} shows the maximum electric field at one point in time \textbf{A}, then the maximum magnetic field at a time \textbf{B} one quarter-cycle (90 degrees or $\frac{\pi}{2}$ radians) later.
		
	\section{Cavity Q and Ferrite Bias Tuning}
		RF cavities must be very efficient oscillators to keep power loss as low as possible and to maximize the field delivered to the beam. However, this means that they only work at frequencies very close to their resonance. For synchrotrons, where the RF frequency increases throughout the acceleration process, we must be able to change the resonant frequency of the RF cavity as needed.
		
		The \textit{quality factor} or ``Q'' provides a quantitative definition of the energy efficiency of an oscillator, and is defined as the resonant frequency $f_0$ multiplied by the ratio of stored energy to lost power \cite{circuits}:
		
		\begin{equation} \label{eq:Qenergy}
			Q = 2\pi f_0 \frac{energy\:stored}{power\:lost}
		\end{equation}
		
		RF cavities are high-$Q$ resonators, because there is very little power lost in the cavity. An alternate expression for Q shows how this is related to the frequency width of the cavity response. \textit{Equation~\ref{eq:Qfreq}} expresses Q as the ratio of the resonant frequency $f_0$ to the frequency width at half-maximum power $\Delta f$. 
		
		\marginpar{
			\centering
			\includegraphics[width=1.0\linewidth]{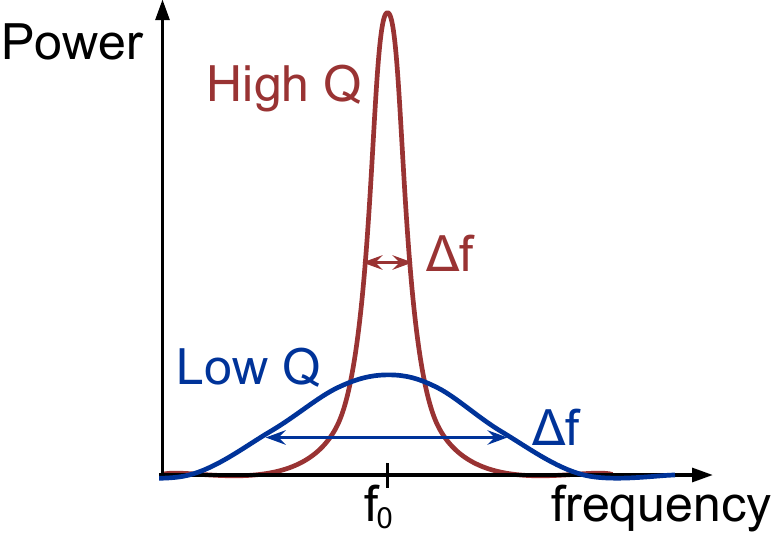}
			\captionof{figure}{Resonator quality factor, high vs. low Q}
			\label{fig:highQlowQ}
		}
		
		\begin{equation} \label{eq:Qfreq}
			Q = \frac{f_0}{\Delta f}
		\end{equation}
		
		This means that high-Q resonators like RF cavities have a narrow frequency response. In other words, RF cavities only resonate efficiently in a very limited frequency range.  If we try to drive an RF cavity at a frequency that is not close to its resonance, it will take much more power to achieve a strong field. \textit{Figure~\ref{fig:highQlowQ}} shows that a high-Q response in frequency space has a correspondingly lower frequency width.
		
		This trade-off between power efficiency and narrow frequency range is acceptable for linear accelerators, where the cavity frequency does not change. However, synchrotrons require the RF frequency to increase with the beam energy, so we must have a way of changing the resonant frequency of the cavities. 
		
		In the Main Injector and \index{Booster}Booster, we accomplish this by attaching small coaxial RF transmission lines to the RF cavities that are loaded with ferromagnetic material. These \keyterm{\glspl{ferritetuner}} are coupled to the cavity so they become part of the resonating volume. The ferrite that fills the tuners effectively changes the inductance of the entire system, thus altering the resonant frequency of the cavity. We apply a large amount of current, usually several thousand amps, through the center of the tuner. This creates a biasing field that changes the magnetic permeability of the ferrite depending on how much current we apply. By changing the permeability of the ferrite, we change the inductance of the resonant volume in the cavity and tuner. 
		
		\marginpar{\keyterm{\Glspl{ferritetuner}} are small coaxial transmission lines loaded with ferrite that change the resonant frequency of RF cavities in synchrotrons.}
		
\begin{figure}[!htb]
		\centering
			\includegraphics[width=.7\linewidth]{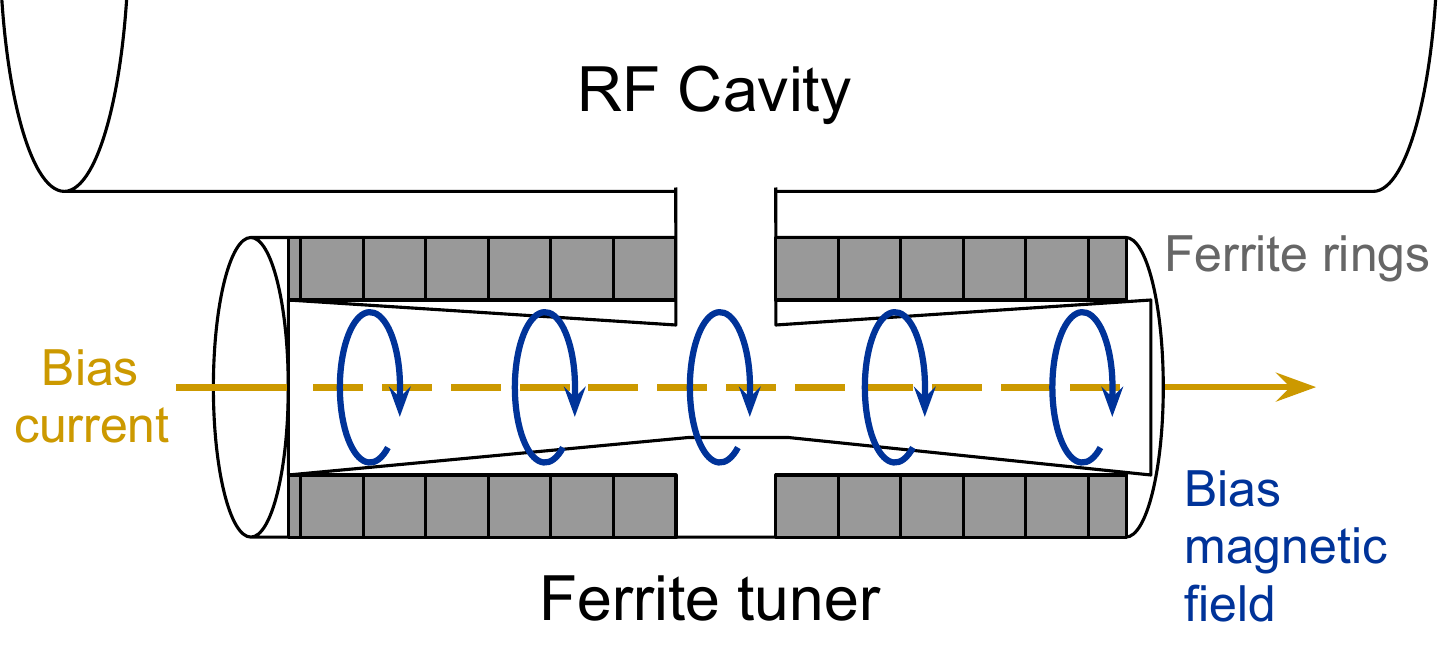}
			\caption{RF cavity with attached ferrite tuner.}
			\label{fig:biastuner}
		\end{figure}

		Therefore, by changing the biasing current applied to the tuners, we can actively control the resonant frequency of the entire RF cavity, which allows for the changing RF frequency we need in a synchrotron. \textit{Figure~\ref{fig:biastuner}} shows a model of a ferrite tuner waveguide attached to an RF cavity.
		
	\section{RF Phase and Synchronicity}
\begin{figure}[!htb]
		\centering
			\includegraphics[width=.8\linewidth]{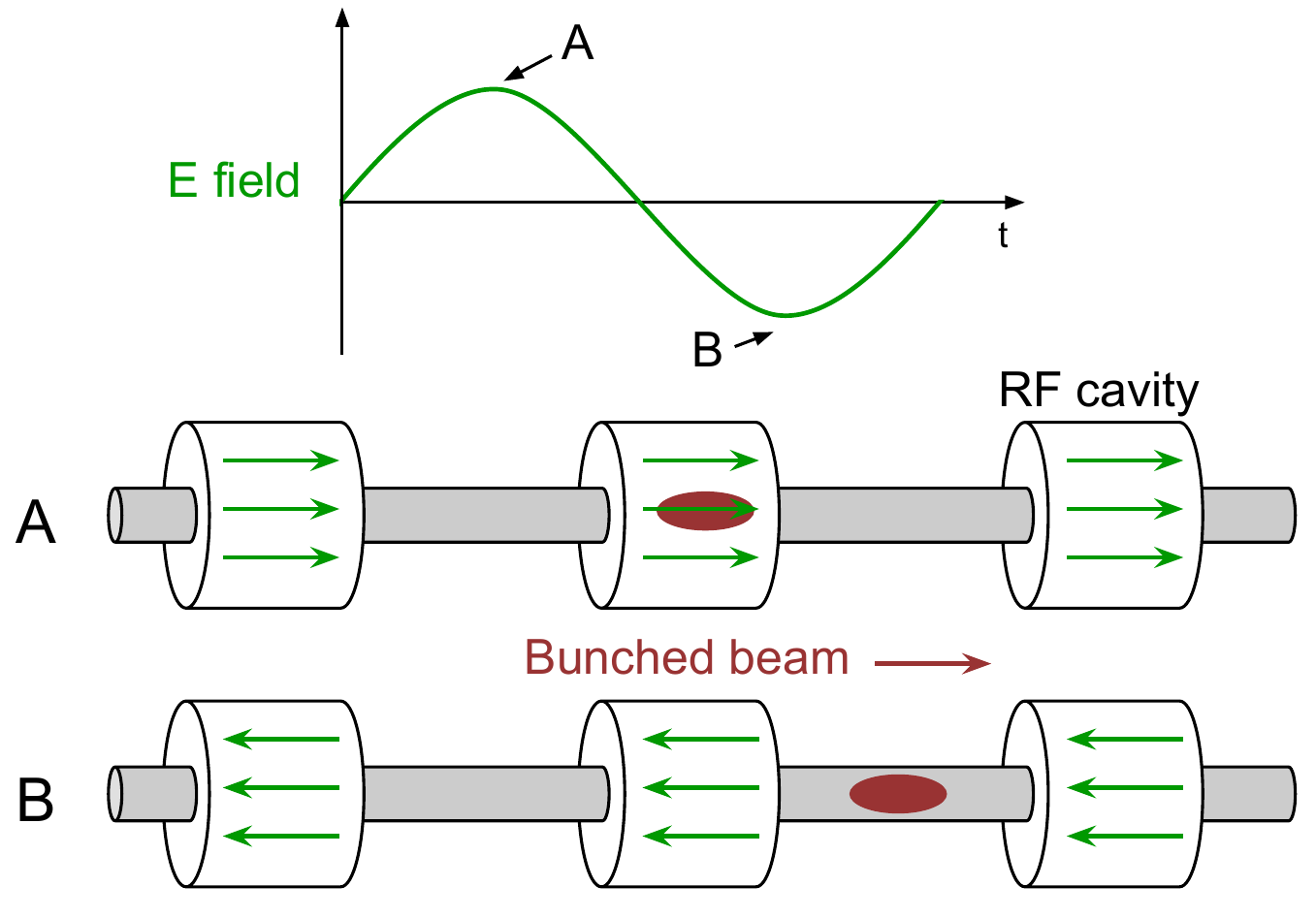}
			\caption{Simple linear accelerator and the importance of RF phase.}
			\label{fig:linacmodel}
		\end{figure}

		The RF oscillations must match precisely with the periodic arrival of beam in the cavity to achieve stable acceleration. Since the cavity's fields oscillate in time, their force direction changes throughout the cycle. Even though the electric field is parallel to the beam velocity, it points in the opposite direction during half the cycle. Thus the timing of the RF oscillations must match with the arrival of particles so they only see the electric field when it is pointing forward. Particle arrival time with respect to the RF cycle is known as the RF \keyterm{\gls{phase}}. By controlling the RF phase, we ensure that the forward-pointing electric field accelerates the beam; we also avoid deceleration during the other half of the RF cycle from backward-pointing field. For example, consider the model of a simple linear accelerator shown in \textit{Figure~\ref{fig:linacmodel}}: the RF frequency and distance between cavities have been chosen to prevent beam exposure to the electric field when it is pointing backwards. 
		
		\marginpar{The RF \keyterm{\gls{phase}} is the timing of the RF cycle with respect to particle arrival time in the cavity.}
		
		The example of \textit{Figure~\ref{fig:linacmodel}} shows the need for a particle to arrive at the cavity only when the electric field is pointing forward. For this to occur, the following relationship must be met between the RF frequency $f_0$, the distance between cavities $L$, and the particle velocity $v$, and integer $n$ \cite{introRF}:
		
		\begin{equation} \label{eq:synchronicity}
			f_0 L=nv
		\end{equation}
		
		In other words, the product of the RF frequency and the distance between cavities must be an integer-multiple $n$ of the particle velocity. \textit{Equation~\ref{eq:synchronicity}} is known as the ``synchronicity condition.'' If this condition is met, it ensures that the particle will always be between cavities when the RF field is pointing backwards, and thus the particle will only see accelerating fields. 
		
		The particle that meets the synchronicity condition of \textit{Equation~\ref{eq:synchronicity}} is known as the \keyterm{\gls{synchronousparticle}}, and it receives the correct energy increase from the RF as per the design of the accelerator. The phase at which the synchronous particle arrives in the cavity is known as the \keyterm{\gls{synchronousphase}}, and provides a reference point in RF system timing.
		
		\marginpar{The \keyterm{\gls{synchronousparticle}} arrives in the RF cavity at the \keyterm{\gls{synchronousphase}}, and is accelerated perfectly according to the accelerator design.}
		
		As the particle velocity increases, the product $f_0L$ must increase to maintain synchronicity. For a linear accelerator like \textit{Figure~\ref{fig:linacmodel}}, the distance between accelerating cavities $L$ progressively increases to follow the increase in particle momentum. However, circular accelerators use successive passes through the same RF cavities to accelerate beam; instead, the RF frequency $f_0$ increases to keep pace with the accelerating beam particles.
		
		Note that any slight deviation in particle momentum, and thus velocity, will violate the synchronicity condition; it is also inevitable that beam will have a non-zero energy and phase spread. We show in the next section that the RF is still able to maintain stability for non-synchronous particles, because the shape of the RF waveform can adjust particles toward the synchronous phase.
		
		\section{Phase Focusing and Stability}
		So far we have only considered the cavity electric field at maximum and minimum values, but its sinusoidal shape is important for beam stability and manipulation. Beam particles in a group do not all have the exact same momentum; in other words, we say that the beam has a non-zero \textit{momentum spread}. The RF must be able to maintain particle stability even with these small perturbations.
		
		Because of this momentum spread, each particle arrives in the cavity at a slightly different time, which corresponds to different points along the RF waveform. Thus, each particle will experience a different electric field strength. By controlling the RF phase, we can provide a restoring force that encourages errant particles to move toward the momentum of the synchronous particle. 
		
		\marginpar{
			\centering
			\includegraphics[width=.9\linewidth]{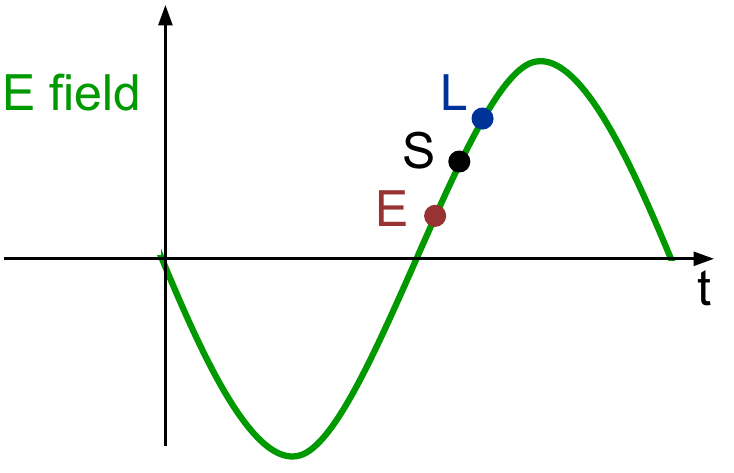}
			\captionof{figure}{Particle arrival time and phase focusing}
			\label{fig:ESL}
		}
		
		Consider \textit{Figure~\ref{fig:ESL}}, which shows how particles with different momenta will feel a different electric field magnitude based on their arrival time. \textit{Figure~\ref{fig:ESL}} shows an RF phase that simultaneously provides acceleration and longitudinal focusing to maintain stability. An early-arriving particle \textbf{E} has a higher momentum than the synchronous particle \textbf{S}, and will see less of an electric field; this encourages the particle to lag behind others, moving it toward the synchronous phase. Similarly, a late-arriving particle \textbf{L} has lower momentum, and sees a higher electric field that helps it catch up to the synchronous particle. Thus particles with too much momentum start lagging behind, and those with too little momentum speed up; this encourages all particles to move toward the synchronous phase in a longitudinally-focusing process known as \keyterm{\gls{phasefocusing}}. This process allows particles with momentum deviations to be corrected slightly with each pass through an RF cavity. 
		
		\marginpar{\keyterm{\Gls{phasefocusing}} is the process where the RF waveform shape creates longitudinal particle stability.}
		
		Note that in this example, every particle sees a positive electric field, so there is a net acceleration of the entire beam. The small adjustments mentioned above for phase focusing adjust the beam momentum for stability, but overall all particles are accelerated together by the electric field.
		
		The phase focusing process causes all non-synchronous particles to oscillate about the synchronous phase. Phase focusing corrects a momentum error, which leads to a phase error, which leads to another momentum error, and so on. We call this longitudinal motion ``synchrotron oscillation,'' and we describe this effect in detail in \textit{Chapter~\ref{chap:con_physics}}.
		
	\section{Buckets and Bunching}
		\marginpar{The \keyterm{\gls{bucket}} is the stable space created by the RF that can capture and accelerate beam.}
		%\marginpar{The RF \keyterm{\gls{bucket}} is the set of deviations in energy and phase from the synchronous particle that can still be phase focused.}
		
		There are limits to the phase focusing ability of the RF: a particle that deviates sufficiently from the synchronous phase or momentum cannot be pushed enough to maintain stable oscillation. We define this stable RF space that can phase focus the beam as the \keyterm{\gls{bucket}}\footnote{As we will show in \textit{Chapter~\ref{chap:con_physics}}, the bucket can be expressed as an area in ``phase space''.}. If a particle's energy and phase are within the bucket, then it can be phase focused into stable motion. \textit{Figure~\ref{fig:ESL}} corresponds to an \textit{accelerating bucket}, since every particle receives some positive kick from the electric field, and thus an increase in kinetic energy.
		
		\marginpar{The \keyterm{\gls{harmonicnumber}} for a synchrotron is the number of RF oscillations completed in the time it takes a particle to traverse one orbit. Thus the harmonic number is the maximum number of bunches a synchrotron can accelerate.}
		
		Consider the RF system of a synchrotron: the number of RF buckets in a machine is limited, because there are only so many RF oscillations in the time it takes a particle to make one orbit. The number of RF buckets for a given synchrotron is called the \keyterm{\gls{harmonicnumber}}, which is equal to the number of RF oscillations per revolution period.
		
		%The harmonic number is also the maximum number of bunches a synchrotron can accelerate at once.
		
		We can formulate a relationship between the harmonic number $h$, the RF frequency $f_0$, and the transit time $t$ (the time it takes a particle to make one lap around the accelerator). The units of $h$ are $\frac{RF cycles}{lap}$ and the units of $f_0$ are $\frac{RF cycles}{second}$. The transit time $t$ has units of $\frac{seconds}{lap}$, so \textit{Equation~\ref{eq:hnumber}} shows how to calculate $t$ using $h$ and $f_0$.
		
		\begin{equation} \label{eq:hnumber}
			h=tf_0=\frac{seconds}{lap}*\frac{RF cycles}{second}=\frac{RF cycles}{lap}
		\end{equation}
		
		\marginpar{\keyterm{\Glspl{bunch}} are discrete groups of beam created by interaction with the RF buckets.}
		
\begin{figure}[!htb]
		\centering
			\includegraphics[width=.5\linewidth]{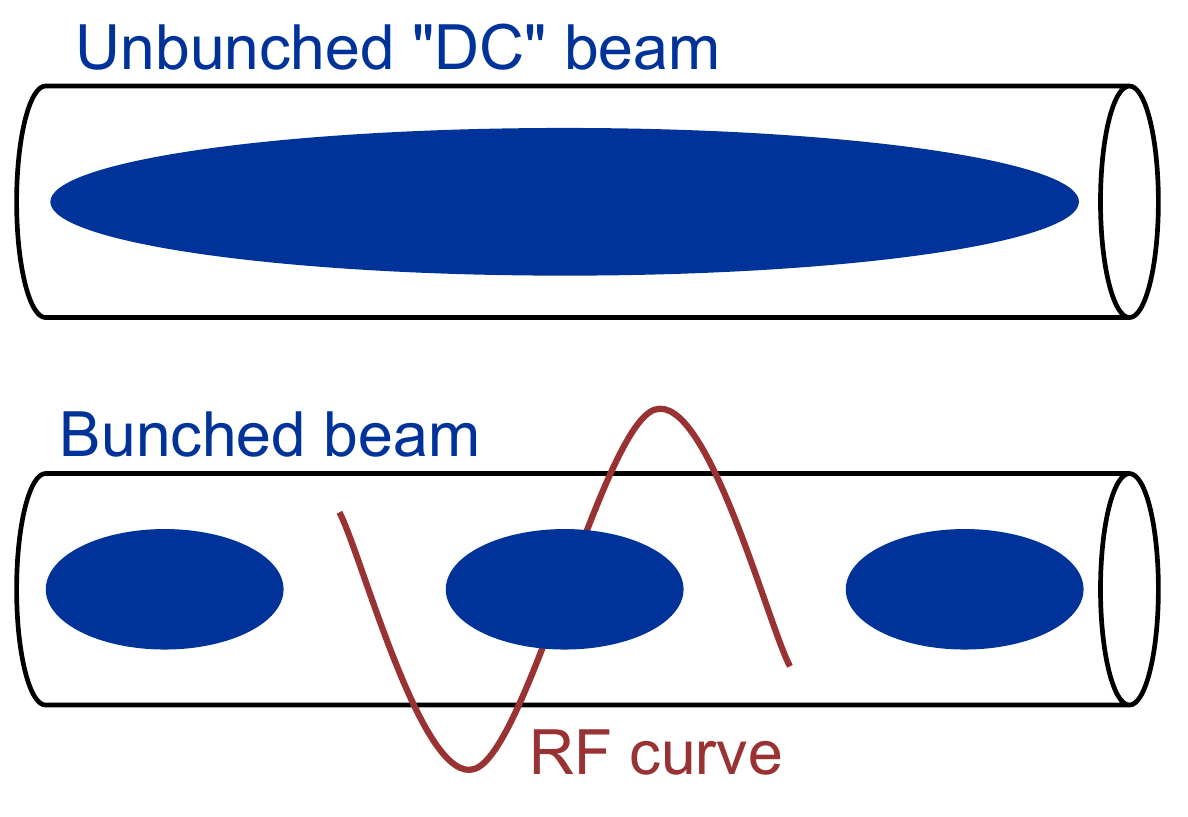}
			\caption{Beam bunching due to RF phase focusing.}
			\label{fig:bunching}
		\end{figure}

		The phase focusing process causes beam to collect into discrete packets, known as \keyterm{\glspl{bunch}}. \textit{Figure~\ref{fig:bunching}} shows the qualitative difference between unbunched (called ``DC'') beam, and beam bunched by interaction with the RF waveform. Since the bucket area is the only stable place for beam to exist, it is often said that the bunches fill the buckets. 
		
		Bunching occurs during normal acceleration, but \textit{Figure~\ref{fig:stationaryESL}} shows the particular synchronous phase for which there is no net beam acceleration, only bunching. This is typical when injected beam needs to be captured by the RF, or extracted to another machine. \textit{Figure~\ref{fig:stationaryESL}} illustrates this non-accelerating synchronous phase using the familiar three-particle example: notice that the early particle with too much momentum now sees a negative electric field that decelerates it slightly to match the synchronous particle. The bucket that corresponds to the synchronous phase in \textit{Figure~\ref{fig:stationaryESL}} is known as a \textit{stationary bucket}, as it provides no net beam acceleration.
		
		\marginpar{
			\centering
			\includegraphics[width=.9\linewidth]{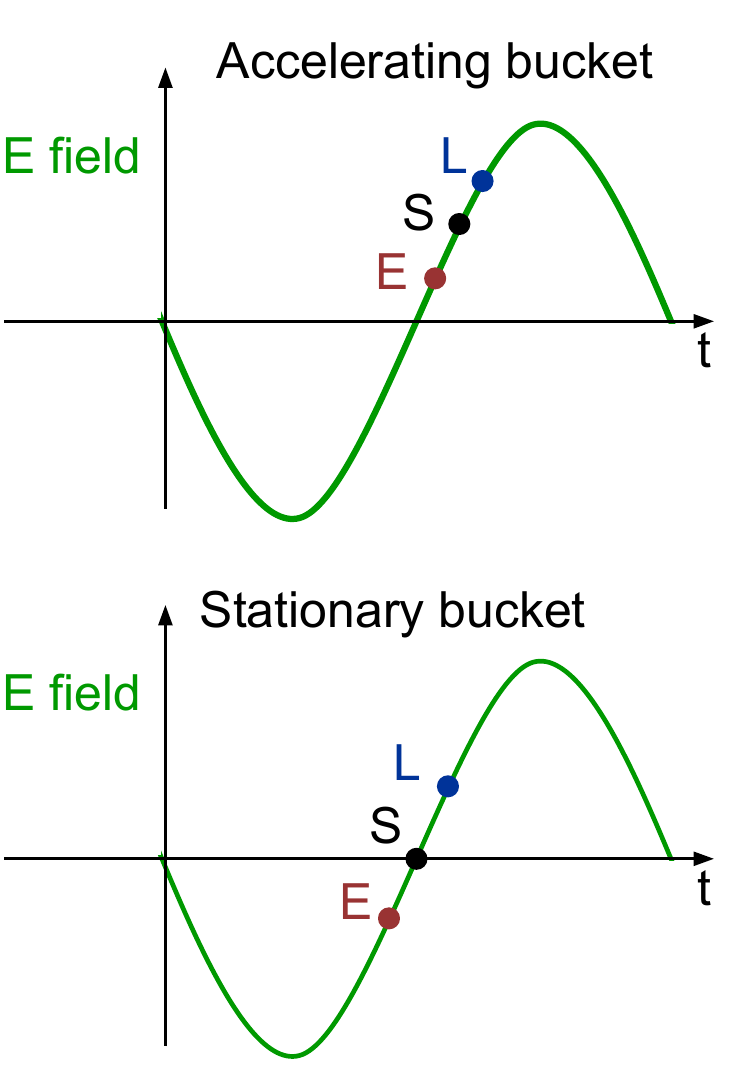}
			\captionof{figure}{Accelerating vs. stationary bucket}
			\label{fig:stationaryESL}
		}
	
	\section{RF Signal Generation and Amplification}
\begin{figure}[!htb]
		\centering
			\includegraphics[width=.8\linewidth]{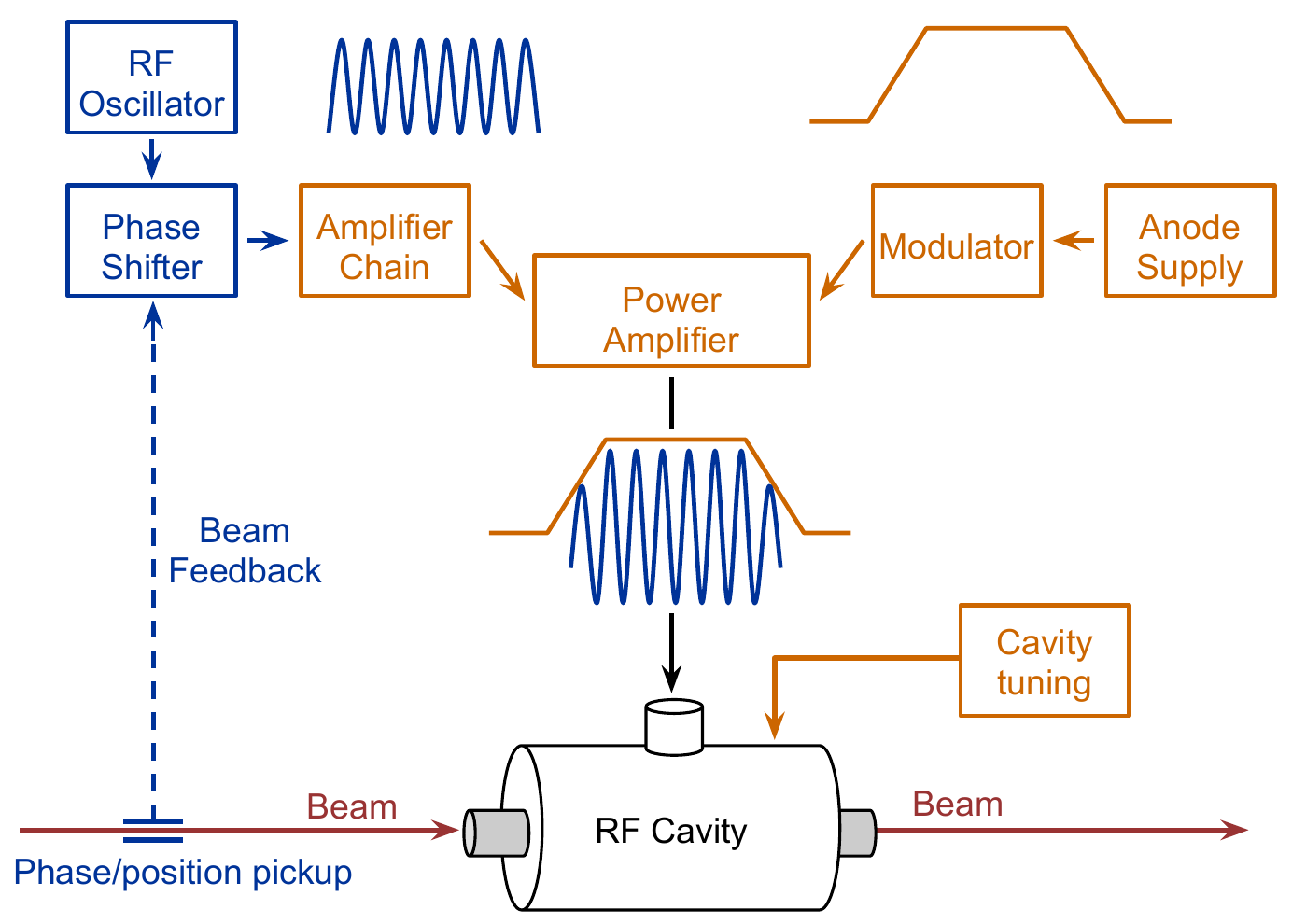}
			\caption{RF system block diagram}
			\label{fig:RFblock}
		\end{figure}

		The system that provides the high-power RF signal to the cavities can be split into two categories, known as the \gls{llrf} and \gls{hlrf} systems. An overview of the entire RF system is shown in \textit{Figure~\ref{fig:RFblock}}.
		
		The \gls{llrf} system consists of electronics and software responsible for generating the high-frequency electrical signal. It uses feedback from the beam to precisely control the RF signal's amplitude, phase, and frequency to meet the needs of the accelerator. The \gls{hlrf} system provides high-power amplification of the \gls{llrf} signal, and consists of an amplifier chain that leads to the cavity.
		
		\subsection{Low-Level RF}
			The fundamental parts of the \gls{llrf} system are pictured as the blue sections of \textit{Figure~\ref{fig:RFblock}}. The \textit{RF Oscillator} provides the high-frequency electrical signal that will be amplified and sent to the cavity; this device is usually a \gls{dds}. The \textit{phase shifter} module keeps the timing of the signal locked to the arrival of the beam for precise control of the synchronous phase. 
			
			A phase pickup measures the arrival time of the beam and provides precise feedback to the phase shifter for quick error correction.	Synchrotron accelerators also have a synchronous phase program input to the phase shifter to allow the phase to match the beam throughout acceleration.
			
			There is also a feedback system that measures the beam's transverse position to help with energy matching between the magnet system and the RF; this feedback is known as \gls{rpos} feedback. \Gls{rpos} feedback works because of an effect known as ``dispersion,'' explained in \textit{Chapter~\ref{chap:con_magnets}}. Basically, particles have a different radial position depending on their momentum, and the \gls{llrf} uses this information to monitor the beam energy. If there is a mismatch between the \gls{rf} frequency and the magnetic bending field, it will manifest as a position offset at the \gls{rf} detector; the \gls{llrf} system can then adjust the \gls{phase} to compensate.
		
		\subsection{High-Level RF}
			The High-Level RF system is pictured in orange on \textit{Figure~\ref{fig:RFblock}}, and provides the power to drive the \gls{rf} cavities. An amplifier chain applies the high-frequency \gls{llrf} signal to the \gls{pa}, which is the final stage of amplification before the cavity. The \gls{pa} is a very high-power vacuum tube that combines the high-voltage from its \textit{Anode Supply} with the high-frequency signal from the \gls{llrf}.  To save on power and equipment wear, a \textit{Modulator} switches the high-power signal on and off to the cavity; the modulator basically shapes the anode supply's power into a pulse to switch the \gls{rf} signal on before beam arrives in the cavity. \textit{Figure~\ref{fig:RFblock}} shows how the blue high-frequency signal from the \gls{llrf} and the orange high-power pulse from the \gls{hlrf}s modulator combines at the \gls{pa} to form a pulsed high-power oscillating signal for the cavity.
		\chapter{Fermilab Accelerators} \label{chap:con_accels}
This chapter begins with the layout of the accelerator complex, and then moves on to introducing each individual machine. We will continue to mention important concepts regarding the machines in this chapter. 

	\section{Overview}
	
	\begin{figure}[!htb]
		\includegraphics[width=0.85\fp]{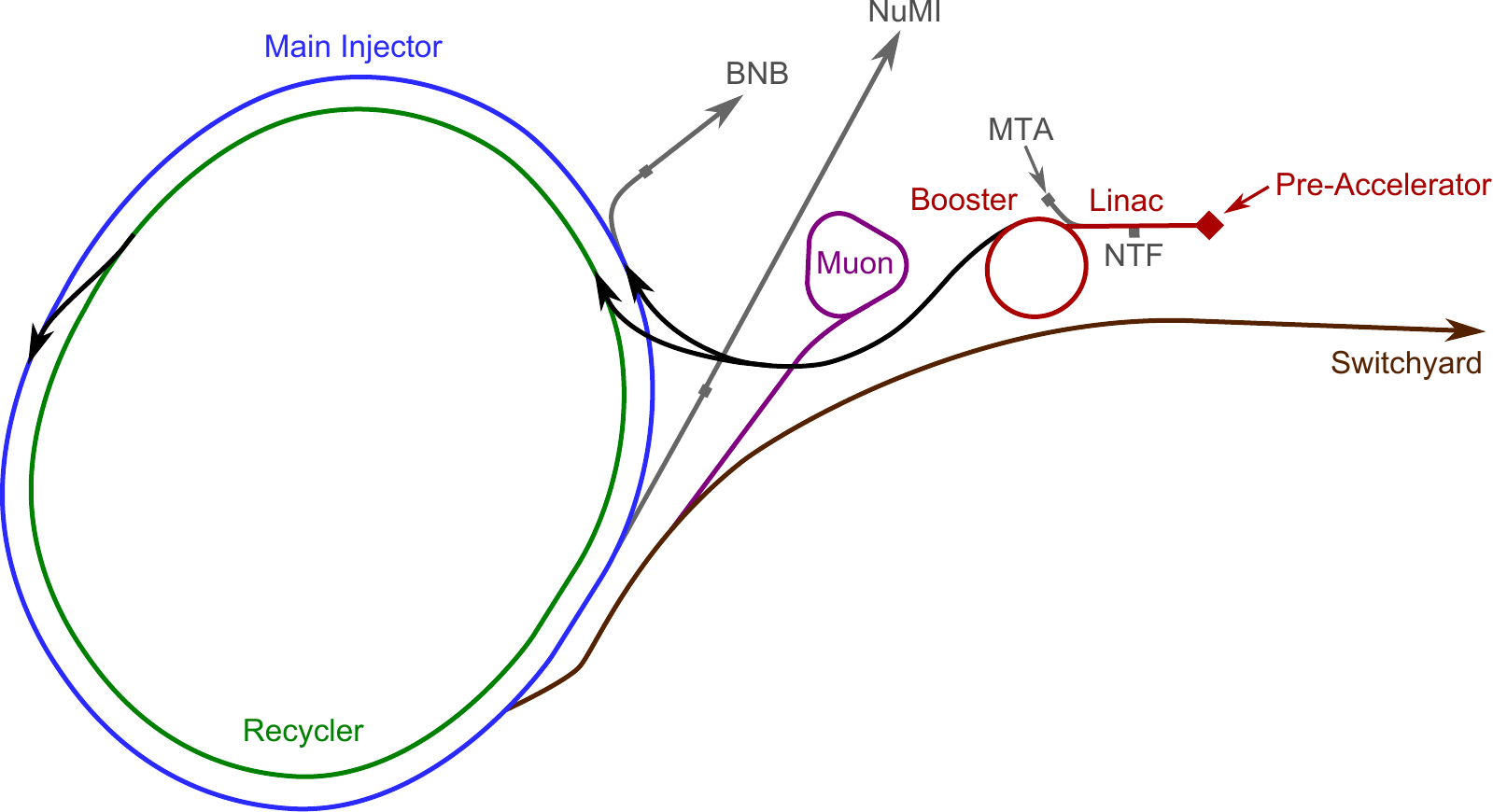}
		\caption{Layout of the accelerators at Fermilab}
		\label{fig:acc-map}
	\end{figure}
	
	The Fermilab accelerator complex consists of the following machines:
	
	\begin{itemize}
		\item Proton Source
		\begin{itemize}
			\item Pre-Accelerator
			\item \index{Linac}Linac
			\item Booster
		\end{itemize}
		\item Main Injector
		\item Recycler
		\item External Beamlines
		\begin{itemize}
			\item NuMI (Neutrinos at the Main Injector)
			\item BNB (Booster Neutrino Beam)
			\item Switchyard
			\item MTA (Muon Test Area)
		\end{itemize}
		\item Muon Campus
	\end{itemize}

	The layout of the machines, including the beamlines that connect them to one another, should be studied in detail. \figref{acc-map} contains a minimal amount of information, but more will be added as the machines are introduced. 

	\section{Proton Source}
	
	The Fermilab Proton Source consists of the \index{Pre-Acc}Pre-Accelerator, the \index{Linac}Linac, the \index{Booster}Booster synchrotron, and the beginning part of the MI-8 Line. The purpose of the Proton Source is to provide a proton beam through the MI-8 Line to the Recycler, Main Injector, and the Booster Neutrino Beam (BNB) Line. The Proton Source also provides H- ion beams to the Neutron Therapy Facility (NTF) and the \gls{mta}. 
	
		\subsection{Pre-Accelerator}
		The Fermilab \index{Pre-Acc}Pre-Accelerator contains the following:
		
		\begin{itemize}
			\item Two H- ion sources
			\item Low Energy Beam Transport (LEBT)
			\item Radio-frequency Quadrupole (RFQ) cavity
			\item Medium Energy Beam Transport (MEBT)
		\end{itemize}
		
		\marginpar{
			\centering
			\includegraphics[width=3cm]{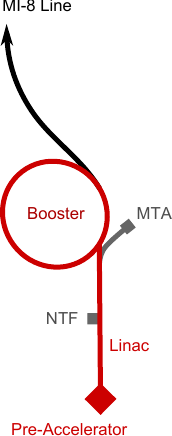}
			\captionof{figure}{Layout of the Fermilab Proton Source}
			\label{fig:acc-map-ps}
		}
		
		\begin{figure}[!htb]
			\centering
				\includegraphics[width=0.9\linewidth]{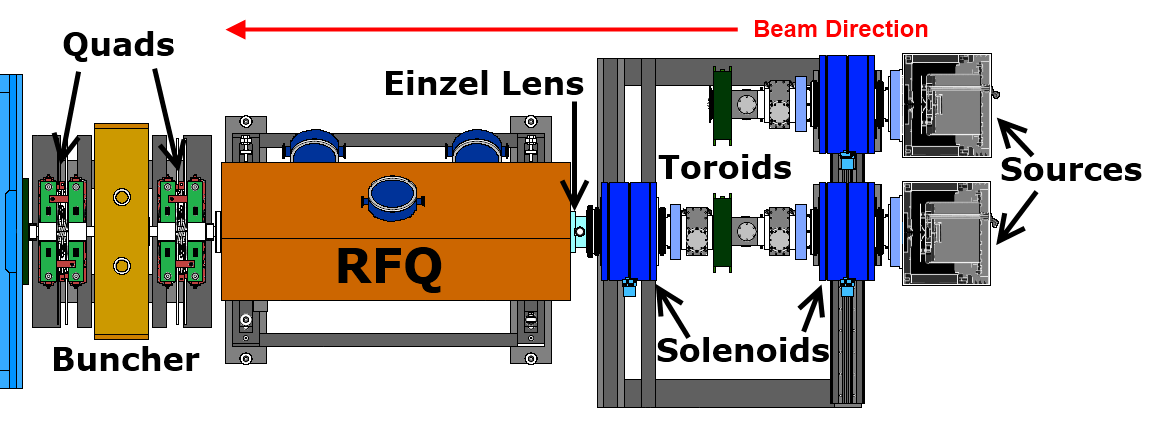}
				\caption{Drawing of the Fermilab Pre-Accelerator}
				\label{fig:preacc}
		\end{figure}
			
			\subsubsection{Source and LEBT}
			The Pre-Accelerator begins with two beam sources, each generating a \kev{35} H- ion beam. The sources are identical and either can provide beam to the accelerator complex. Redundancy is required here because it can take days to recover from the failure of a source. Having a running spare significantly minimizes the amount of time without beam following a failure. The H- source produces beam in a pulse that is approximately \us{100} long at a rate of \hz{15}. Note that we do not use every single beam pulse that the source produces. 
			
			The source directs the H- ion beam into the \gls{lebt}. The \gls{lebt} contains solenoid magnets for transverse focusing of the beam, as well as dipole trims for controlling the trajectory of the beam through the \gls{lebt}. 
			
			%\subsubsection{Typical Source Beam Pulse}
			%\comment{[Add a figure and quick explanation of a typical source pulse.]}

			\subsubsection{Einzel Lens as a Beam Chopper}
			An Einzel lens located at the end of the LEBT is used as a \keyterm{beam \gls{chopper}}, controlling the portion of the beam pulse that passes through to the next stage of the accelerator complex. Chopping the beam is necessary because we do not want the entire beam pulse from the source to get accelerated by the \gls{rfq} and the \index{Linac}Linac. Beam choppers give us control over the portion of the beam pulse that we want to use and over the length of that beam pulse. 
			
			\begin{figure}
				\centering
					\includegraphics[width=0.8\linewidth]{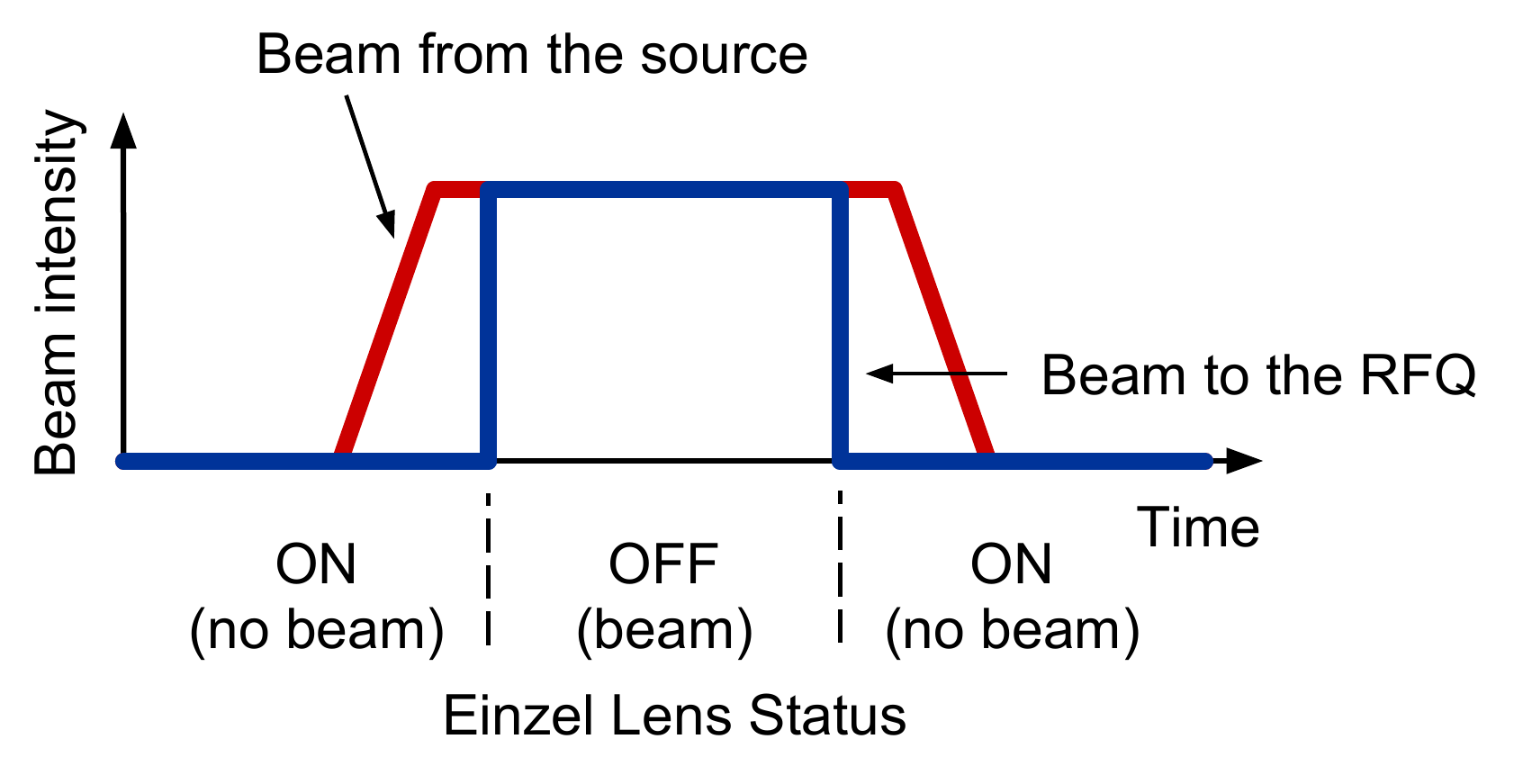}
					\caption[Beam chopping]{Beam chopping: a specific portion of the incoming beam is selected to pass through the chopper.}
					\label{fig:einzel-time}
			\end{figure}
						
			\marginpar{A \keyterm{beam \gls{chopper}} is a device that controls the portion of a beam pulse that is passed through to the next machine.}
			
			\marginpar{
				\centering
					(a) ON - beam blocked
					\includegraphics[width=4.5cm]{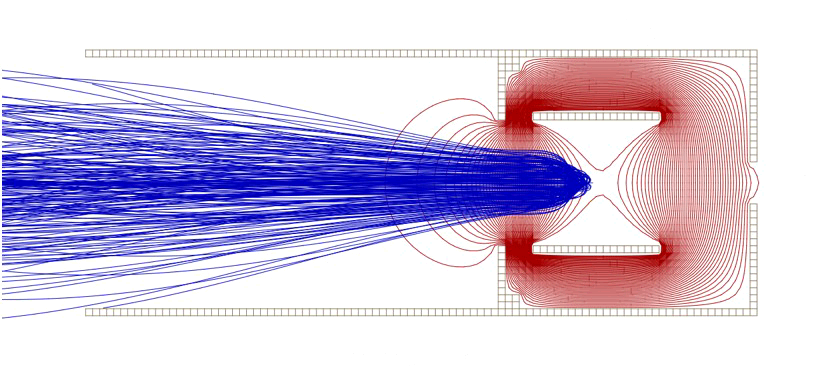}
					
					(b) OFF - beam transmission
					\includegraphics[width=4.5cm]{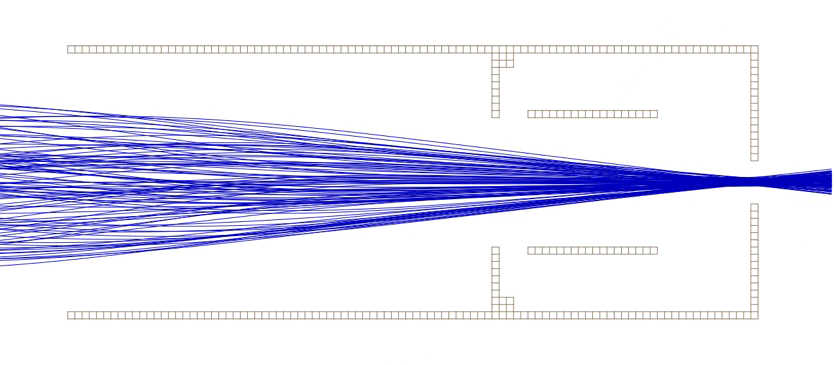}
					\captionof{figure}[Einzel lens electric field]{The electric field of the Einzel lens when it is ON and OFF. Beam is shown in blue.}
					\label{fig:einzel-beam}
			}
			
			Fast switches are used to turn the lens ON or OFF quickly. \textit{Figure~\ref{fig:einzel-time}} shows how the chopper selects a portion of the source pulse. Initially, the chopper is ON and it deflects beam from the source. At time T1, the lens turns OFF and beam passes through the lens and into the \gls{rfq}. The lens turns back ON at time T2 to end the pulse. The time difference from T1 to T2 is the Linac beam pulse width. 

			\subsubsection{Radio-frequency Quadrupole}
			The repulsive force between the particles in low-energy beams is very strong, but this repulsion can be minimized by increasing the kinetic energy of the beam as quickly as possible. \glspl{rfq} are useful as initial beam accelerators because they focus, bunch, and accelerate the beam all at once. Beam is focused transversely in the \gls{rfq} via an electric field rather than a magnetic field. The Fermilab \gls{rfq} accelerates beam from the LEBT to \kev{750} and directs it into the MEBT. 
			
			\begin{figure}[!htb]
				\centering
					\includegraphics[width=1.0\linewidth]{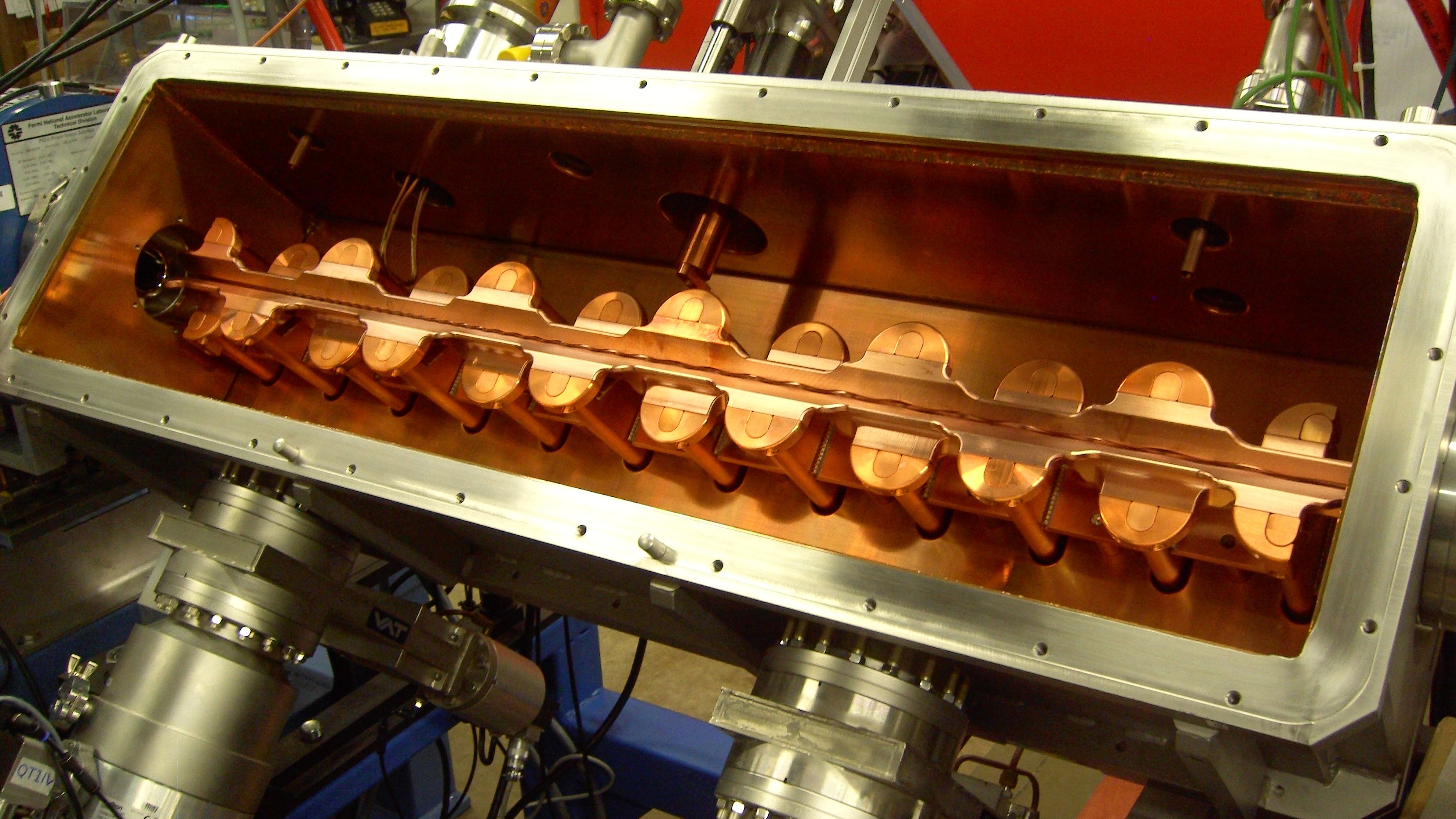}
					\caption[The inside of the RFQ]{The RFQ is a specialized RF cavity containing ``rod'' electrodes that shape the fields to provide simultaneous bunching and acceleration of beam}
					\label{fig:rfq}
			\end{figure}
			
			\subsubsection{MEBT}
			The Medium Energy Beam Transport (MEBT) is composed of two quadrupole doublets and an RF buncher cavity. \footnote{An RF buncher is an RF cavity that initializes or reinforces beam's bunch structure. Typically, bunchers do not provide any net acceleration.}A focusing quad and a defocusing quad are packaged together to form a quadrupole doublet. Small dipole trim magnets are embedded in the quadrupole doublets to control the position of the beam. The RF buncher helps keep beam from debunching before it enters the \index{Linac}Linac. This increases the amount of beam that is captured and accelerated by the RF systems in \index{Linac}Linac. Thus, the buncher increases the acceleration efficiency of the entire Proton Source. 
			
			\subsubsection{Pre-Accelerator Machine Parameters}
			\begin{table}[H]
				\centering
				\begin{tabular}{| l | c |}
					\hline
					Source Output Particle Beam & Negative Hydrogen ions (H-) \\ \hline
					Source Output Energy & \kev{35} \\ \hline
					Source Output beam current & \ma{65} \\ \hline
					RFQ Input Energy & \kev{35} \\ \hline
					RFQ Output Energy & \kev{750} \\ \hline
					RFQ RF frequency & \Mhz{201.24} \\ \hline
					Beam Current into Linac & \ma{40} \\ \hline
					Device prefix & L: \\ \hline
				\end{tabular}
				\caption[Pre-Accelerator Machine Parameters]{}
				\label{tab:preacc-params}
				\vspace{-3em}
			\end{table}
			
			%\subsubsection{Typical Pre-Accelerator Beam Pulse}
			%\comment{[Add a figure and quick explanation of a typical beam pulse.]}
			
		\subsection{Linac}
		The Fermilab \index{Linac}Linac accelerates H- ions from \kev{750} to \Mev{400}. The Linac contains two sections: the Low Energy Linac and the High Energy Linac, as shown in \figref{linac}. The Low Energy Linac is a Drift Tube Linac that accelerates the beam from \kev{750} to \Mev{116.5}. The High Energy Linac is a Side-Coupled Cavity Linac that accelerates the beam from \Mev{116.5} to \Mev{400}. The Low Energy section and the High Energy section will be discussed separately because they are different types of linacs\footnote{Originally, the Linac was just a drift tube linac. The downstream section of the Linac was upgraded in 1991-92 to a side-coupled cavity linac.}. A transition section matches the beam longitudinally to the High Energy Linac. The devices in the \Mev{400} Area direct the H- ion beam to one of several destinations. 
		
			\subsubsection{LE Linac/Drift Tube Linac}
\begin{figure}[!htb]
				\centering
				\begin{minipage}{0.48\textwidth}
					\centering
					\includegraphics[width=0.95\textwidth]{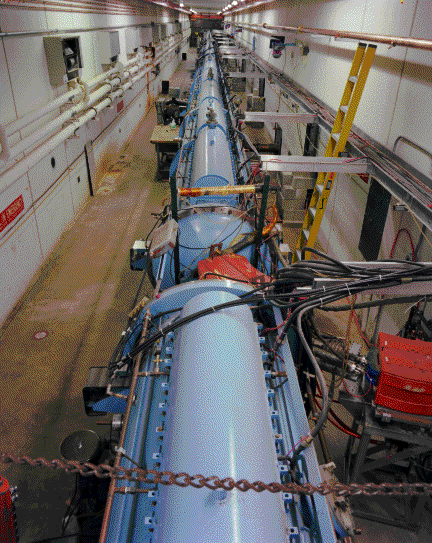}
					
					(a) From above
				\end{minipage}
				\begin{minipage}{0.48\textwidth}
					\centering
					\includegraphics[width=0.91\textwidth]{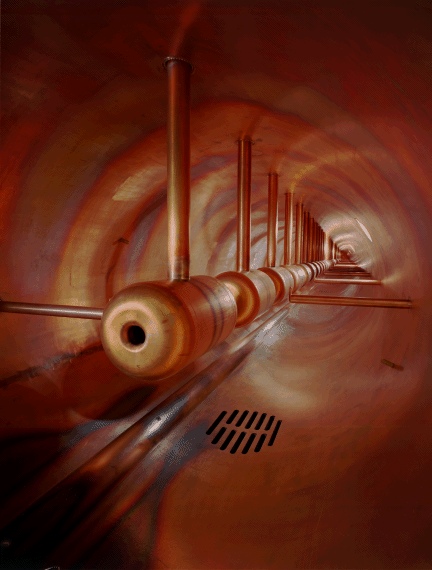}
					
					(b) On the inside
				\end{minipage}
				\caption[Drift Tube Linac]{The \gls{dtl} tanks along with the drift tubes placed along the center axis of the tanks. Beam travels down the center of the drift tubes.}
				\label{fig:dtl}
			\end{figure}

			The Low Energy (LE) \index{Linac}Linac is composed of large RF cavities, called ``tanks.'' Conducting tubes in the cavity shield the beam from unwanted portions of the electric field. Beam acceleration occurs only in the gaps located between these tubes. The tubes are called ``drift tubes'' because the beam is not accelerated while passing through them, but it ``drifts'' instead. This type of Linac is called a \gls{dtl}. \textit{Figure~\ref{fig:dtl}} shows the inside and outside of a \gls{dtl} tank. 
			
			The LE Linac contains five \gls{dtl} cavities that operate at a resonant frequency of \Mhz{201.24} and accelerate the H- ion beam from \kev{750} to \Mev{116.5}. The LE Linac is segmented by these RF cavities into what are called ``Linac RF stations,'' or simply LRF stations. The term ``station'' refers to all of the associated equipment required to operate one cavity. 
			
			In addition to the cavities, the LE Linac contains dipole magnets for controlling the trajectory of the beam and quadrupole magnets for focusing the beam. Dipoles are in the space available between RF tanks, and quadrupoles are inside the drift tubes. Thus, drift tubes perform two functions: they shield particles from unwanted electric fields and they contain the quadrupoles which focus the beam. 
			
			\marginpar{
				\centering
				\includegraphics[width=3cm]{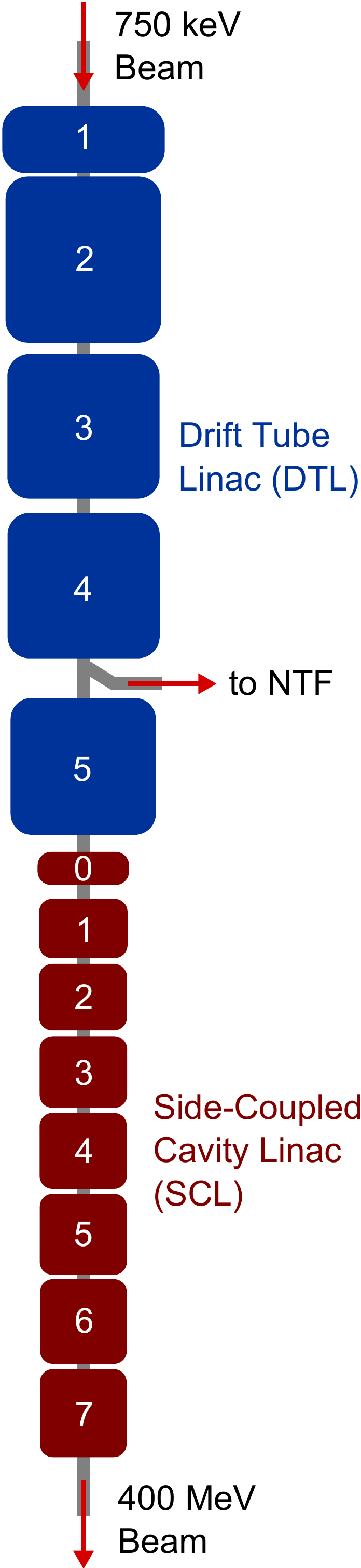}
				\captionof{figure}{Layout of the Fermilab Linac}
				\label{fig:linac}
			}
			
			\subsubsection{Neutron Therapy Facility}
			The Neutron Therapy Facility (NTF) is a treatment facility that uses high energy neutrons to treat cancer. The first three LRF tanks accelerate the H- ions to \Mev{66} and then it drifts through tank 4. Magnets located between tanks 4 and 5 direct the beam into the NTF target, and the interaction between H- beam and the target creates a beam of neutrons. The Neutron beam treats malignant tumors. A freight elevator in the \index{Linac}Linac gallery houses the NTF treatment room which lowers to the beamline level. The Operations Department is not involved in most aspects of the process except for working with the Linac group to provide stable beam to NTF while patient treatment is taking place. 
			
			\subsubsection{Transition Section}
			The transition section follows the LE Linac and is composed of two RF stations, the ``buncher'' and the ``vernier.'' The beam must be matched longitudinally to the High Energy Linac because it operates at a different RF frequency than the LE Linac. The transition section performs the longitudinal matching using side-coupled cavity structures, which are discussed in the following section. The ``0'' in \figref{linac} represents the transition section.
			
			\subsubsection{HE Linac/Side-Coupled Linac}
			The High Energy (HE) \index{Linac}Linac contains ten RF stations: seven main stations along with three low-power ones. The seven main RF stations accelerate the H- ion beam from \Mev{116.5} to \Mev{400}. Two of the low-power stations are the previously discussed buncher and vernier. The third low-power RF station, called the ``debuncher,'' is located in the beamline that connects the Linac to the Booster. 
			
\begin{figure}[!htb]
				\centering
					\includegraphics[width=0.9\textwidth]{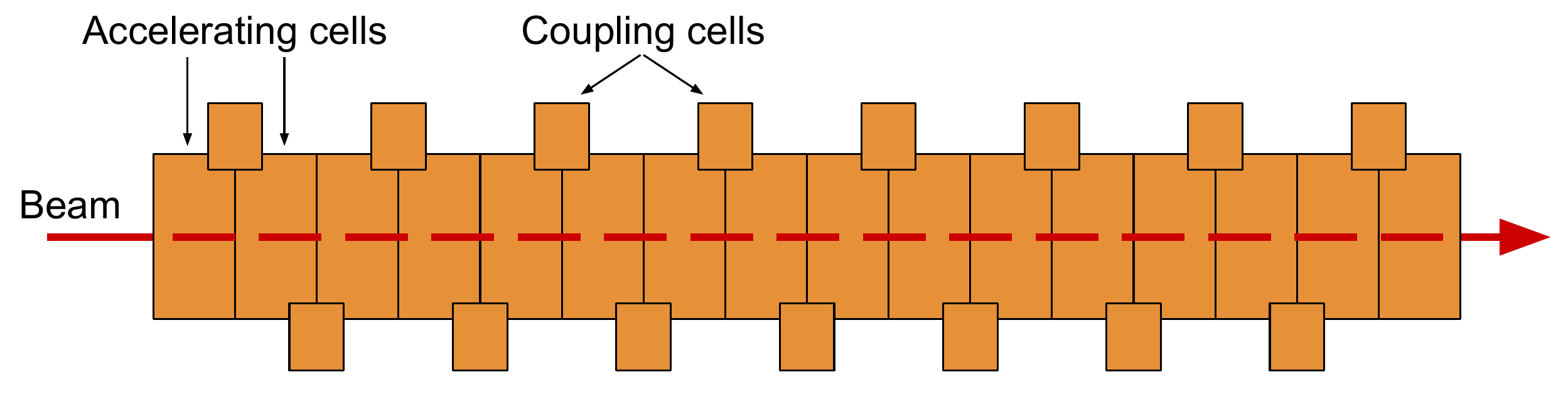}
					\caption[Side-view of an SCL cavity module]{Side-view of an \gls{scl} cavity module with 16 accelerating cells and 15 coupling cells}
					\label{fig:scl-cells}
			\end{figure}

			Two types of disk-shaped RF cavities, accelerating cells and coupling cells, are used in the HE Linac. The accelerating cells each generate a longitudinal electric field near the radial center of the cavity. The coupling cells provide electromagnetic coupling between adjacent accelerating cells. As shown in \textit{Figure~\ref{fig:scl-cells}}, accelerating cells are placed inline along the beam path while the coupling cells are offset from the beam path. A group of accelerating and coupling cells forms a cavity module. This type of accelerating structure is called a \gls{scl} due to the function and position of the coupling cells. Each of the cells, and the cavity modules as a whole, operate at a resonant frequency of \Mhz{804.96}. 
			
			The main RF stations contain four modules that each have sixteen accelerating cells. The buncher, vernier, and debuncher each contain only a single SCL cavity module with a smaller number of accelerating cells when compared to the main HE stations. Dipole and quadrupole trims are placed in between the modules to steer and focus the beam.
			
			\begin{figure}[!htb]
				\centering
				\begin{minipage}{0.48\textwidth}
					\centering
					\includegraphics[width=0.95\textwidth]{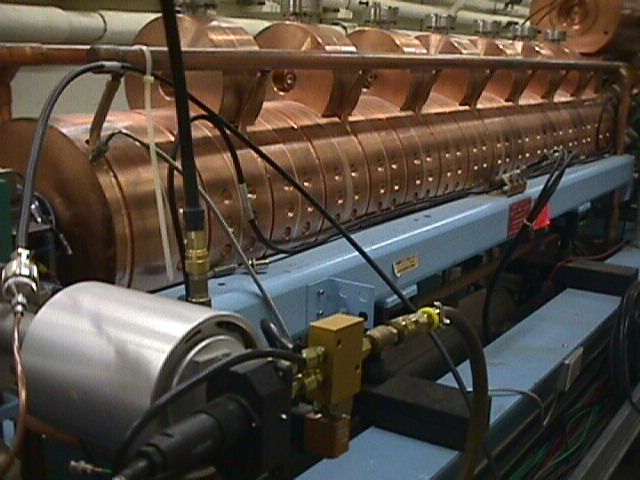}
					
					(a) One cavity module
				\end{minipage}
				\begin{minipage}{0.48\textwidth}
					\centering
					\includegraphics[width=0.95\textwidth]{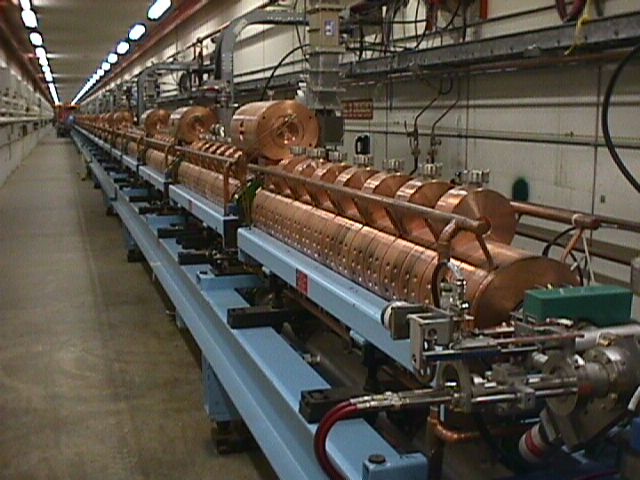}
					
					(b) Multiple modules
				\end{minipage}
				\caption{Fermilab SCL cavity modules}
				\label{fig:scl}
			\end{figure}
			
			High gain amplifiers called ``klystrons'' generate the high power signal sent into the cavity modules. For this reason, the HE Linac stations are called \Gls{krf} Stations. High-power klystrons are used in the main HE Linac RF stations, while low-power klystrons feed the buncher, vernier, and debuncher. 
			
			\subsubsection{Linac Machine Parameters}
			
			\begin{table}[H]
				\centering
				\begin{tabular}{| l | c |}
					\hline
					Input Energy & \kev{750} \\ \hline
					LE to HE \index{Linac}Linac Energy & \Mev{116.5} \\ \hline
					Output Energy & \Mev{400} \\ \hline
					Length & \m{150}  \\ \hline
					Output Current (nominal) & \ma{34} \\ \hline
					\multirow{2}{*}{LE \index{Linac}Linac Nomenclature} & Drift Tube Linac (DTL) \\
					   & Linac RF station (LRF) \\ \hline
					LE \index{Linac}Linac RF frequency & \Mhz{201.24} \\ \hline
					Number of LRF stations & 5 \\ \hline
					\multirow{2}{*}{HE \index{Linac}Linac Nomenclature} & Side-Coupled Cavity Linac (SCL) \\
					   & Klystron RF Station (KRF) \\ \hline
					HE \index{Linac}Linac RF frequency & \Mhz{804.96} \\ \hline
					Number of KRF stations & 7 \\ \hline
					Device prefix & L: \\ \hline
				\end{tabular}
				\caption[Linac Machine Parameters]{}
				\label{tab:linac-params}
				\vspace{-3em}
			\end{table}
			
		\subsection{400 MeV Area}
		The \Mev{400} H- ions exiting the \index{Linac}Linac are directed to one of the following destinations:
		
		\begin{enumerate}
			\item Linac straight ahead dump
			\item Linac momentum dump
			\item Muon Test Area
			\item Booster
		\end{enumerate}
		
\begin{figure}[!htb]
			\centering
				\includegraphics[width=1.0\linewidth]{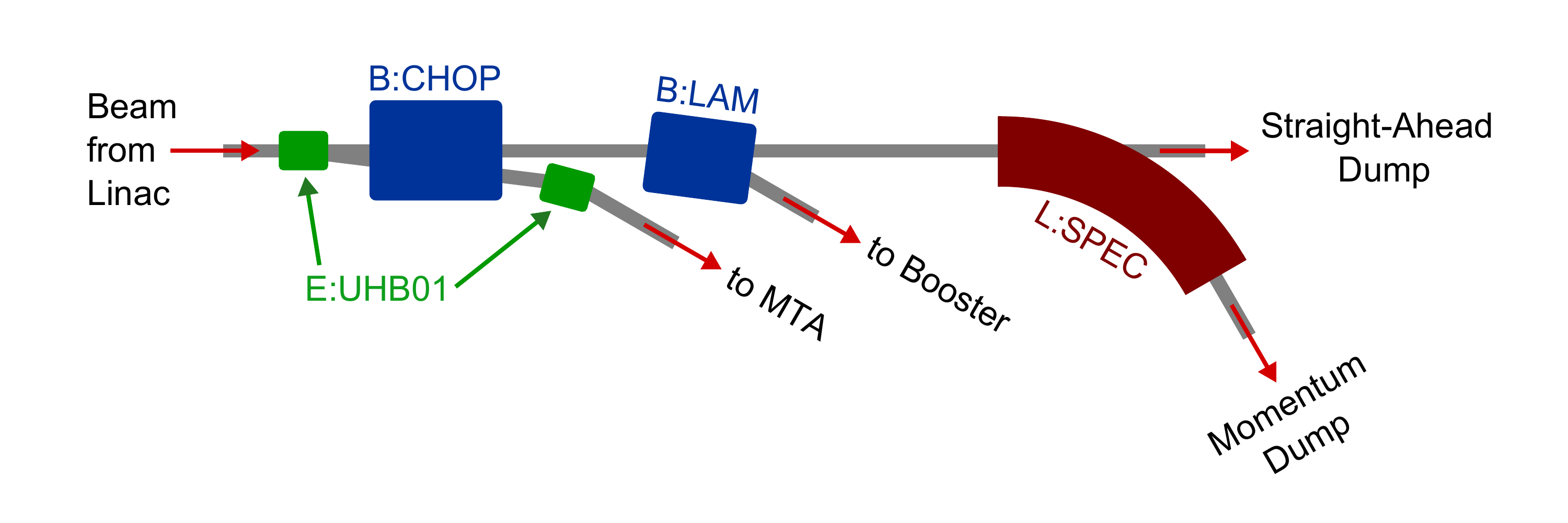}
				\caption{Layout of the 400 MeV area}
				\label{fig:400mev}
		\end{figure}

		The components (and devices) which determine the destination of the beam are the Linac spectrometer magnet (L:SPEC), the Booster chopper (B:CHOP), the Booster Lambertson (B:LAM), and the \gls{mta} C-magnets\footnote{C-magnets are dipoles that are used in locations with limited space.} (E:UHB01). Figure~\ref{fig:400mev} contains a basic overview of the \Mev{400} Area. The rest of this section details the configuration these devices need to be in to send beam to each of the four destinations. 
		
			\subsubsection{Linac Beam Dumps}
			When the \gls{mta} C-magnets and the Booster chopper are not powered, beam travels to the \index{Linac}Linac spectrometer magnet\footnote{The Linac spectrometer magnet is just a dipole magnet with a large bend angle. It gets its name from the dispersion of the beam caused by the large bend field, just as a prism does with light.}. The state of this component determines which Linac beam dump is currently in use. If L:SPEC is off, beam goes to the straight-ahead dump. If L:SPEC is on, beam is deflected by the spectrometer magnet towards the momentum dump. The momentum dump is used operationally due to its larger beam absorber. The straight ahead dump is only used during beam studies or commissioning. Thus, L:SPEC is on during normal beam operations. 
			
			\subsubsection{\gls{mta} Line}
			Located just upstream of the Booster chopper, the first C-magnet bends beam into the \gls{mta} line. The second C-magnet is located at the beginning of the \gls{mta} beamline. The rest of the line contains dipoles and quadrupoles for steering and focusing the beam towards the experiment in the \gls{mta} experimental hall. 
			
			\subsubsection{\Mev{400} Line}
			A chopper/Lambertson combination is used to send beam into the \Mev{400} Line and towards the \index{Booster}Booster. When powered on, the Booster chopper deflects beam into the field region of the Booster Lambertson, which bends the beam into the \Mev{400} Line. This beamline directs beam out of the \index{Linac}Linac enclosure, through the Booster enclosure, and into the accelerator. 
			
			\marginpar{
				\centering
					\includegraphics[width=4.0cm]{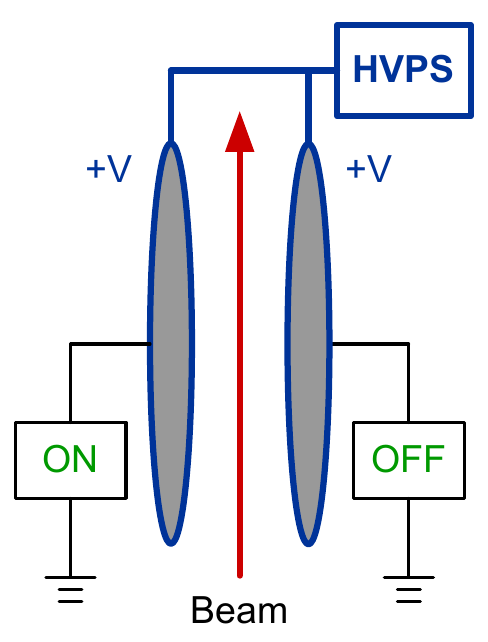}
					
					(1) The HVPS charges the electrodes
					\includegraphics[width=4.0cm]{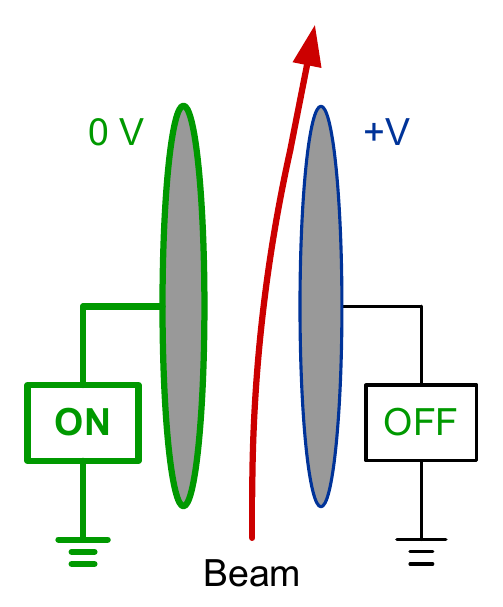}
					
					(2) Fire ON switch, beam deflected
					\includegraphics[width=4.0cm]{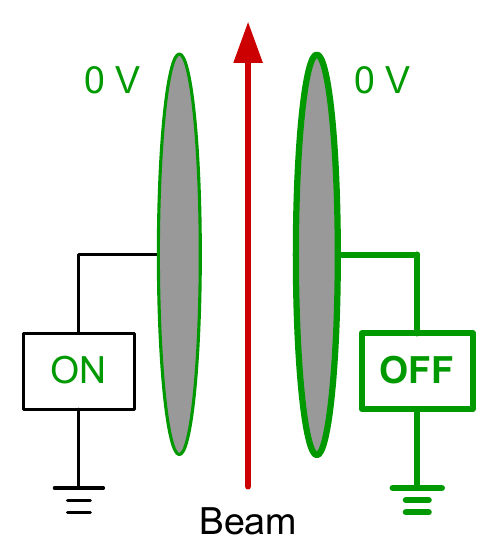}
					
					(3) Fire OFF switch
					\captionof{figure}{Booster chopper operation}
					\label{fig:b-chop}
			}
			
			\subsubsection{\Mev{400} Chopper and Booster Turns}
			As you may recall from the discussion on the Einzel lens, beam choppers are devices that select a specific portion of the beam pulse for transmission to the next stage of the accelerator complex. The \index{Booster}Booster chopper is a pulsed electrostatic deflector made up of two electrodes that are equally spaced from the center of the beam pipe in the vertical direction. The potential difference between these electrodes determines whether or not the chopper deflects beam. A high voltage power supply charges the electrodes and fast-acting switches discharge the electrodes to ground. 
			
			The operation of the chopper is described below and shown in \textit{Figure~\ref{fig:b-chop}}. 
			
			\begin{enumerate}
				\item The high voltage power supply charges both electrodes to the same voltage. The beam passing through is not deflected because there is no potential difference between the two electrodes.
				\item \textbf{ON:} The `ON' switch is fired and it grounds one of the electrodes. The potential difference across the chopper electrodes deflects the beam.
				\item \textbf{OFF:} After a designated amount of time, the `OFF' switch is fired to ground the other electrode. Passing beam is no longer deflected.
				\item Steps 1-3 are repeated as needed.
			\end{enumerate}

			The \Mev{400} chopper controls which portion of the \index{Linac}Linac beam pulse is sent to the \index{Booster}Booster by triggering the chopper `ON' and `OFF' switches appropriately. The length of the stream of beam sent to the Booster depends on \emph{t}, the time difference between the `ON' and `OFF' triggers. This length is referred to in ``turns,'' where a \keyterm{\gls{turn}} is equal in length to the circumference of the Booster. Since the revolution period of Booster at injection is \(2.2 \mu s\), \emph{t} must be \(2.2 \mu s\) to generate one turn of beam. Similarly, \emph{t} must be \(n*2.2 \mu s\) to generate \emph{n} turns of beam. 
						
			\marginpar{A \keyterm{\gls{turn}} is a stream of beam that is equal in length to the circumference of the Booster.}
			
			\begin{figure}[!htb]
				\centering
					\includegraphics[width=0.6\linewidth]{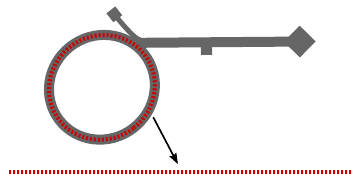}
					\caption[Booster turn]{One turn of beam is equal in length to the circumference of the Booster}
					\label{fig:booster-turns}
			\end{figure}
			
		\subsection{Booster}
		The Fermilab \index{Booster}Booster\footnote{Booster accelerators are used at most accelerator facilities for ``boosting'' the energy of the beam from the \index{Linac}Linac stage to an acceptable input energy for the final accelerator. At Fermilab, the final accelerator is the Main Injector.} accelerator is the final stage of the Proton Source. The Booster is a rapid-cycling synchrotron that accelerates protons from 400 MeV to 8 GeV at a rate of \hz{15}. It is made up primarily of magnets and RF cavities, and it is divided into 24 equal-length ``periods.'' 
		
			\subsubsection{Injection into Booster}
			Beam injection into Booster consists of passing the H- ions (injected beam) through a stripping foil to remove the electrons. The resulting proton beam circulates in the Booster accelerator. 
			%Look into changing this sentence
			Additional turns of beam inject in the same manner by placing the injected beam on the exact same trajectory as circulating beam. Passing the combined beam through the stripping foil leaves a circulating proton beam. This method of injection is called ``Charge Exchange Injection.'' 
			
			\marginpar{\keyterm{\Gls{paraphasing}} is the process by which unbunched beam is captured by the Booster RF system.}
			
			After injection is complete, the Booster RF system must capture the beam in order to accelerate it. Since the HE \index{Linac}Linac and the Booster operate at different RF frequencies, the bunch structure of injected beam from Linac doesn't match that of Booster. The beam is captured by \keyterm{\gls{paraphasing}}, a process in which mostly DC beam is captured into RF buckets. The Booster RF system is split into two groups, A and B stations. Initially, the A and B stations are out of phase with one another. The net acceleration at this point is zero, as the beam is accelerated by some RF cavities but decelerated by others. To capture the beam, the phases of the A and B stations are slowly aligned, capturing beam in RF buckets before acceleration. 
			
			The bunch structure and momentum spread of the incoming beam have a significant impact on the efficiency of paraphasing. Paraphasing is most efficient when the beam is completely unbunched and mono-energetic. The Debuncher, an \gls{scl} RF cavity in the \Mev{400} Line, minimizes the momentum spread of the beam. De-bunching occurs naturally as beam circulates in the Booster at \Mev{400}, since there are no RF buckets to maintain the bunch structure. 
		
			\subsubsection{Booster Magnets}
			The main magnet system consists of combined-function magnets, which provide both a quadrupole field as well as a dipole field. These magnets are named like regular quadrupoles, `F' for focusing and `D' for defocusing. The lattice is FOFDOOD, where `O' represents a short straight section and `OO' represents a long straight section. RF cavities, injection components, and extraction components are located in the long straight sections. Beam diagnostics and correction element packages are located in both the long and short straight sections. The correction element packages are made up of six different circuits that generate dipole, quadrupole, and sextupole fields, giving fine control over the beam. 
			
			\subsubsection{Booster RF System}
			%The \index{Booster}Booster RF system, often called ``BRF,'' is composed of nineteen stations\footnote{One of the stations remains off during normal %operations. It is a spare that can be turned on quickly if another station fails.}. The RF system increases the beam energy from \Mev{400} to \gev{8}. %Since Booster is a synchrotron, the resonant frequency of the cavities increases throughout the cycle to match the change in the beam velocity. Throughout %the acceleration process, the frequency sweeps from \Mhz{37.8} to \Mhz{52.8}, and the beam revolution period goes from \us{2.2} at injection to \us{1.6} at %extraction. 
			
			The \index{Booster}Booster RF system, often called ``BRF,'' is composed of nineteen stations\footnote{One of the stations remains off during normal operations. It is a spare that can be turned on quickly if another station fails.} that increase the beam energy from \Mev{400} to \gev{8}. Since Booster is a synchrotron, the cavity resonant frequency sweeps from \Mhz{37.8} to \Mhz{52.8} as the beam revolution period decreases from \us{2.2} at injection to \us{1.6} at extraction. 
			
			\marginpar{\keyterm{Bunch rotation} is a Booster RF beam manipulation that reduces beam energy spread before extraction.}
			
			At the end of the acceleration cycle, the Booster RF performs a beam manipulation called \keyterm{bunch rotation}. The field amplitude of all the RF stations oscillate at just the right frequency to drive beam synchrotron oscillation. This reduces the energy spread of the beam to facilitate later manipulation by the Recycler's RF system, known as ``slip-stacking.'' Bunch rotation in Booster before extraction makes the slip-stacking process in the Recycler more efficient. 
			
			\marginpar{A \keyterm{\gls{batch}} is a string of 84 bunches.} 
			
			\marginpar{\keyterm{\Glspl{partialbatch}} is the process of delivering less than a full batch.} 
			
			\marginpar{\keyterm{\Glspl{multibatch}} is the process of sending two or more consecutive batches to Recycler or Main Injector.} 
			
			The harmonic number for the Booster is 84. All 84 bunches are extracted from the Booster at once, and each string of 84 bunches is called a Booster \keyterm{\gls{batch}}. The Booster, and therefore, the entire Proton Source, provides a batch of protons at a rate of \hz{15} to the MI-8 Line\footnote{The maximum average rate is currently about \hz{8}.}. The process of sending two or more consecutive batches to Recycler/Main Injector is called \keyterm{\glspl{multibatch}}. Often, we wish to send less than a full batch to Recycler, Main Injector, or BNB. \keyterm{\Glspl{partialbatch}} is the act of sending just a portion of a batch down the rest of the MI-8 Line and kicking the rest of the batch into the Booster dump. 
			
			% \subsubsection{Typical Booster Beam Cycle}
			
			\subsubsection{Booster Machine Parameters}
			
			\begin{table}[H]
				\centering
				\begin{tabular}{| l | c |}
					\hline
					Injection Energy & \Mev{400} \\ \hline
					Extraction Energy & \gev{8} \\ \hline
					Circumference & \m{474.2} \\ \hline
					Periods & 24 \\ \hline
					Lattice & FOFDOOD \\ \hline
					Revolution Period (at \Mev{400}/\gev{8}) & \(2.2 \mu s\)/\(1.6 \mu s\) \\ \hline
					RF Stations & 19 \\ \hline
					RF Frequency & \Mhz{37.77} to \Mhz{52.8} \\ \hline
					Harmonic Number & 84 \\ \hline
					RF Voltage Gain Per Turn & \kV{920}/turn \\ \hline
					Device prefix & B: \\ \hline
				\end{tabular}
				\caption[Booster Machine Parameters]{}
				\label{tab:booster-params}
				\vspace{-3em}
			\end{table}
			
		\subsection{MI-8 Line}
		The MI-8 Line delivers the proton beam from the Proton Source to the \index{Booster}Booster dump, the Main Injector, the Recycler, or the BNB beamline. The MI-8 Line is composed mostly of permanent magnets, but the beginning and end of the line contain powered magnets. The Booster dump is the first possible destination. For the other destinations, the MI-8 Line curves around the Muon Campus on its way into the Main Injector enclosure and sends the beam into the BNB beamline, the Recycler, or the Main Injector. 
		
		\begin{figure}[!htb]
			\centering
				\includegraphics[width=0.7\linewidth]{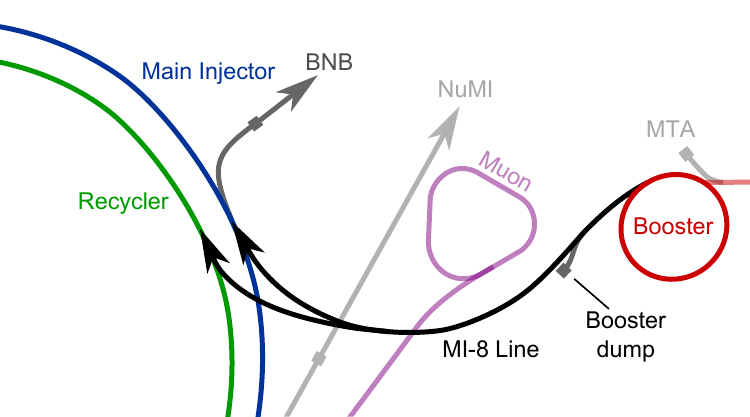}
				\caption[Layout of the MI-8 line]{The MI-8 Line sends beam from the Booster to the dump, the BNB beamline, the Main Injector, and the Recycler}
				\label{fig:mi8}
		\end{figure}
		
		\subsection{Proton Source Modes of Operation}
		The \index{Pre-Acc}Pre-Accelerator, \index{Linac}Linac, and \index{Booster}Booster work together to provide beam to one of several different destinations on each machine cycle\footnote{Recall that the machine cycle rate is \hz{15}.}. The following is a list of operational modes, which are often called ``states,'' that determine where the beam is headed. The following cycles determine where the beam is destined on each machine cycle. 
		
		The \index{Pre-Acc}Pre-Accelerator, \index{Linac}Linac, and \index{Booster}Booster work together to provide beam to one of several different destinations on each machine cycle\footnote{Recall that the machine cycle rate is \hz{15}.}. The Pre-Accelerator produces a beam pulse at a rate of 15 Hz, but the Linac and Booster are not always used to accelerate a particule pulse. The following cycles determine whether or not the Linac and Booster accelerate the beam pulse. The type of cycle that is used depends on the destination of that particular beam pulse. 
		
		\begin{itemize}
			\item High Energy Physics (HEP) cycle:
			
			The Linac and Booster accelerate a beam pulse and send it down the MI-8 Line where it is destined for the BNB beamline, the Recycler, or the Main Injector. 
			
			\item Studies cycle:
			
			The studies cycle is used for performing machine studies, or for tuning. The Linac and the Booster each have their own studies cycle where beam is sent to their respective beam dumps. For Linac studies cycles, beam is accelerated by the Linac, but not the Booster. For Booster studies, beam is accelerated by all three machines and is then directed into the Booster dump.
			
			\item NTF cycle:
			
			Part of the LE Linac accelerates a pulse of H- ions that are directed towards NTF.
			
			\item \gls{mta} cycle:
			
			The Linac accelerate a pulse of H- ions that is destined for the \gls{mta} beamline.
			
			\item Standby cycle:
			
			Standby cycles occur whenever part of the Proton Source does not accelerate beam on a particular machine cycle. This is the default cycle for the Linac and the Booster. With no beam being requested from the Proton Source, the machine cycle happens without the presence of beam.
			
			\begin{itemize}
				\item Linac Standby cycles occur whenever there is no beam request. In this case, the Source generates a pulse of beam, but it is not accelerated past the \gls{rfq}. 
				\item Booster Standby cycles occur whenever beam is going to a destination that is upstream of the Booster (such as NTF and \gls{mta}), or whenever there is no beam request for a particular machine cycle. 
			\end{itemize}
		\end{itemize}

	\section{Main Injector and Recycler}
	The Recycler and the Main Injector are the largest operational machines at Fermilab. They share the same enclosure and are just over two miles in circumference. The Main Injector is located on the bottom of the enclosure, with the Recycler 47 inches above it. The Recycler and Main Injector are integral parts of the Fermilab accelerator complex,  providing beam for NuMI , Switchyard, and the Muon Campus. A large majority of the beam generated by the Proton Source goes through to the Recycler and the Main Injector.
	
	\begin{figure}[!htb]
		\centering
			\includegraphics[width=0.7\linewidth]{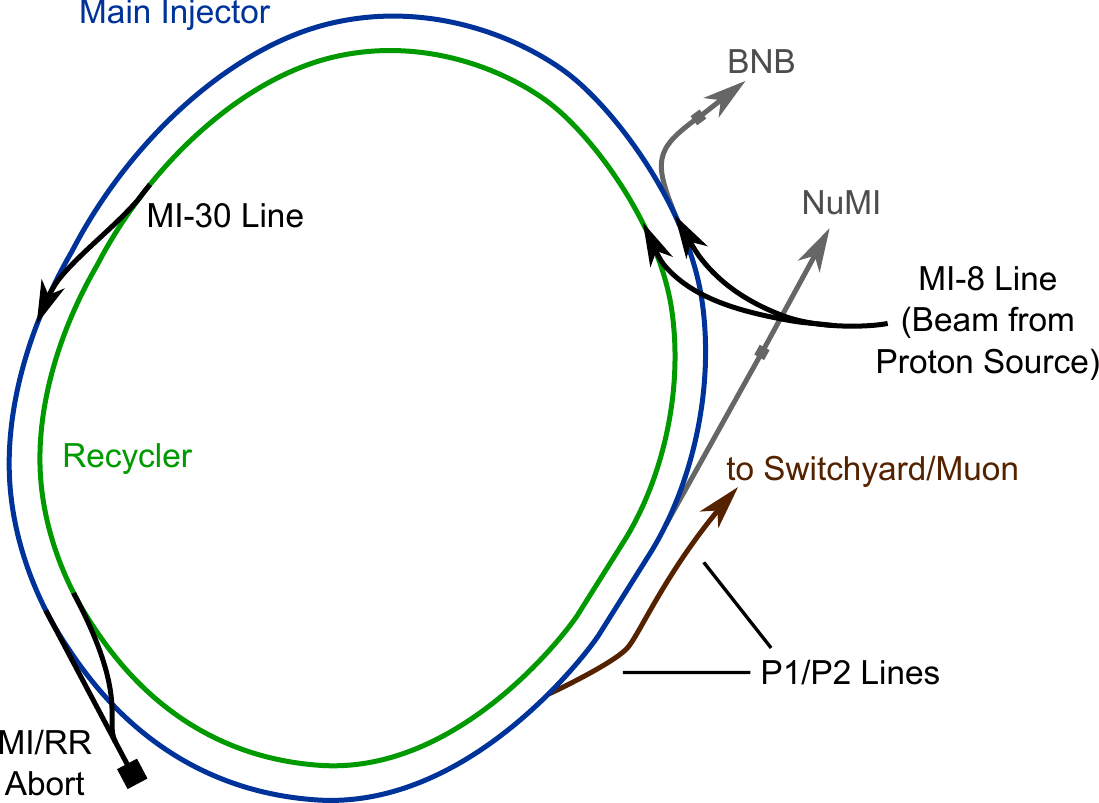}
			\caption[Layout of the MI/RR]{The layout of the Main Injector and the Recycler, along with all associated beamlines.}
			\label{fig:map-mirr}
	\end{figure}

	The machine cycle rate for the Recycler and Main Injector varies depending on the current operational mode and other factors. A typical Main Injector machine cycle takes on the order of a second or two, which is large compared to the \ms{67} cycle of the Proton Source\footnote{For example, a typical Main Injector machine cycle that delivers \gev{120} protons to the NuMI beamline is \s{1.33} long.}.
		
		\subsection{Main Injector}
		The Main Injector (MI) is a proton synchrotron with an injection energy of \gev{8} and an extraction energy of \gev{120}. In typical operations, the Main Injector delivers \gev{120} protons to the NuMI beamline, the Switchyard beamline, or the Muon campus. The Main Injector can also deliver \gev{8} protons to the Muon campus. 
		
		The Main Injector is made up mostly of dipoles, quadrupoles, and RF cavities. There are eight straight sections inside the Main Injector which are used for RF cavities, injection devices, and extraction devices. The quadrupoles are arranged in a FODO lattice. The main magnet systems are split into the following circuits: 
		
		\begin{itemize}
			\item Bend Magnet Circuit
			\item Focusing Quad Circuit
			\item Defocusing Quad Circuit
			\item Horizontal Sextupole Circuit
			\item Vertical Sextupole Circuit
		\end{itemize}
		
		Separate circuits for each type of magnetic component allow for independent control of the strengths of the bend, focusing, and defocusing fields. Other magnets in the Main Injector include corrector dipoles, corrector quadrupoles, and octupoles. The corrector dipoles and quadrupoles allow for fine control\footnote{Correctors are complementary to the coarse control offered by the main magnet systems.} of the beam trajectory and tune at each location around the ring. 
		
			\subsubsection{MI Magnet Cycle}
\begin{figure}[!htb]
				\centering
				\includegraphics[width=0.6\linewidth]{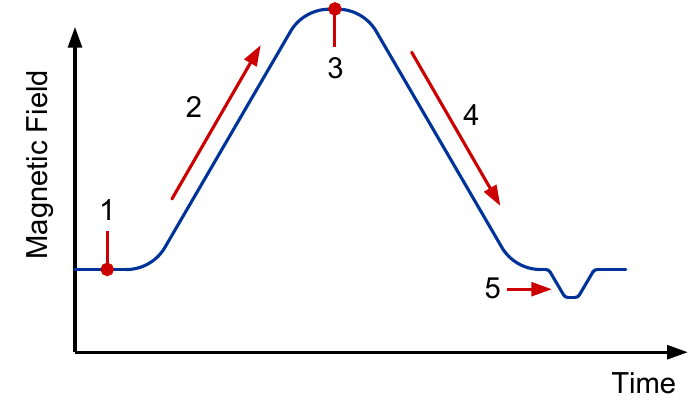}
				\caption{Typical magnet current ``ramp''}
				\label{fig:ramp}
			\end{figure}

			As mentioned in \textit{Chapter~\ref{chap:con_intro}}, electromagnets are particularly useful because their magnetic field can change to track beam energy. In a typical synchrotron, the current in the magnet system follows a ``ramp,'' as shown in \textit{Figure~\ref{fig:ramp}}. 
			
			In the MI, the ramp can be described as follows:

			\begin{enumerate}
				\item The bend field starts at an \gev{8} level and remains there while beam injections are completed.
				\item The bend field begins to increase, slowly at first, as the beam accelerates.
				\item When the beam energy has reached \gev{120}, the bend field stops increasing and remains at this level while the beam is extracted from the machine. This step is called ``flattop.''
				\item With no beam in the machine, the bend field is lowered back down to the 8 GeV level.
				\item The ramp undergoes a small dip, called a ``reset,'' to undo any hysteresis in the magnets. 
				\item Proceed to next ramp (typically another \gev{120} ramp).
			\end{enumerate}

			The Main Injector completes a typical ramp, such as the one that delivers \gev{120} beam to NuMI, in just 1.33 seconds. Once the entire cycle is completed, the next one can begin. This process is repeated over and over again.
			
			\subsubsection{Main Injector RF}
			The Main Injector RF system, often called ``MIRF,'' is made up of 20 RF cavities. During the acceleration from \gev{8} to \gev{120}, the RF frequency sweeps from \Mhz{52.8} to \Mhz{53.1}. This frequency sweep is much smaller than that of the \index{Booster}Booster\footnote{Consider why the RF frequency is swept by a much smaller amount when going from \gev{8} to \gev{120} as compared to the RF frequency sweep completed when going from \Mev{400} to \gev{8}. See \chapref{con_physics} for more information.}. 
			
			\marginpar{A \keyterm{\gls{btobtransfer}} is the transfer of beam from one machine directly into the RF buckets of another machine.} 

			Notice that the MIRF injection frequency of \Mhz{52.8} is equal to the BRF extraction frequency. This enables the ability to perform \keyterm{\glspl{btobtransfer}}, where injected beam is placed directly into RF buckets in the MI. The phases of the two RF systems must be aligned for efficient transfers as well as to place beam in the correct RF buckets in MI. This phase matching process is called \keyterm{\gls{transfercogging}}. The beam's bunch structure remains the same throughout the process.
			
			\marginpar{\keyterm{\Gls{transfercogging}} is when a downstream machine matches its RF phase to an upstream machine during bucket-to-bucket transfers}
			
			\subsubsection{Resonant Extraction}
			\index{Resonant Extraction}The Main Injector can function in a special mode for Switchyard that allows it to extract beam in a long low-intensity pulse. Known as \keyterm{\gls{resonantextraction}}, this mode uses the main quadrupoles to induce unstable transverse beam oscillations. Specialized quadrupole trims known as ``harmonic quads'' provide a driving force to encourage these oscillations, causing beam to fall out slowly toward extraction to Switchyard. As a fine-tuning mechanism, a system known as ``QXR'' smooths the extraction rate by monitoring MI beam intensity and adjusting current through a special quadrupole corrector. This entire process is colloquially known as ``slow-spill'' extraction. 
			
			\marginpar{\keyterm{Resonant extraction} is a process that causes beam to fall out of the Main Injector in a slow controlled extraction, providing a long pulse of low-intensity beam for Switchyard.}
			
			\subsubsection{Main Injector Machine Parameters}
			\begin{table}[!htb]
				\centering
				\begin{tabular}{| c | c |}
					\hline
					Injection Energy & \gev{8} \\ \hline
					Extraction Energy & \gev{8} or \gev{120} \\ \hline
					Circumference & \m{3319.4} \\ \hline
					Sectors & 6 \\ \hline
					Lattice & FODO \\ \hline
					Revolution Period & \(11.1 \mu s\) \\ \hline
					RF Stations & 20 \\ \hline
					RF Frequency & \Mhz{52.8} to \Mhz{53.1} \\ \hline
					Harmonic Number & 588 \\ \hline
					RF Voltage Gain per Turn & \Mv{3.75}/turn \\ \hline
					Device prefix & I: \\ \hline
				\end{tabular}
				\caption[Main Injector Machine Parameters]{}
				\label{tab:mi-params}
				\vspace{-3em}
			\end{table}
		
		\subsection{Recycler}
		
		\marginpar{\keyterm{\Gls{slipstacking}} is an RF beam manipulation performed to combine bunches in order to effectively double beam intensity.}
		
		The Recycler Ring (RR) is an 8 GeV proton machine that is mostly composed of permanent magnets. The current role of the Recycler is to facilitate proton injection for the Main Injector. The Recycler receives beam from the Proton Source, performs an RF manipulation called ``slip-stacking,'' and then extracts the beam directly to the Main Injector. The Recycler RF system consists of two stations, each with an RF cavity that resonates at \Mhz{52.8}\footnote{Notice that this is the same frequency as BRF at extraction and MIRF at injection.}. The Recycler, like the Main Injector, has a harmonic number of 588. 
		
		\keyterm{\Gls{slipstacking}} is the process of injecting pairs of batches into the Recycler and then merging the pairs to form double-intensity batches. The Recycler is capable of slip-stacking up to twelve batches, which results in six double-intensity batches for extraction to MI. This must occur at a fixed energy because \gev{8} beam is continually injected throughout the slip-stacking process.
		
\begin{figure}[!htb]
			\centering
				\includegraphics[width=1.0\linewidth]{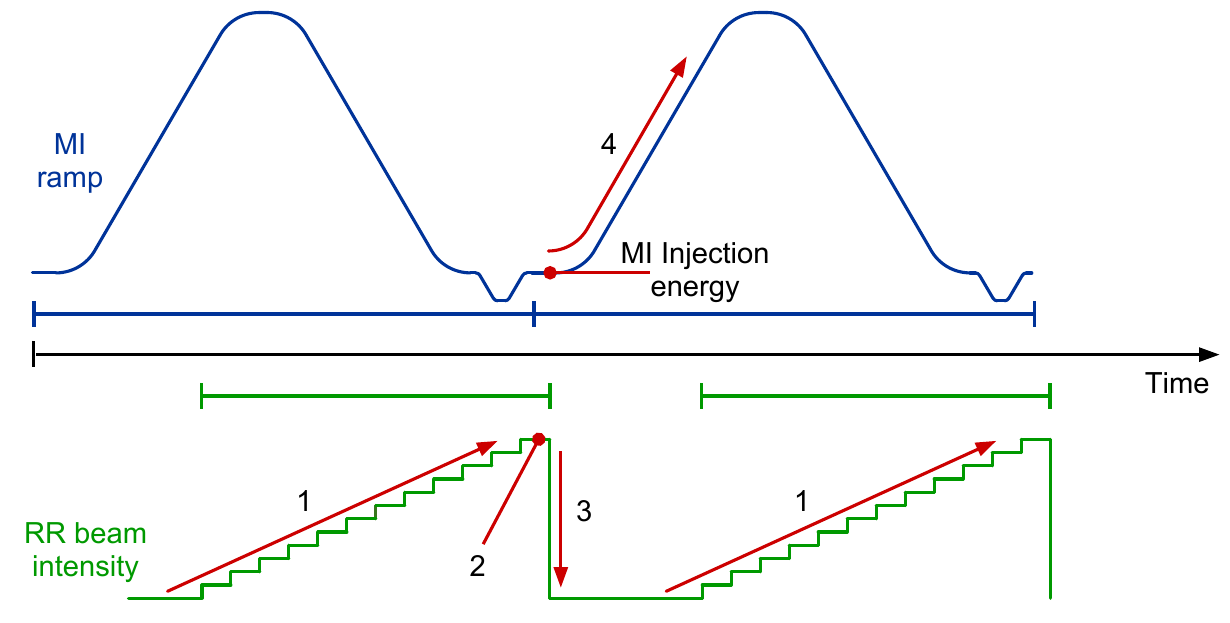}
				\caption[The slip-stacking process]{The slip-stacking process in the Recycler and Main Injector.}
				\label{fig:mirr-ss}
		\end{figure}

		The process of slip-stacking takes place simultaneously with the acceleration of beam in MI, as explained below and in Figure ~\ref{fig:mirr-ss}:
		
		\begin{enumerate}
			\item Slip-stacking begins with the injection of 12 batches into the Recycler
			\item With beam injections into Recycler completed, the beam is merged into 6 batches
			\item Beam is extracted to MI
			\item MI captures and accelerates the beam
			\begin{enumerate}
				\item Step 1 begins again in Recycler while MI accelerates
			\end{enumerate}
		\end{enumerate}
		
			\subsubsection{Recycler Machine Parameters}
			\begin{table}[H]
				\centering
				\begin{tabular}{| c | c |}
					\hline
					Beam Energy & \gev{8} \\ \hline
					Circumference & \m{3319.4} \\ \hline
					Lattice & FODO \\ \hline
					Revolution Period & \(11.1 \mu s\) \\ \hline
					RF stations & 2 \\ \hline
					RF frequency & \Mhz{52.8} \\ \hline
					Harmonic Number & 588 \\ \hline
					Device prefix & R: \\ \hline
				\end{tabular}
				\caption[Recycler Machine Parameters]{}
				\label{tab:rr-params}
				\vspace{-3em}
			\end{table}
			
		\subsection{MI/RR Beam Destinations}
		The beam that enters the Main Injector can be directed to one of three locations: the MI/RR abort beamline, the NuMI beamline, or the P1/P2 Lines. 
		
			\subsubsection{Abort Line}
			The Main Injector and Recycler abort beamlines send beam into the absorber located near MI-40. The RR-40 line and the MI-40 line merge together to send beam to the absorber. 
			
			\subsubsection{NuMI Beamline}
			The NuMI beamline directs \gev{120} protons from the Main Injector to the NuMI target located underground at MI-65. This beamline begins in the Main Injector in the MI-60 straight section. 
			
			\subsubsection{P1/P2 Lines}
			The P1 Line begins in the Main Injector at the MI-52 straight section. Beam destined for the Switchyard beamline or the Muon campus is extracted from the Main Injector by the P1 Line. The P1/P2 Lines are able to transport protons with a kinetic energy of either \gev{8} or \gev{120}.
			
		\subsection{MI/RR Modes of Operation}
		The Recycler and Main Injector have flexible operating modes, but there are two basic categories of machine cycles: HEP and studies. The type of cycle determines the potential destination of the beam.
		
			\subsubsection{HEP Cycles}
			On typical HEP cycles, the Recycler receives beam from the Proton Source and then sends it to the Main Injector. The beam is accelerated by the Main Injector to \gev{120} and extracted to the NuMI beamline or the P1 line. Note that beam on HEP cycles does not always make it to the intended destination. The beam can be aborted if there are problems mid-cycle. 
		
			\subsubsection{Studies Cycles}
			For Recycler studies cycles, the Recycler sends beam down the abort line after finishing the cycle. For Main Injector studies, the Main Injector receives beam from the Recycler and then sends it down the abort line after finishing the cycle. 
		
	\section{External Beamlines}
	External beamlines transport beam from the accelerator complex to the experiments. The Accelerator Division External Beamlines Department is in charge of the \Gls{numi}, \Gls{bnb}, Switchyard, and \gls{mta} beamlines. 
		
		\subsection{Neutrinos at the Main Injector}
		The NuMI beamline transports \gev{120} protons to a target in the underground areas near MI-65. The protons that interact with the target material create secondary particles that decay into neutrinos. A large majority of the neutrinos have the same trajectory as the primary proton beam. The following neutrino experiments use the neutrino beam provided by NuMI: \gls{minos}, \gls{minerva}, and \gls{nova}. 

		It is essential to keep beam losses at a minimum throughout the NuMI beamline because it passes through ground water. Beam loss that occurs upstream of, or in the middle of, the ground water can create tritium. The inhalation or ingestion of tritium is potentially dangerous. The amount of tritium created in the water must be kept to a minimum, which is accomplished by carefully monitoring the beam in the NuMI beamline. 
		
		\subsection{Booster Neutrino Beamline}
		The \Gls{bnb} begins towards the downstream end of the MI-8 Line inside the Main Injector enclosure. This beamline directs \gev{8} protons into the MiniBooNE target. This interaction creates a beam of secondary particles which eventually decay into neutrinos. The MiniBooNE experiment and the MicroBooNE experiment are the users of this neutrino beam. 
		
		\subsection{Switchyard/Fixed Target}
		The Switchyard beamlines begin at the end of the P2 Line and transport beam to the Neutrino-Muon Line and/or the Meson Line. The Neutrino-Muon Line provides beam to the SeaQuest experiment, and the Meson Line provides beam to the MTest experiment\footnote{The Meson Line can also provide beam to the MCenter experiment, but that experiment is not currently operational.}.
		
		\subsection{Muon Test Area}
		The \gls{mta} beamline begins in the \Mev{400} Area of the \index{Linac}Linac enclosure. The H- ions coming from the \index{Linac}Linac are directed towards the \gls{mta} enclosure hall, which is the location of \gls{mta} experiment.
		
	\section{Muon Campus}
	The Muon Campus consists of the Delivery ring, associated beamlines, the target, the g-2 experiment, and the Mu2e experiment. The five beamlines, designated M1 through M5, direct beam to the Delivery ring and from the ring to the experiments. The path that the beam takes through the Muon campus depends on which experiment is taking beam. For this reason, only one experiment is run at a time. Note that the Muon campus is not yet operational.
	
	\begin{figure}[!htb]
		\centering
			\includegraphics[width=1.0\linewidth]{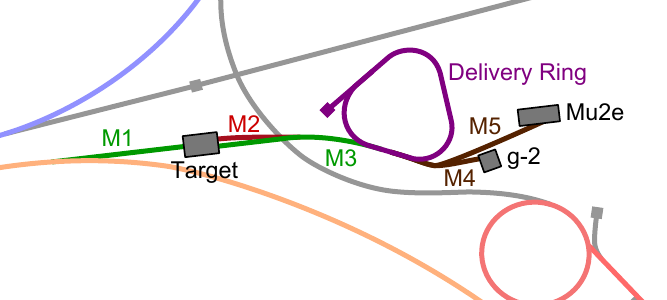}
			\caption{Layout of the Muon Campus}
			\label{fig:acc-map-muon}
	\end{figure}
	
	\section{Review}
\begin{figure}[!htb]
		\includegraphics[width=1.35\linewidth]{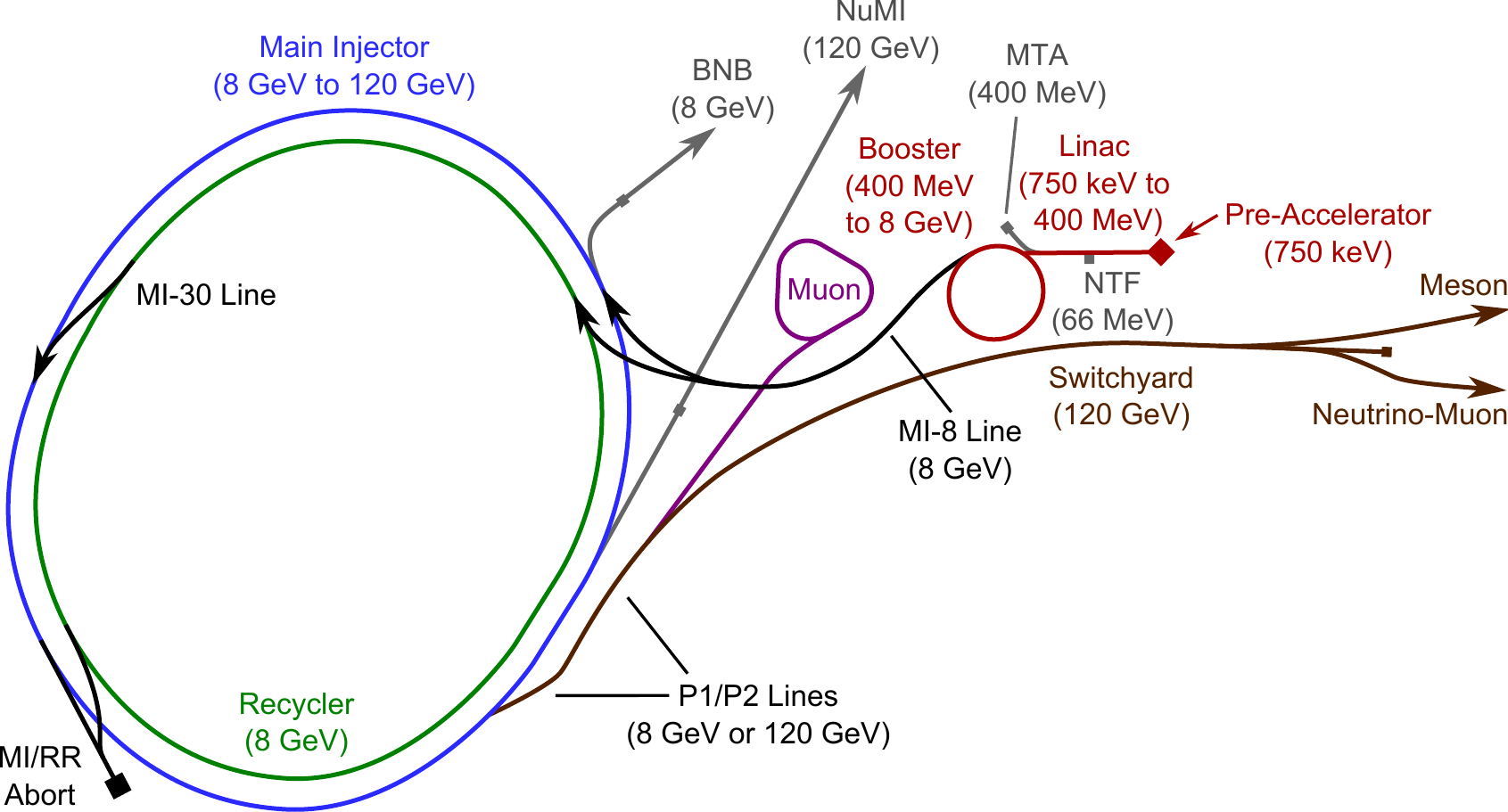}
		\caption{More detailed layout of the Fermilab accelerators}
		\label{fig:acc-map-full}
	\end{figure}

	\textit{Figure~\ref{fig:acc-map-full}} shows the accelerators with some extra details that were not included in the original figure. It is useful to try to replicate this map of the accelerators on paper, at least until it becomes part of your memory. It may also be useful to follow the beam's path throughout the accelerator chain. For example, we will follow a pulse of beam from the source to the \gls{nova} detector.
	
	The beam begins when the source creates a pulse of \kev{35} H- ions and directs it towards the \gls{rfq}. The Einzel lens chopper selects a portion of the beam to continue into the \gls{rfq}. The \gls{rfq} bunches and accelerates the beam to \kev{750}, and the LE \index{Linac}Linac and the HE \index{Linac}Linac accelerate the beam to \Mev{400}. The Booster chopper selects a portion of the beam equal in length to the desired number of turns, and that beam enters the \Mev{400} line and heads towards the \index{Booster}Booster. A stripping foil removes the electrons from the H- ions and proton beam begins to circulate in the Booster accelerator. Beam accelerates to \gev{8} as it makes over 20,000 revolutions around the Booster before being extracted into the MI-8 Line. The devices in the MI-8 line direct the beam into the Recycler, where pairs of bunches are slip-stacked together. The higher-intensity bunches extract to the Main Injector and accelerate to \gev{120}. The NuMI beamline then directs the \gev{120} protons to the \gls{nova} target. The interaction between the proton beam and the target creates a beam of secondary particles (pions, muons, electrons, etc.) which decay into neutrinos. Finally, the beam of neutrinos passes through the Earth and towards the \gls{nova} detector. 
	
%	\begin{figure}[!htb]
%		\noindent\makebox[\textwidth][c]{
%			\begin{minipage}{\fp}
%			\includegraphics[width=0.85\fp]{acc-map-full}
%			\caption{More detailed layout of the Fermilab accelerators}
%			\label{fig:acc-map-full}
%			\end{minipage}}
%	\end{figure}
	
		\subsection{Beam Velocity Throughout the Accelerator Chain}
		The following chart shows the beam kinetic energy and velocity at distinct points in the Fermilab accelerator chain. Notice how the velocity quickly jumps to very near the speed of light in the Proton Source. Due to special relativity, the velocity increases very little in the Main Injector, even though the kinetic energy is increased by a factor of fifteen. Other notable machine kinetic energies are included at the end. 
		
			\begin{table}[!htb]
				\centering
				\begin{minipage}{\textwidth}
					\centering
					\begin{tabular}{| c | c | c | c |}
						\hline
						\textbf{Location} & \textbf{Kinetic Energy} & \textbf{Velocity (m/s)} & \textbf{Velocity (c)} \\ \hline
						Source Output & \kev{35} & \e{2.6e6} & 0.0086 \\ \hline
						RFQ Output & \kev{750} & \e{1.2e7} & 0.04 \\ \hline
						NTF Beam & \Mev{66} & \e{1.07e8} & 0.357 \\ \hline
						LE to HE Linac & \Mev{116.5} & \e{1.37e8} & 0.457 \\ \hline
						Linac Output & \Mev{400} & \e{2.14e8} & 0.713 \\ \hline
						Booster Output & \gev{8} & \e{2.98e8} & 0.994 \\ \hline
						MI Operational Output & \gev{120} & \e{2.9978e8} & 0.999969 \\ \hline
						MI Maximum Output & \gev{150} & \e{2.9978e8} & 0.999981 \\ \hline
						Main Ring Output (1976) & \gev{500} & \e{2.9979e8} & 0.9999982 \\ \hline
						Tevatron\footnote{No longer operational} & \gev{980} & \e{2.9979e8} & 0.99999954 \\ \hline
						LHC\footnote{The Large Hadron Collider is located at CERN in Geneva, Switzerland} & \tev{8} & \e{2.9979e8} & 0.9999999931 \\ \hline
					\end{tabular}
					\caption[Beam Velocity Throughout the Accelerator Chain]{}
					\label{tab:beam-velocity}
					\vspace{-3em}
				\end{minipage}
			\end{table}
		\chapter{Accelerator Physics}\label{chap:con_physics}
	This chapter is an introduction to some of the core concepts in accelerator physics. Keep in mind that the equations themselves are not as important as the concepts that underlie them. The math in this chapter should serve as a supplement to the qualitative description of beam dynamics thus far described in this book. 
	
	We begin by defining the curvilinear coordinate system used in this chapter, and look at the relationship between energy and momentum for very fast (relativistic) particles. We then describe betatron and synchrotron oscillations in more detail, using the analogy of a one-dimensional mass on a spring system. This leads to a conceptual tool known as ``phase space,'' which gives us a new way to think about concepts like phase focusing, buckets, and particle oscillation. Finally, we go into more detail about synchrotron transition, and give a mathematical justification for the required synchronous phase flip. 
	
\begin{figure}[!htb]
		\centering
			\includegraphics[width=.8\linewidth]{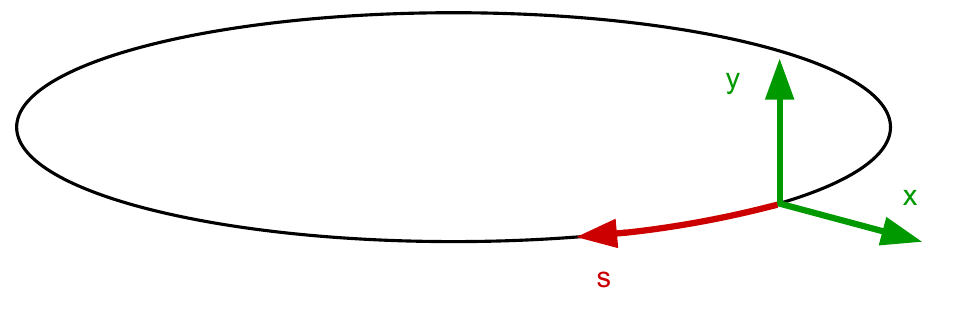}
			\caption{Coordinate system for transverse (x, y) and longitudinal (s) dimensions.}
			\label{fig:xys}
		\end{figure}

	\textit{Figure~\ref{fig:xys}} shows the coordinate system that we will use in the mathematics of beam motion. The variables $x$ and $y$ correspond to horizontal and vertical displacement in the transverse plane, and the variable $s$ refers to longitudinal displacement. Notice that $s$ can describe a curved trajectory if necessary, like the circular beam path of a synchrotron. 
	
	\section{Energy and Momentum}
		Since the purpose of an accelerator is to increase the kinetic energy of charged particles, it is useful for us to look at relativistic momentum-energy relationships. Every particle has a non-zero amount of energy when it is not moving, known as the ``rest energy'' $E_{rest}=mc^2$. For example, the rest energy of a proton is \Mev{938.26}. Thus the total energy of a proton including its kinetic energy is: 
		
		\begin{equation} \label{eq:Etotal}
			E_{total} = E_{rest} + E_{kinetic}
		\end{equation}
		
		Due to the very high particle energies relevant to our accelerators, we must adopt the language and notation of \textit{Special Relativity} to accurately describe the accelerated particles. Particle velocity is described as a fraction of the speed of light in vacuum $c$; this is because, according to \textit{Special Relativity}, nothing can travel faster than $c$. For a particle with velocity $v$, we use $\beta = \frac{v}{c}$ to denote its relativistic velocity\footnote{Accelerator physicists also like to use the letter $\beta$ to describe transverse particle oscillation; be careful to notice the context in which $\beta$ is used to determine if it is the relativistic velocity or the betatron amplitude function.}. Energy and momentum scale using the relativistic factor $\gamma = \frac{1}{\sqrt{1-\beta^2}}$, which gives us a different way to express the total energy: 

		\begin{equation} \label{eq:Etotalgamma}
			E_{total} = \gamma mc^2
		\end{equation}
		
		Similarly, the relativistic momentum is expressed as $p = \gamma mc$. Notice that the momentum and total energy only differ by a factor of the speed of light: $p \propto \frac{E_{total}}{c}$. Thus the relativistic momentum is usually expressed in units of \textit{eV/c}.
		
\begin{figure}[!htb]
		\centering
			\includegraphics[width=.8\linewidth]{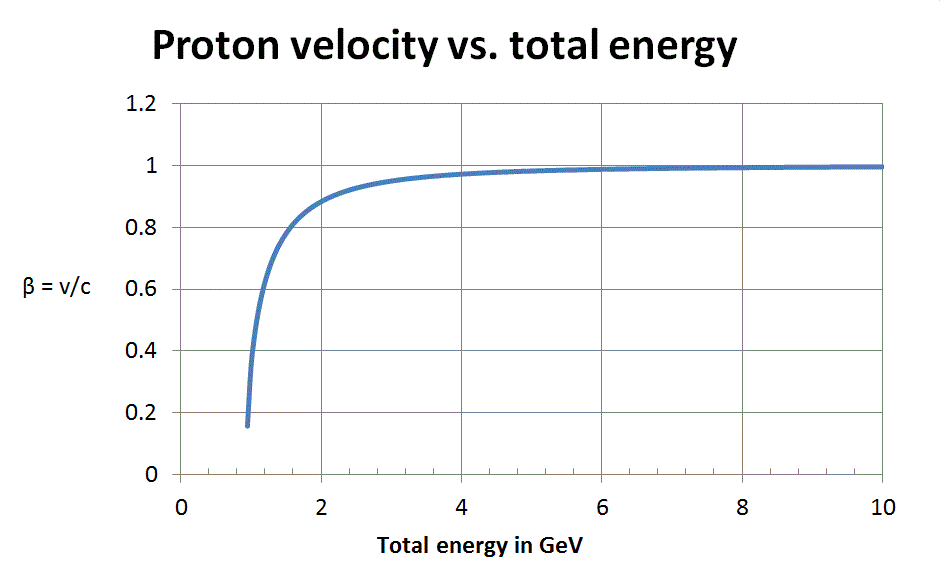}
			\caption[Relativistic particle velocity vs. energy]{Due to relativistic effects, a particle's velocity asymptotically approaches the speed of light in vacuum as its total energy increases.}
			\label{fig:protonEvsV}
		\end{figure}

		As it turns out, only massless particles like the photon can travel at the speed of light. Particles with mass, like the proton, can be accelerated until they are very close to the speed of light, but they can never actually reach that velocity; in mathematical terminology, we say that the velocity ``asymptotically'' approaches $c$. By rewriting \textit{Equation~\ref{eq:Etotalgamma}} to solve for the velocity $v$, we can plot this asymptotic behavior in \textit{Figure~\ref{fig:protonEvsV}} by showing what happens to $v$ as $E_{total}$ increases.
		
		Given the total beam energy, we can now calculate how close to the speed of light the particles are moving. Rewrite the definition of $\gamma$ to solve for $\beta = \sqrt{1-\frac{1}{\gamma^2}}$. From \textit{Equation~\ref{eq:Etotalgamma}}, we can see that $\gamma = \frac{E_{total}}{E_{rest}}$. So for a proton extracted from the Main Injector at \gev{120.00}, $\gamma = \frac{\gev{120.00}}{\Mev{938.26}}$, so $\beta = \sqrt{1-(\frac{938.26*10^6}{120.00*10^9})^2}$, and therefore $\beta = 0.99997$. This means that the Main Injector accelerates protons to 99.997\% the speed of light.
	
	\section{Betatron Motion}
		Recall that the transverse particle motion about the closed orbit is known as \keyterm{\gls{betatronoscillation}}. This motion arises due to the force on the particle from the quadrupole magnets that make up the accelerator's lattice. Accelerators are designed such that the synchronous particle travels along the closed orbit; particles with different momentum than the synchronous particle will undergo betatron oscillation about the closed orbit.
		
		\marginpar{\keyterm{\Gls{betatronoscillation}} is the periodic transverse motion of the beam particles due to the restoring force of the quadrupole magnetic fields.}
		
		Betatron motion is analogous to the simple harmonic oscillation of a mass on a spring. Ignoring gravity and friction for simplicity, the total force exerted on the mass \m{} is both linear and restoring; in other words, $F=-kx$ for displacement $x$ and spring constant $k$. Applying Newtons Second law $F=ma$, inserting the spring force to get $-kx=ma$, and realizing that $a=\frac{d^2x}{dt^2}$, we arrive at the following differential equation of motion:
		
		\begin{equation} \label{eq:springmassdiffeq}
			\frac{d^2x}{dt^2}+\frac{k}{m}x=0
		\end{equation}
		
		\textit{Equation~\ref{eq:springmassdiffeq}} has the following general solution, where $A$ and $\phi$ are determined by particle initial conditions:
		
		\begin{equation} \label{eq:springmasssoln}
			x(t)=A\cos(\sqrt\frac{k}{m}t+\phi)
		\end{equation}
		
		Shown in \textit{Equation~\ref{eq:springmasssoln}}, the solution represents simple harmonic oscillation whose frequency depends on the mass and the spring constant.
		
		Betatron motion follows analogously from the above example, because the magnetic quadrupole force is linearly-proportional to particle displacement (i.e. $F\propto-kx$). If we look at the force the quadrupole magnets exert in the horizontal direction and plug this force into \textit{Newton's Second Law}, we get the following differential equation of motion for particle charge $e$, momentum $p$, and curvature radius $r$.\cite{syphers}\cite{uspas2012}\footnote{The $\frac{1}{r^2}$ term is a centripital effect that arises from our choice of curvilinear coordinates.}:
		
		\begin{equation} \label{eq:beamdiffeq}
			\frac{d^2x}{ds^2}+[\frac{1}{r^2} + \frac{e}{p}\frac{\partial B_x(s)}{\partial y}]x=0
		\end{equation}
		
		Our independent variable is the longidutinal coordinate $s$, becuase we are interested in particle motion as it moves through the accelerator (i.e. as $s$ increases). Notice that the gradient function $B_x(s)$ describes the field contributions from every quadrupole all around the circular accelerator. The solution to this equation tells us the motion of the particle in the x-direction as it moves through the beam pipe (in other words, as a function of $s$), where $A$ and $\phi$ are arbitrary constants determined by initial conditions of the particle: \footnote{\textit{Equation~\ref{eq:beamdiffeq}} has a form that is known as ``Hill's Equation'', which looks like $\frac{d^2x}{ds^2}+K(s)x=0$. The solution to Hill's Equation has been studied extensively in mathematics, which is why we are able to jump to a solution.}
		
		\begin{equation} \label{eq:hillsoln}
			x(s)=A\sqrt{\beta_x(s)}\cos(\psi_x(s)+\phi)
		\end{equation}
		
		\textit{Equation~\ref{eq:hillsoln}} shows that betatron motion is an oscillation not in time, but in beam path $s$. The \textit{beta function} $\beta_x(s)$ describes how the amplitude of the particle oscillation varies throughout the beam trajectory. The \textit{phase function} $\psi_x(s)$ describes the change in phase of the oscillation throughout the beam trajectory. Both $\beta_x(s)$ and $\psi_x(s)$ are properties of the lattice itself. Note that this analysis proceeds in the same way for the vertical plane, just replace $x$ with $y$; therefore, there exists a beta and phase function for each transverse plane\footnote{Note, however, that \textit{Equation~\ref{eq:hillsoln}} is neglecting \textit{dispersion}. If we take this into account, the solution to the horizontal Hill's equation becomes $x(s)=A\sqrt{\beta_x(s)}\cos(\psi_x(s)+\phi)+\frac{\Delta p}{p}D(s)$ where $D(s)$ is the function that describes the dispersion along the beam path and $\frac{\Delta p}{p}$ is the beam momentum spread. In the absence of vertical or skew dipoles, there is no dispersion term in the vertical solution to Hill's equation.}. 
		
		The betatron phase advance $\psi$ is the amount of oscillation period the particle has gone through in a given arc along $s$. In other words, the phase advance is calculated by adding up changes in the inverse of the beta function over a region of the beam path. \textit{Equation~\ref{eq:psi}} expresses the phase advance along the path from $s_1$ to $s_2$ \cite{uspas2012}. This equation is fully derived in \apref{con_apndx_a}.
		
		\begin{equation} \label{eq:psi}
			\psi_{s_1s_2}=\int_{s_1}^{s_2}\frac{ds}{\beta(s)}
		\end{equation}
		
		\marginpar{The \keyterm{\gls{tune}} is the number of betatron oscillations per beam revolution, and is equivalent to the total phase advance around one orbit.}
		
		If we choose to calculate the phase advance around the entire circular beam path, we arrive at what is known as the \keyterm{\gls{tune}}, represented by the symbol $\nu$. In other words, the tune is the number of full betatron oscillations per revolution around the machine. \textit{Equation~\ref{eq:tune}} defines the tune as the total phase advance for a complete closed path divided by the phase advance of a full period $2\pi$:
		
		\begin{equation} \label{eq:tune}
			\nu=\frac{1}{2\pi}\oint\frac{ds}{\beta(s)}
		\end{equation}
		
		Note from \textit{Equation~\ref{eq:tune}} that if the tune is 1, the phase advance around the entire machine will be $2\pi$; in other words, a tune value of 1 means that the particle undergoes a full period of betatron oscillation in a single revolution. 
		
		The tune is a very important beam parameter because certain tune values cause unbounded oscillations in the beam. For example, an integer tune value ($\nu = 1,2,3...$) means that a particle will have the same transverse position every time it passes a given point. This means that a perturbation from an ideal dipole field will push the particle the same way on every pass, leading to an uncontrolled growth in deviation from the closed orbit. Thus we avoid integer tune values operationally because they can cause betatron resonance with dipole field imperfections, leading to beam loss against the beam pipe walls. 
		
\begin{figure}[!htb]
		\centering
			\includegraphics[width=.8\linewidth]{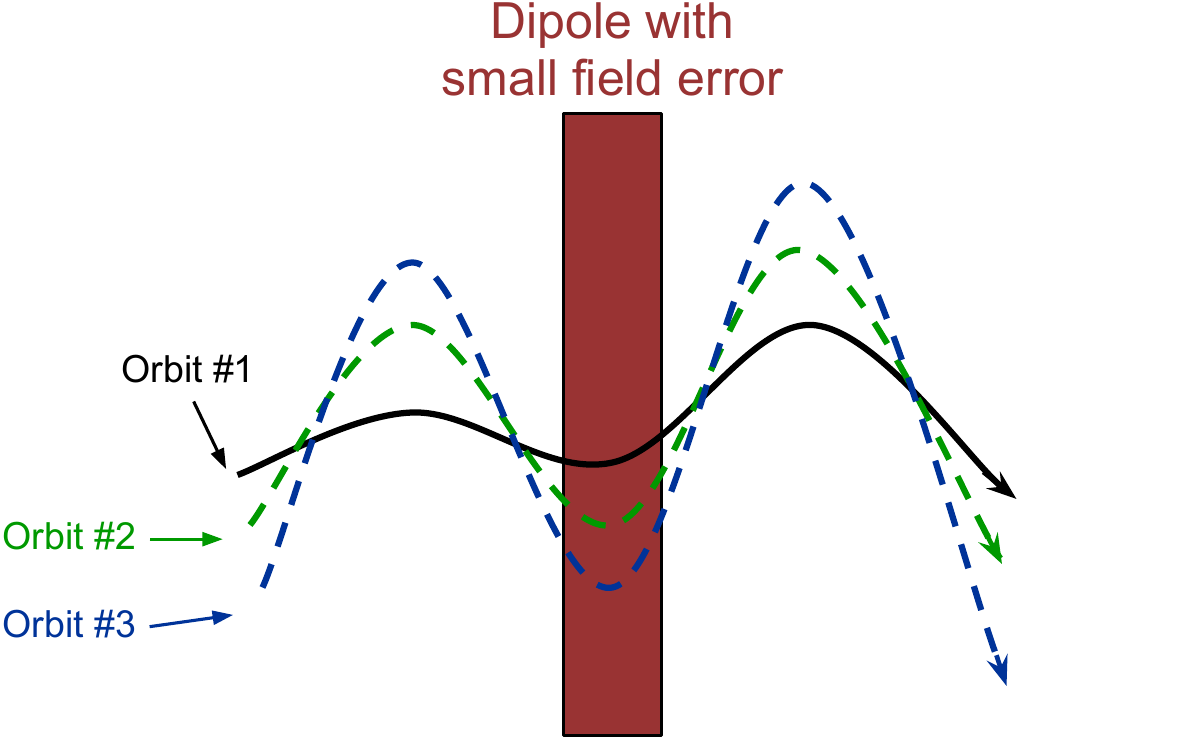}
			\caption[First-order betatron resonance]{First-order betatron resonance due to dipole error and integer tune.}
			\label{fig:imperf}
		\end{figure}

		\textit{Figure~\ref{fig:imperf}} shows how a dipole field error and integer tune leads to an unbounded increase in oscillation amplitude. Each pass through the dipole error results in a larger particle oscillation amplitude. The integer tune allows kicks from the dipole to add up from each pass and cause resonance, which is very bad for beam stability.
		
		Similarly, half-integer tune values ($\nu = \frac{1}{2}, \frac{3}{2}, etc.$) allow for quadrupole field errors to provide a resonant driving force on the beam that leads to beam loss. Third-integer fractional tunes ($\nu = \frac{1}{3}, \frac{2}{3}, etc.$) allow for sextupole resonance conditions, and so on. This means that the operational tune of an accelerator needs to be chosen to avoid resonances; numerically, this means avoiding tune values where the tune $\nu = \frac{k}{n}$, where $k$ is an integer, and $n$ is the integer order of a resonance. First-order betatron resonance, which comes from the combination of dipole field errors and integer tune, are the most destructive to the beam. Higher-order field error resonances from the quadrupoles and sextupoles, etc., are progressively less-destructive to the beam as the order increases. For both planes, unstable resonance occurs when the sum of both tune values make an integer. In other words: $AQ_x + BQ_y = P$, where $A$, $B$, and $P$ are integers and $Q_x$ and $Q_y$ are the horizontal and vertical tunes. The order of the resonance is $A+B$.
			
\begin{figure}[!htb]
		\centering
			\includegraphics[width=.7\linewidth]{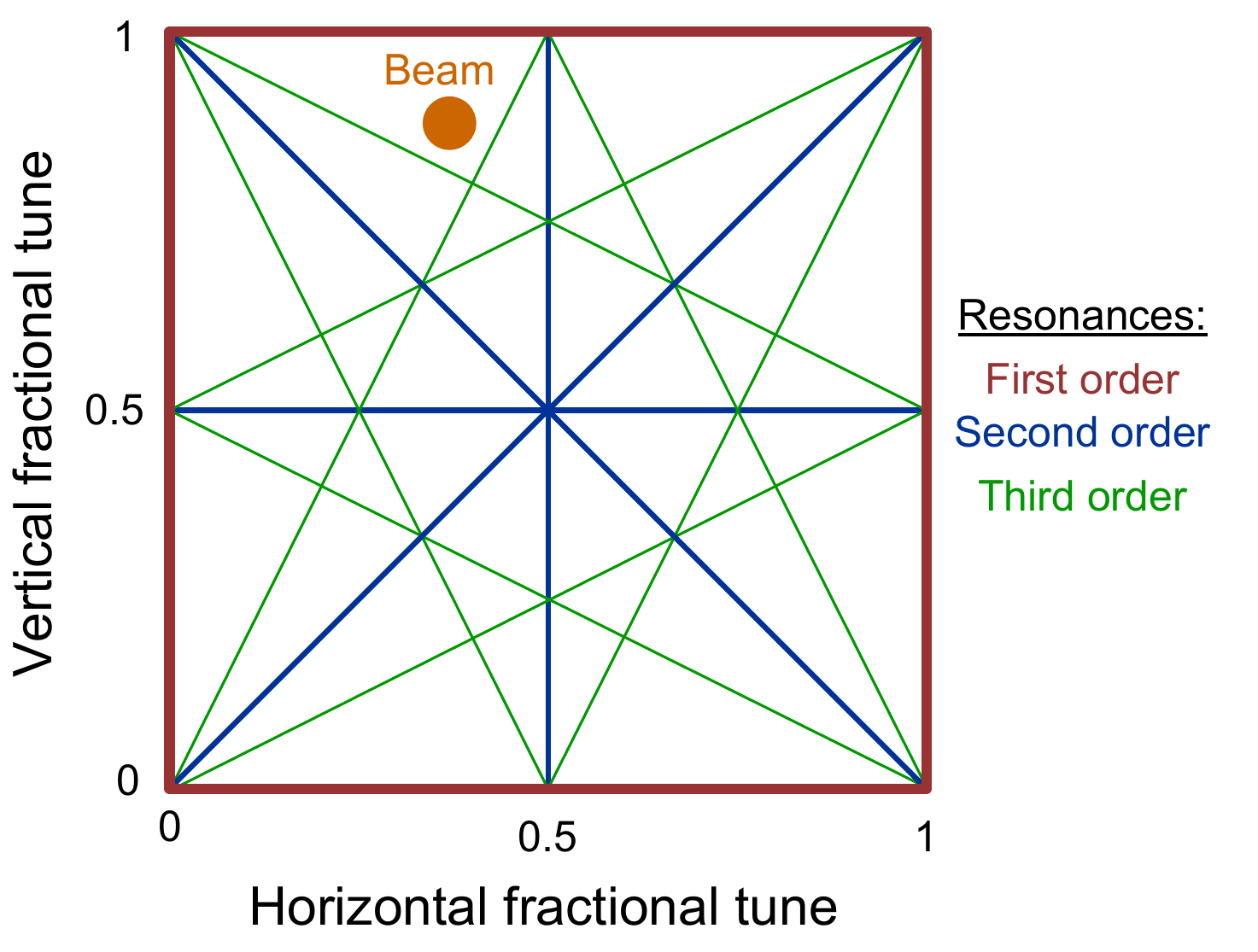}
			\caption{Instability lines in operational tunespace.}
			\label{fig:tunelines}
		\end{figure}

		 To aid in visualizing possible resonances for both horizontal and vertical tune, we use a diagram such as \textit{Figure~\ref{fig:tunelines}} to show lines where instabilities will exist. Accelerator designers must choose an operational tune that does not cross one of the resonant lines in the diagram. In reality, the billions of particles that make up the beam all have slightly different tune values, so there is a non-zero tune spread; thus the operational point on the tune diagram is actually a circular distribution instead of a single dot. The operational tune must be chosen to account for a sufficiently large tune spread as well, to prevent any of the particles in the distribution from crossing a resonance line.
		
	\section{Synchrotron Oscillation}
		Just as quadrupole focusing leads to transverse oscillation, RF phase focusing leads to longitudinal oscillation. The process of phase focusing causes non-synchronous particles to oscillate about the synchronous phase, and this motion is known as \keyterm{\gls{synchrotronoscillation}}. This motion is an inevitable result of RF phase focusing. Whereas a mass on a spring will alternate between maximum displacement and velocity, a particle will alternate between maximum energy and phase deviations from the synchronous particle. The mass is subject to the restoring force of the spring, and errant particles are subject to the restoring force of the RF electric field.
		
		\marginpar{\keyterm{\Gls{synchrotronoscillation}} is the longitudinal motion of the particles in the bucket, caused by RF phase focusing.}
		
\begin{figure}[!htb]
		\centering
			\includegraphics[width=1.0\linewidth]{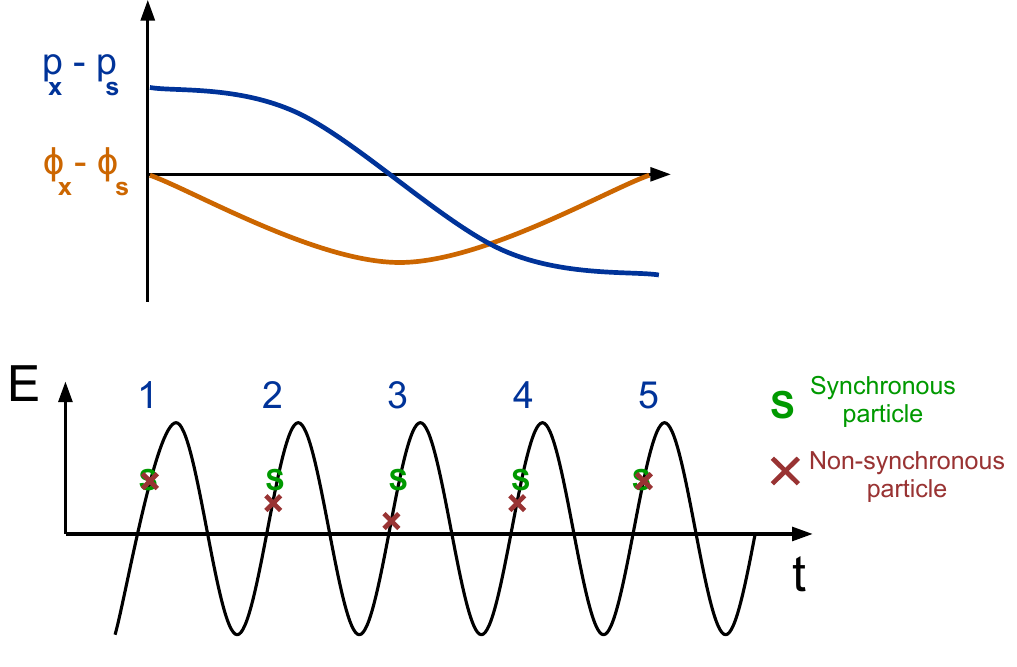}
			\caption{Synchrotron oscillation about the synchronous particle.}
			\label{fig:synchosc}
		\end{figure}

		Consider \textit{Figure~\ref{fig:synchosc}}, which shows the phase and momentum evolution of a non-synchronous particle \textbf{\textcolor{nal-red}{X}} as compared to the synchronous particle \textbf{\textcolor{nal-green}{S}}. The figure steps along successive passes through either the same RF cavity in a circular accelerator, or a chain of many cavities in a linear accelerator \cite{introRF}.
		
		\begin{itemize}
			\item \textit{Pass 1:} \textcolor{nal-red}{X} matches the phase of \textcolor{nal-green}{S}, but has too much energy.
			\item \textit{Pass 2:} \textcolor{nal-red}{X} arrives earlier than \textcolor{nal-green}{S} due to its higher momentum, receiving less of a kick, thus reducing the momentum difference between \textcolor{nal-red}{X} and \textbf{S}.
			\item \textit{Pass 3:} \textcolor{nal-red}{X}'s momentum now matches \textcolor{nal-green}{S}, but \textcolor{nal-red}{X} is now at a lower phase than \textcolor{nal-green}{S}. In other words, even though \textcolor{nal-red}{X} has been slowed down to match the velocity of \textcolor{nal-green}{S}, \textcolor{nal-red}{X} is now arriving much earlier than \textcolor{nal-green}{S}.
			\item \textit{Pass 4:} Now \textcolor{nal-red}{X} once again sees less of an electric field than \textcolor{nal-green}{S}, causing \textcolor{nal-green}{X} to arrive later on the next pass, thus reducing the phase difference between \textcolor{nal-red}{X} and \textcolor{nal-green}{S}.
			\item \textit{Pass 5:} \textcolor{nal-red}{X} once again matches the phase of \textcolor{nal-green}{S}, except now \textcolor{nal-red}{X} has a lower momentum than \textcolor{nal-green}{S}.
		\end{itemize}
		
		This process continues until \textcolor{nal-red}{X} once again arrives at the same phase as \textcolor{nal-green}{S} but with too much momentum, and the process repeats again. The result is that \textcolor{nal-red}{X} oscillates between having too much or too little momentum, and arriving too early or too late. The important thing to note is that this oscillation is stable, as long as the maximum phase and momentum deviations of \textcolor{nal-red}{X} remain within the bucket. Any deviations that are too large will cause oscillations that keep increasing with amplitude, and the particle will be lost.

		 As long as momentum or phase offsets from the synchronous particle are within the stable space of the RF, the beam can oscillate longitudinally without being lost. We call this stable space the ``RF bucket.'' In the next section on Phase Space, we introduce a useful way of visualizing the bucket.
		
	\section{Phase Space}
		A useful way to see how an oscillating system changes is by plotting the oscillation amplitude and its derivative. This type of plot is known as a ``phase space'' plot, which is a way to characterize the periodic motion of any general system. Phase space plots show the oscillation amplitude on one axis and its derivative on the other, plotting each as a dot on the graph at different times in the cycle. For example, the phase space plot for a mass on a spring shows its displacement $x$ and velocity $v$ (the time derivative of $x$) for multiple points in time.
		
		For betatron oscillation, phase space involves transverse particle displacement $x$ and its derivative with respect to the beam path $\frac{dx}{ds}$, plotted for different values of $s$ along the accelerator. In the context of beam synchrotron oscillation, we plot the particle phase $\phi$ and energy $E$ at different points in time. The shape that the beam traces in phase space provides vital information about beam quality and stability.
		
\begin{figure}[!htb]
		\centering
			\includegraphics[width=.5\linewidth]{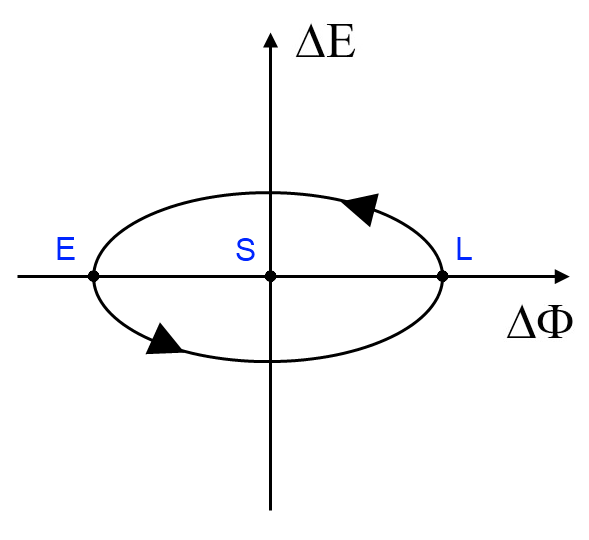}
			\caption{Synchrotron oscillation in phase space.}
			\label{fig:phasesynch}
		\end{figure}

		This method of looking at longitudinal particle motion, pictured in \textit{Figure~\ref{fig:phasesynch}}, helps us see how phase focusing leads to synchrotron oscillation. Phase focusing can correct for the phase error of an early particle at point \textbf{E}, but this reduces the particle's energy. Eventually, the particle arrives too late at point \textbf{L}, because its slightly smaller energy has led to a phase error again. The repetition of fixing one error and leading to the other is precisely what we have defined as synchrotron oscillation.
		
		\marginpar{The RF \keyterm{\gls{bucket}} is the set of energy and phase deviations from the synchronous particle that can be phase focused. This may be mapped as an area in phase space.}
		
\begin{figure}[!htb]
		\centering
			\includegraphics[width=.8\linewidth]{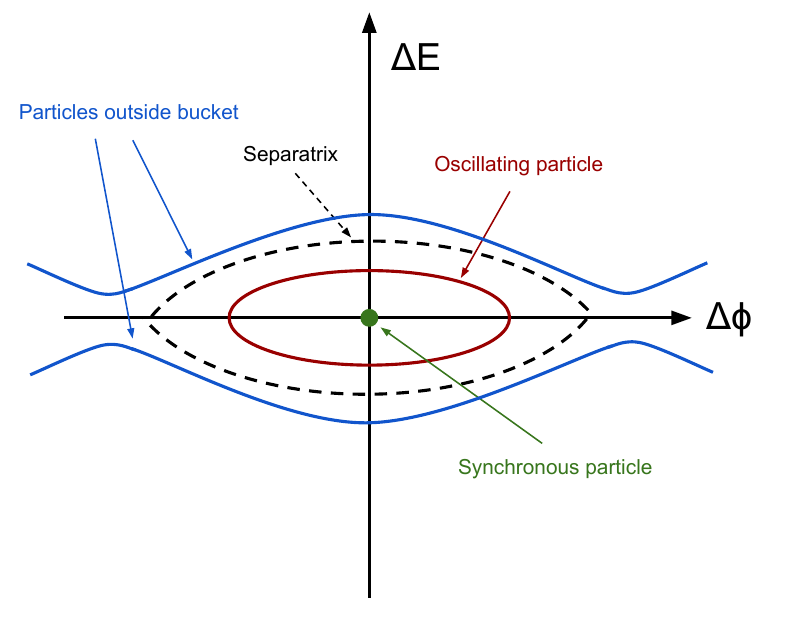}
			\caption{RF bucket in phase space.}
			\label{fig:psbucket}
		\end{figure}

		Recall that particles with sufficiently high phase or energy deviations from the synchronous particle cannot be longitudinally focused by the RF. We have defined the set of such deviations that can still be focused as the RF \keyterm{\gls{bucket}}. Phase space conceptualization gives a more intuitive way of picturing the bucket. The region of stability appears as an area in phase space, the boundary of which is called the ``\gls{separatrix}.'' As long as a particle's deviations $\Delta E$ and $\Delta\phi$ are inside the bucket area in phase space, stable phase focusing is possible. \textit{Figure~\ref{fig:psbucket}} shows the bucket in longitudinal phase space with three kinds of particles: the green synchronous particle lies at the origin, a red particle inside the bucket undergoes synchrotron oscillation, and blue particles outside the bucket undergo unstable motion.
		
		We now look at how transverse oscillations appear in phase space. In the horizontal plane, the phase space diagram of the beam's betatron motion involves particle position $x$ and divergence $\frac{dx}{ds}$ ($s$ being the curvilinear coordinate for the beam path); this motion also appears as an elliptical contour in phase space. Think of $\frac{dx}{ds}$ as the angle of the particle's trajectory during betatron oscillation. The vertical plane is similar, and uses $y$ and $\frac{dy}{ds}$ for its phase space axes. 
		
\begin{figure}[!htb]
		\centering
			\includegraphics[width=.45\linewidth]{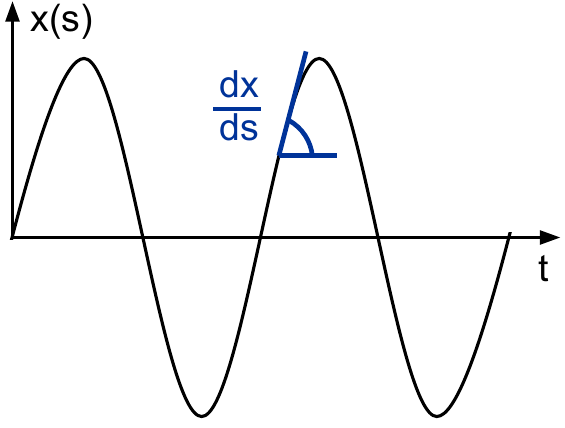}
			\includegraphics[width=.45\linewidth]{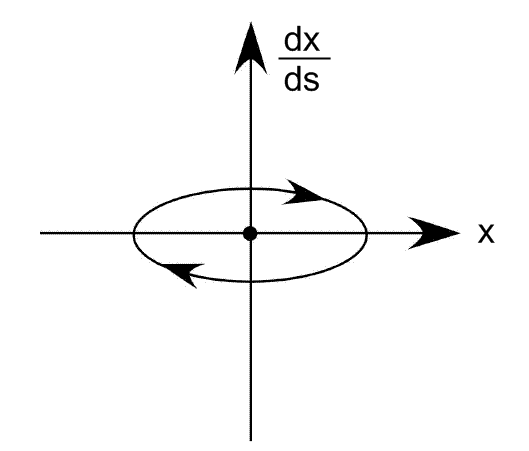}
			\caption[Betatron oscillation]{Betatron oscillation as a function of beam path $s$ and in phase space.}
			\label{fig:phasebeta}
		\end{figure}

		\textit{Figure~\ref{fig:phasebeta}} shows horizontal betatron oscillation in two equivalent ways. The left plot shows the oscillation as a function of $s$, the distance along the beam path, and also illustrates what we mean by the divergence $\frac{dx}{ds}$. The right plot shows the same oscillation in transverse phase space, where the particle oscillation corresponds to tracing through an ellipse.
		
	\section{Emittance}
	
		\marginpar{\keyterm{\Gls{emittance}} is the area that beam occupies in phase space, and shows how much the particles are oscillating in a particular dimension.}
		
		The phase space method of graphing oscillations allows us to define a useful beam parameter known as \keyterm{\gls{emittance}}. For a particle undergoing betatron oscillation, its transverse emittance is defined as the area of the particle's ellipse in transverse phase space.\footnote{The horizontal emittance is the area in horizontal ($x, \frac{dx}{ds}$) phase space, and the vertical emittance is the area in vertical ($y, \frac{dy}{ds}$) phase space.} Similarly, the longitudinal emittance is defined as the area of the particle's ellipse in longitudinal phase space.  The emittance is a convenient number to give us an idea of how much the particle is oscillating, and it is related to the size and quality of the beam. This is an important parameter because a particle with too high an emittance will not remain stable in the accelerator. 
		
		The beam is made up of \e{10e9} to \e{10e11} particles, so it appears as a cloud in phase space; the emittance of the beam as a whole is the area that this entire cloud occupies in phase space. Beam with high transverse emittance will have some particles collide with magnets and beam pipe, causing beam loss and extraneous radiation. Beam with large longitudinal emittance will have some particles with energy or phase errors too large to be phase focused into stable oscillations; in other words, large emittance beam may not all fit into the RF bucket. 
		
\begin{figure}[!htb]
		\centering
			\includegraphics[width=.8\linewidth]{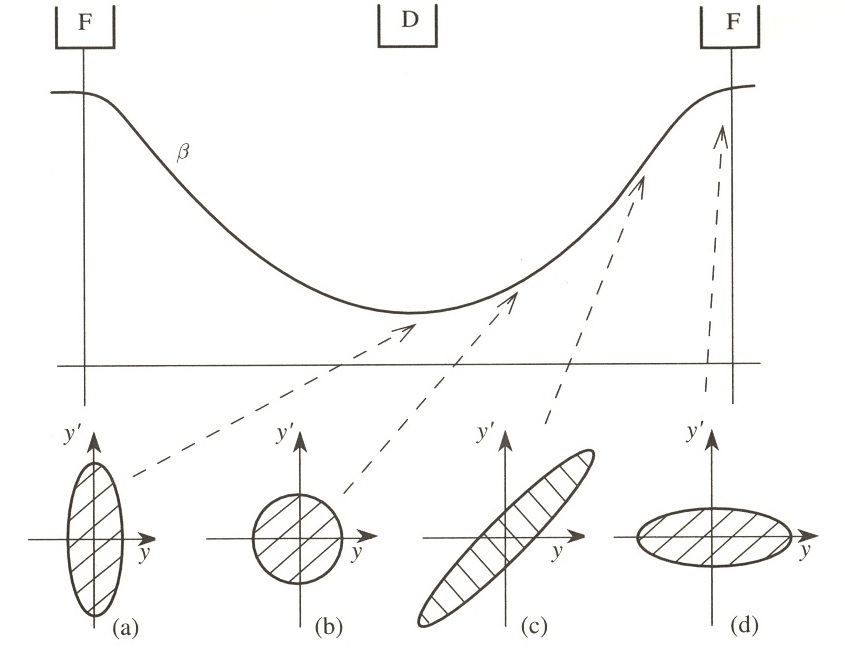}
			\caption[Constant transverse emittance through a lattice]{Transverse emittance is constant while the shape of the phase space distribution changes through the lattice.}
			\label{fig:latticeemittance}
		\end{figure}

		Beam emittance will not change unless the beam is acted upon by a nonconservative external force; we cannot reduce beam emittance without using active cooling systems.\footnote{This is a restatement of Liouville's Theorem, which states that the phase space density of a Hamiltonian system is constant along any trajectory in phase space. Actually, in order to correctly satisfy Liouville's Theorem, we must apply relativistic corrections to our definition of emittance. See \textit{Appendix~\ref{chap:con_apndx_a}} for an explanation of this adjusted emittance.} This is why preservation of low-emittance beam is important: once something causes beam emittance to increase (due to lattice mismatch between machines, intra-beam scattering, etc.), that high emittance will propagate from machine to machine, and it is very difficult to reduce the emittance again. In other words, while it is relatively easy for beam emittance to increase due to some error in the accelerator, it is \textit{not} easy to reduce beam emittance without expensive and complicated ``cooling'' systems. \textit{Figure~\ref{fig:latticeemittance}} shows that the transverse phase space distribution of beam changes throughout the lattice, but the area of the ellipse remains constant \cite{wilson}.
		
		Notice in \textit{Figure~\ref{fig:latticeemittance}} that as the amplitude function $\beta$ changes while beam passes through the lattice, the shape of the beam ellipse in phase space also changes. However the area of the beam's ellipse, the emittance, does not change.
		
	\section{Transition}
		For a synchrotron accelerator, two factors determine how long a particle will take to complete a revolution around the machine: the path length the particle travels, and the velocity at which it travels. Due to the relativistic effects discussed earlier, it becomes more and more difficult to increase the velocity of a particle as its energy increases. At some point during acceleration, an increase in kinetic energy ceases to make much of a difference in particle velocity (refer to \textit{Figure~\ref{fig:protonEvsV}}). Thus there is a particular energy at which the velocity of the particle plays less of a role than the path length in affecting the transit time. This effect is known as \keyterm{\gls{transition}}, because the beam behaves differently on either side of this critical \keyterm{\gls{transitionenergy}}. The transition energy is a characteristic of the accelerator itself, and depends on its design parameters. 
		
		\marginpar{\keyterm{\Gls{transition}} is a relativistic effect in synchrotrons where particle path length begins to have more of an effect on transit time than particle velocity. This occurs at the \keyterm{\gls{transitionenergy}}, and requires a synchronous phase flip to maintain RF phase focusing.}
		
\begin{figure}[!htb]
		\centering
			\includegraphics[width=1.0\linewidth]{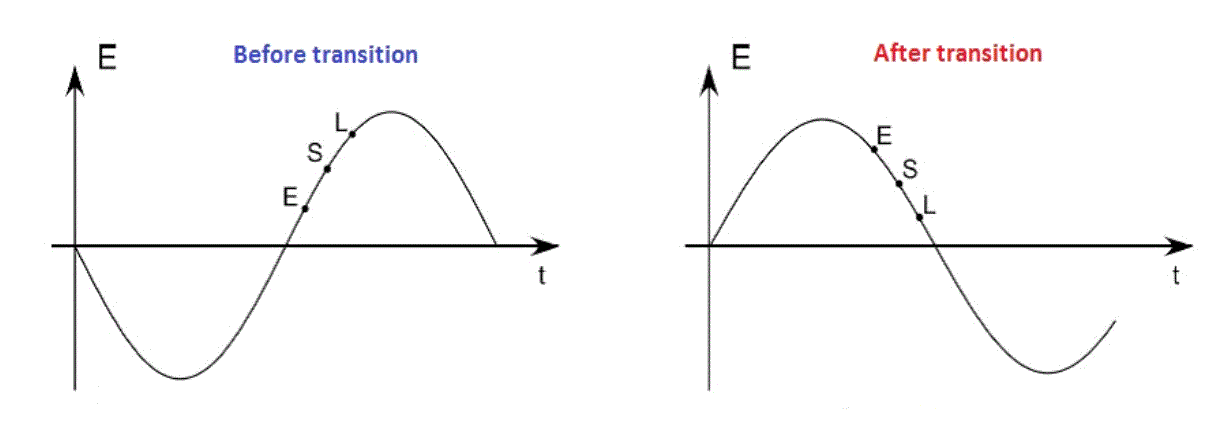}
			\caption[Synchronous phase flip at transition]{The synchronous phase is ``flipped'' to the other side of the RF curve after transition to maintain phase focusing stability.}
			\label{fig:phaseflip}
		\end{figure}

		Phase focusing does not work at the transition energy, because small changes in particle velocity are exactly canceled by small changes in path length. However, by flipping the synchronous phase to the downward-slope side of the RF curve, phase focusing can recover after passing through the transition energy. Fermilab's \index{Booster}Booster and Main Injector accelerate beam through their transition energies, and so must perform this phase flip to move the synchronous phase from the positive to the negative slope of the RF curve. This allows phase stability to recover after the machine has passed through transition. \textit{Figure~\ref{fig:phaseflip}} shows the difference between stable position on the RF curve before and after transition.
		
		To understand how beam behaves on either side of transition, consider \textit{Equation~\ref{eq:tautransition}} that shows the fractional transit time change that results from a particle momentum change. \textit{Equation~\ref{eq:tautransition}} is derived in \textit{Appendix~\ref{chap:con_apndx_a}}. In this expression, we use the familiar relativistic gamma factor $\gamma$ discussed at the beginning of this chapter, which is proportional to the beam energy. We introduce a new variable $\gamma_t$ that is proportional to the transition energy of a particular accelerator. 
		
		\begin{equation} \label{eq:tautransition}
			\frac{\Delta\tau}{\tau} = (\frac{1}{\gamma_t^2} - \frac{1}{\gamma^2}) \frac{\Delta p}{p}
		\end{equation}

		For compactness, we call the coefficient $\frac{1}{\gamma_t^2} - \frac{1}{\gamma^2}$ the \textit{slip factor}, and denote it with the variable $\eta$. \textit{Equation~\ref{eq:tautransitionsimp}} uses this new variable, and shows that when the particle's energy (proportional to $\gamma$) exceeds the transition energy (proportional to $\gamma_t$), $\eta$ changes sign. Before transition, $\eta$ is negative, so a fractional momentum increase corresponds to a transit time decrease, because the particle speeds up. After transition, $\eta$ is positive, so a fractional momentum increase corresponds to a transit time increase. Therefore, the sign of $\eta$ tells us if we are before or after transition.
		
		\begin{equation} \label{eq:tautransitionsimp}
			\frac{\Delta v}{v}=\eta\frac{\Delta p}{p}
		\end{equation}
		
		This sudden sign change has consequences for phase focusing, since we choose the synchronous phase to give a large kick to early particles and a smaller kick to late particles. \textit{Equation~\ref{eq:tautransition}} illustrates the need for the synchronous phase flip at transition to compensate for the sign flip in the slip factor $\eta$. The phase flip allows phase focusing to continue working after transition, and we regain longitudinal stability of the beam.
		
		Transition occurs in the \index{Booster}Booster at \gev{4.2}, or about 17 milliseconds into its cycle, and the Main Injectors transition energy is \gev{18.2}. Notice that only synchrotrons have a transition energy; transition does not occur in the \index{Linac}Linac. 
		\chapter{Instrumentation}\label{chap:con_inst}
	This section describes the accelerator components that provide beam diagnostic information. These instruments measure beam properties such as intensity, transverse position, longitudinal profile, emittances, and secondary radiation. Beam instrumentation is vital to accelerator operations, because it is how we ``see'' what the beam is doing.
		
		\section{Beam Position Monitors}
			A \gls{bpm}, is a passive electromagnetic pickup that measures transverse beam position at a single location. Ideally, BPMs are passive in that they have no effect on the beam. All of our accelerators use BPMs to see the transverse motion of the beam. Pictured in \textit{Figure~\ref{fig:bpmelectrode}}, the \gls{bpm} in its simplest form is a pair of parallel conducting striplines, or ``plates,'' on either side of the beam. As positively-charged beam passes, the electric field induces a voltage on both plates. 
				
			If the beam is in the center of the pipe, the induced voltages on plates A and B will be equal; however, transverse position offsets in the beam will induce uneven voltages on the plates. Thus we can measure transverse beam position by measuring the voltage difference between the two plates. \textit{Figure~\ref{fig:bpmelectrode}} depicts a \gls{bpm} that only measures vertical displacement, but some BPMs are designed with four plates to measure both horizontal and vertical dimensions in one device.
			
			\marginpar{
				\centering
				\includegraphics[width=1.0\linewidth]{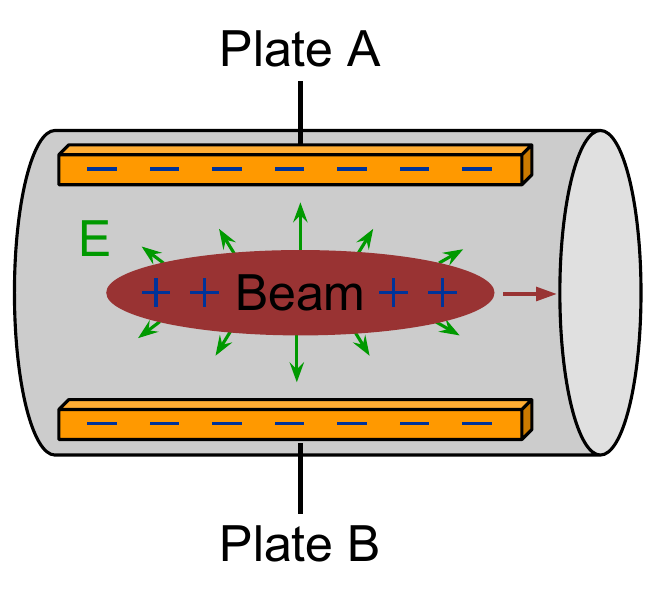}
				\captionof{figure}{Beam position monitor}
				\label{fig:bpmelectrode}
			}
			
			\glspl{bpm} also provide beam momentum information for synchrotron RF systems in the form of ``RPOS feedback''\footnote{Radial Position feedback}. Due to dispersion, particles have a different horizontal position depending on their momentum. The RF system of a synchrotron uses a single horizontal \gls{bpm} in a specific location to monitor the beam momentum; any mismatch between the RF and dipole bending field will manifest as a transverse position offset in the RPOS \gls{bpm}. So the LLRF system changes the RF phase to adjust beam momentum until the RPOS \gls{bpm} shows the beam at the desired transverse position. 
			
		\section{Resistive Wall-Current Monitors}
			A wall current monitor is a passive diagnostic device that is used mostly to measure longitudinal beam profile and emittance. The device measures the current that travels along the wall of the vacuum chamber due to the passing charged beam. 
			
			\marginpar{
				\centering
				\includegraphics[width=1.0\linewidth]{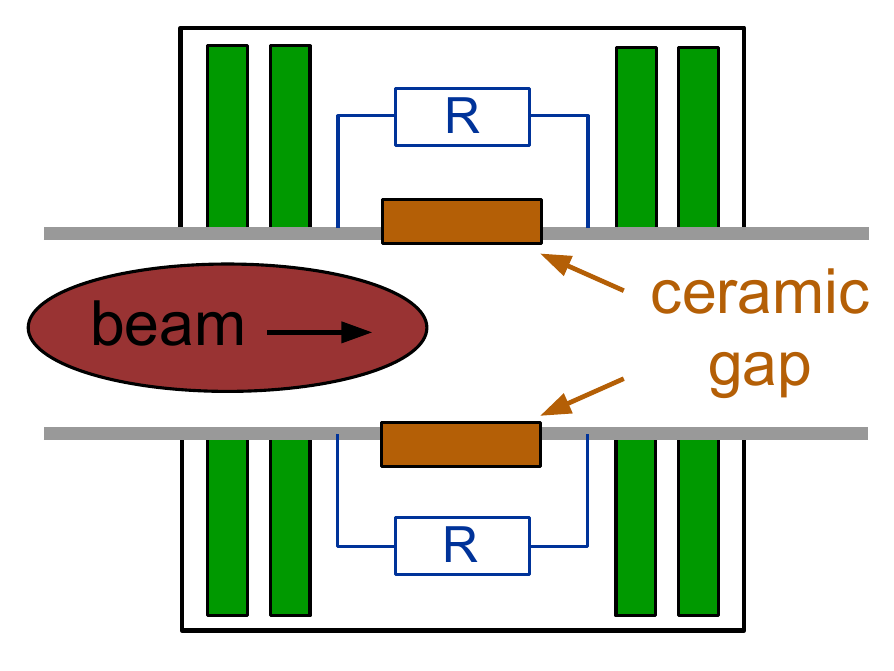}
				\captionof{figure}{Resistive wall-current monitor}
				\label{fig:RWM}
			}
			
\begin{figure}[!htb]
				\centering
				\includegraphics[width=.8\linewidth]{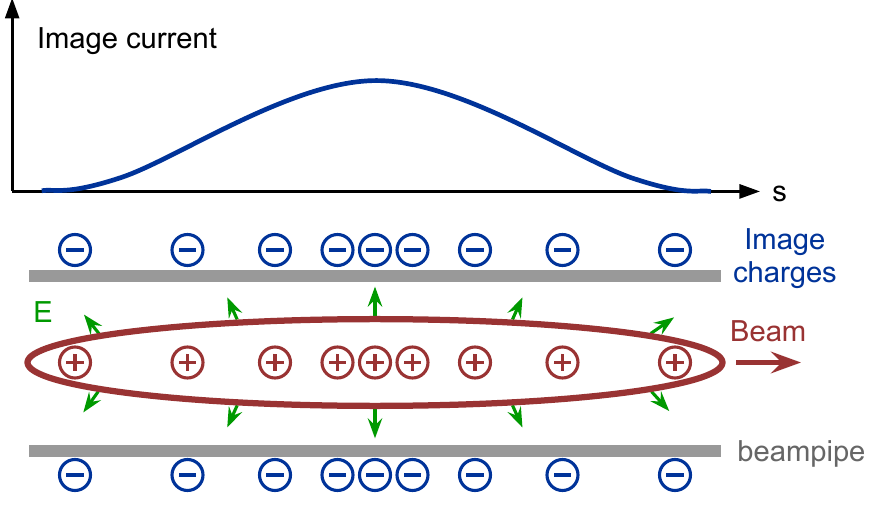}
				\caption{Beam-induced image current on the beam pipe wall.}
				\label{fig:wallcurrent}
			\end{figure}

			As charged particles from the beam pass through, the electric wake field induces image charges in the conducting beam pipe that follow the beam. The passage of image charges in the beam pipe is known as ``image current'' or ``wall current.'' Since this current is proportional to beam intensity, the wall current measured at a single physical point as a function of time is proportional to the longitudinal profile of the beam. \textit{Figure~\ref{fig:wallcurrent}} shows a model of induced image charge and its associated wall current as beam passes.
			
			The detector itself consists of a resistor connected across a gap in the beam pipe, pictured in \textit{Figure~\ref{fig:RWM}}. By measuring the voltage across the resistor, the wall current magnitude can be calculated by Ohm's Law $I_{wall} = \frac{V_{resistor}}{R}$. Ferrite surrounds the gap to adjust the frequency response of the detector. While the measured wall current is dependent on the transverse position of the beam, allowing the wall current monitor to be used like a \gls{bpm}, this feature is typically not desired. To reduce the transverse dependence, the wall current is often measured across resistors in several different locations and averaged so the signal depends purely on the longitudinal structure of the beam\cite{bpm}\cite{bpm2}.
			
		\section{Beam Loss Monitors}
			Beam loss monitors, or ``BLMs,'' are devices that measure ionizing radiation caused by the beam. Beam particles that strike the beam pipe and magnet apertures cause secondary emission of charged particles. This radiation can cause damage to accelerator components, excessive heat load on cooling systems, and personnel hazard due to high background and activation levels. The BLMs provide monitoring and protection by inhibiting beam when loss radiation becomes high enough. They also show areas where beam may be mistuned or mis-steered.
			
\begin{figure}[!htb]
				\centering
				\includegraphics[width=.9\linewidth]{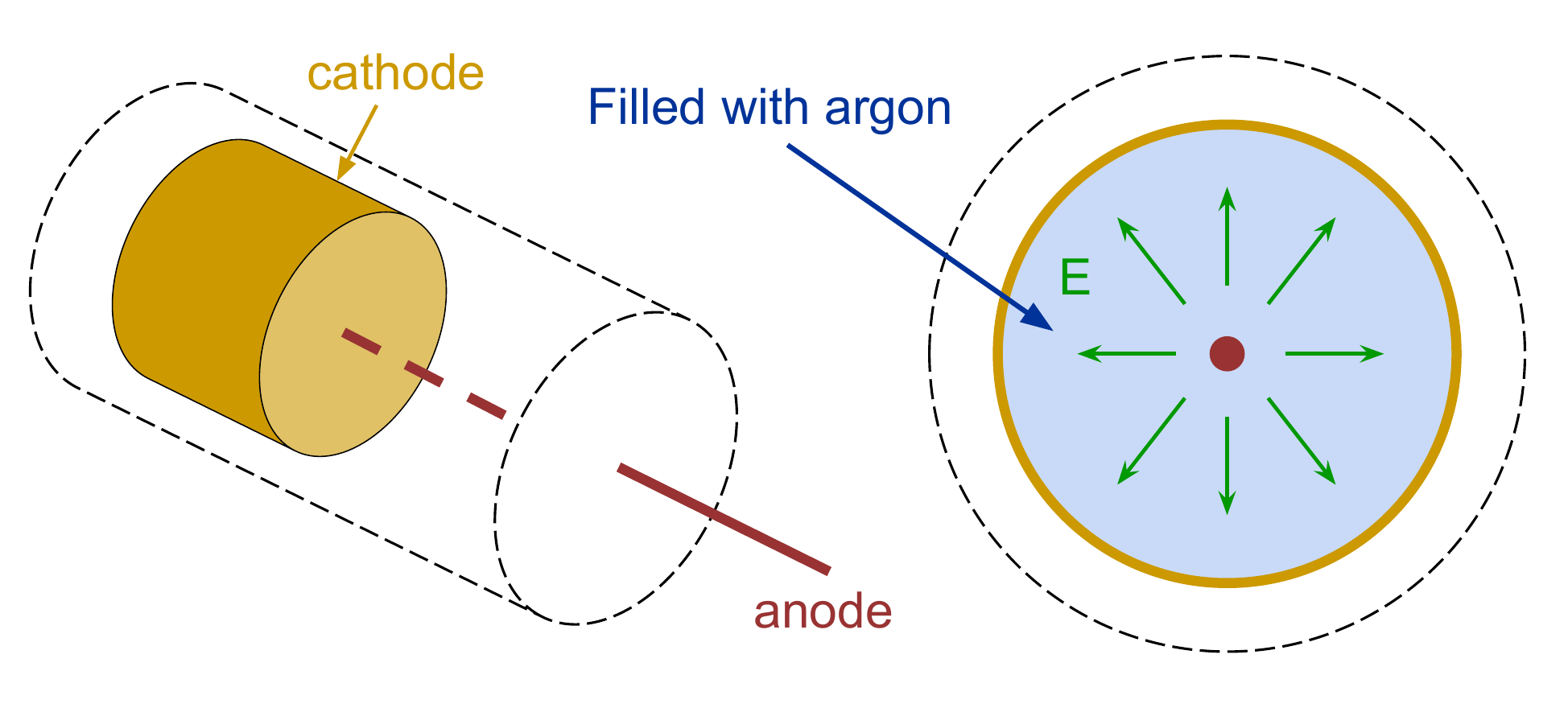}
				\caption{Model of an argon ion chamber beam loss monitor}
				\label{fig:blm}
			\end{figure}

			The most common form of loss monitor is the argon ionization chamber. It is a cylindrical glass cylinder filled with argon with two coaxial electrodes, pictured in \textit{Figure~\ref{fig:blm}}.
			
			Charged radiation particles that pass through the chamber ionize the argon molecules. The high voltage on the outer cathode electrode accelerates the ions toward the anode inner electrode, creating an electrical current. The measured current through the anode is proportional to the number of radiation particles passing through the ion chamber. 
			
		\section{Toroids}
			The toroid measures the amount of beam current that passes through its aperture, and is pictured in \textit{Figure~\ref{fig:toroid}}. Since beam is a collection of moving charged particles, it is conceptually equivalent to an electric current. Thus the beam, like an electric current, generates a circular magnetic field that spirals around the beam's velocity vector. This magnetic field impinges on the ferrite of the toroid, which amplifies its magnitude. This field induces a current in the pickup loop wrapped around the ferrite; the induced current in the pickup loop is directly proportional to the beam current that passed through the toroid.
		
			\marginpar{
				\centering
				\includegraphics[width=1.0\linewidth]{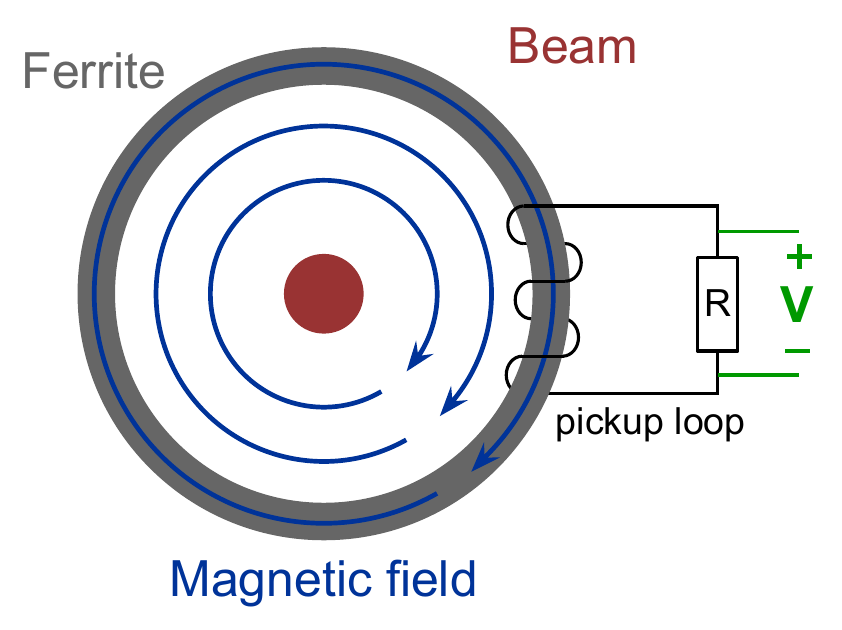}
				\captionof{figure}{Model of a beam toroid}
				\label{fig:toroid}
			}
		
			A convenient way to measure the current in the pickup loop is to connect a known resistance $R$, and measure the voltage across it\footnote{For voltage $V$, current $I$, and resistance $R$, Ohm's Law states that $V=IR$. Thus we can measure the current through the pickup loop by measuring the voltage across the resistor, and calculating $I=\frac{V}{R}$.}. Thus the voltage across the resistor R is proportional to the beam current.
		
			The beam toroid is vital for accelerator operations: it allows us to monitor the amount of beam we send to particular areas, and also shows when beam intensity decreases due to mistuning or component issues. However, because the toroid relies on electromagnetic induction to sense beam current, it cannot see unbunched DC beam\footnote{Faraday's Law of electromagnetic induction states that the electric field through a loop is equal to the rate of change of magnetic flux through the surface of that loop: $\oint \vec{E}\cdot d\vec{l}=-\frac{d}{dt}\iint\vec{B}\cdot d\vec{S}$. If the beam is unbunched, it looks like a steady line of current with an unchanging magnetic field. Thus there is zero change in the magnetic flux through the toroid, and the induced current (proportional to the electric field) is zero. Therefore, a beam toroid cannot measure unbunched DC beam because its unchanging magnetic flux does not induce any current in the pickup loop.}.
		
		\section{DC Current Transformers}
\begin{figure}[!htb]
					\centering
					\includegraphics[width=.45\linewidth]{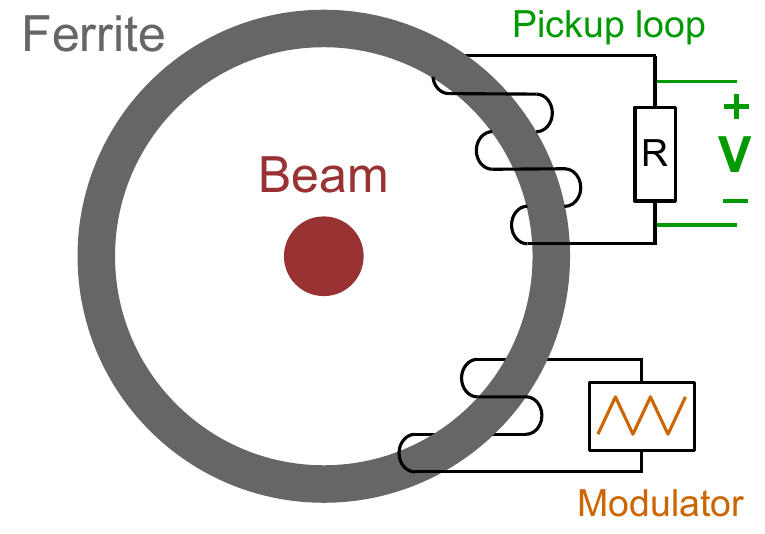}
					\includegraphics[width=.45\linewidth]{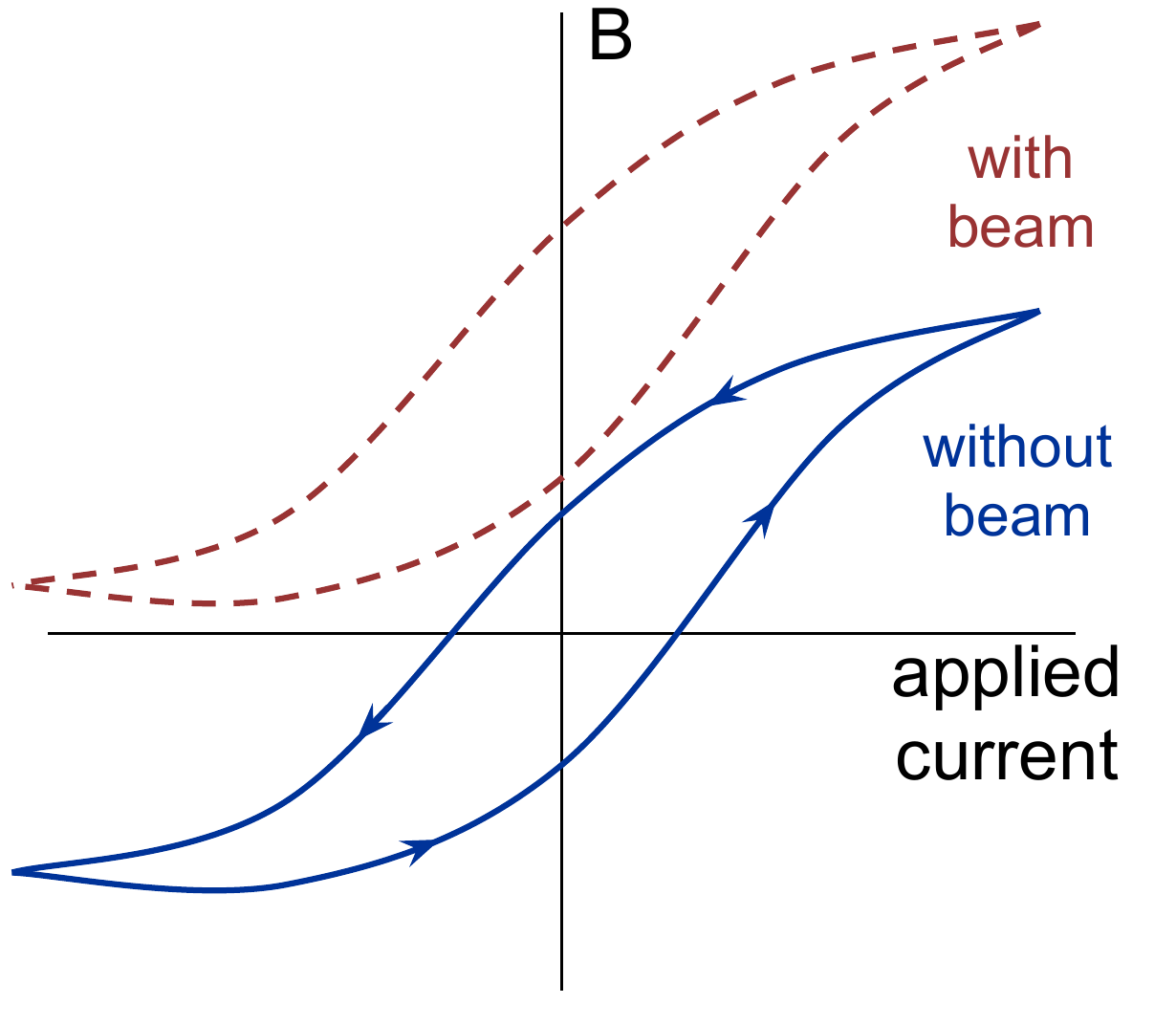}
					\caption[DCCT]{Model of a DCCT and the effect of beam on the ferrite's hysteresis curve}
					\label{fig:DCCT}
			\end{figure}

			The \Gls{dcct} is a device that measures beam intensity like a toroid, but is also able to see unbunched ``DC'' beam. Because the toroid relies on the principle of magnetic induction to generate the measurable signal, it is incapable of detecting unbunched beam. To measure DC beam current, we must rely on a different effect than electromagnetic induction.\textit{Figure~\ref{fig:DCCT}} shows a modification to the beam toroid that makes it a DCCT.\cite{DCCT}
		
			We can apply a changing current to a coil wrapped around the toroid's ferrite, which causes the ferrite to repeatedly cycle through the hysteresis curve mentioned in \textit{Chapter~\ref{chap:con_magnets}}. This ``magnetic modulation'' signal is usually a triangle wave, which is composed of only even harmonics. In the absence of external magnetic fields, the hysteresis curve of the ferrite is centered around zero, and the induced current in the pickup loop contains only odd harmonics. However, the beam's magnetic field biases the ferrite's hysteresis curve, which adds even harmonics to the current induced in the pickup loop\footnote{In general, any external magnetic field will bias the ferrite's hysteresis curve. This type of device is known as a ``flux magnetometer.'' The beam DCCT is just a particular use of a flux magnetometer to measure DC beam current.}. The second harmonic is most affected by the presence of beam; therefore, we can indirectly determine the amount of DC beam passing through the DCCT aperture by measuring the power of the second harmonic in the pickup loop current. \textit{Figure~\ref{fig:ipickup}} shows the pickup loop current in both time and frequency domain and how it changes in the presence of beam.
			
			\marginpar{
				\centering
				\includegraphics[width=1.0\linewidth]{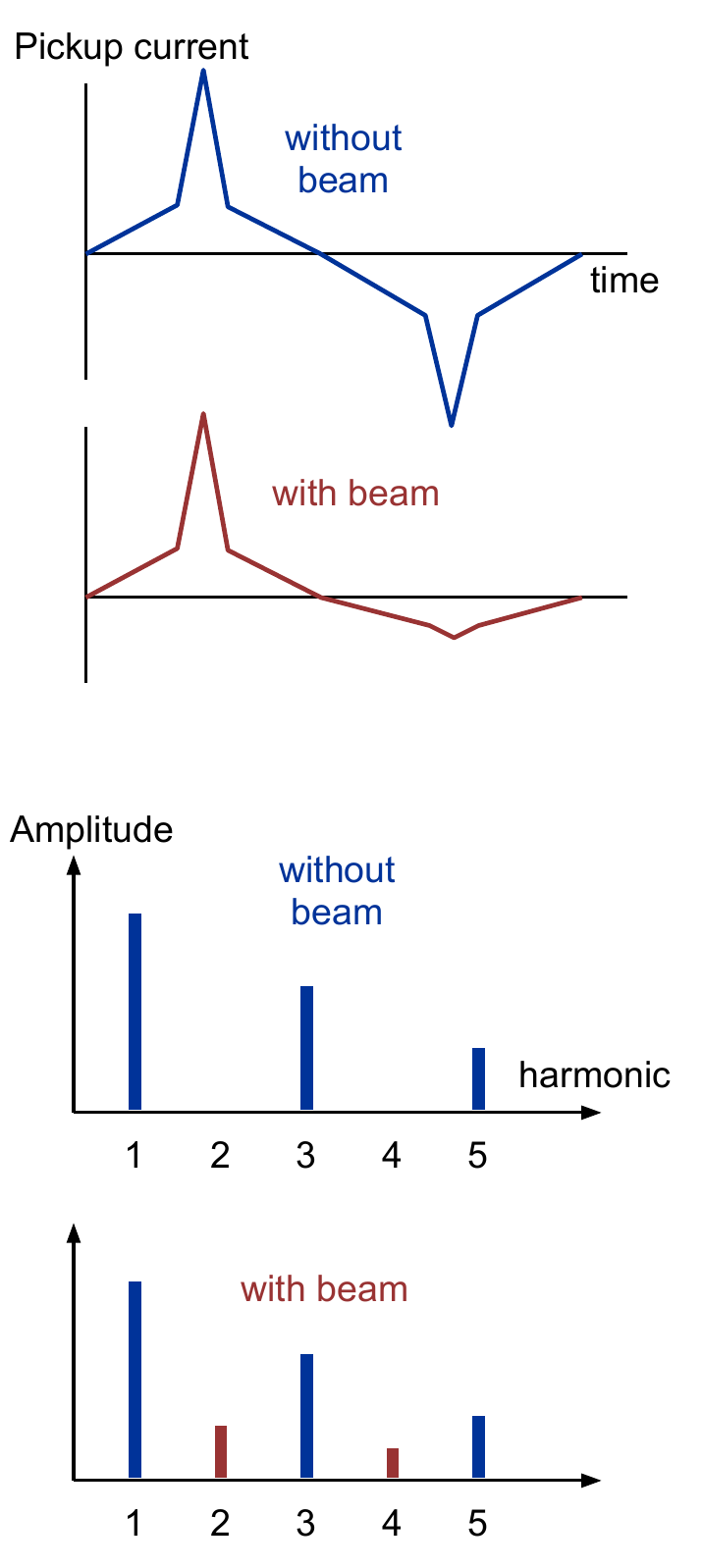}
				\captionof{figure}{Pickup loop current in a DCCT with and without beam}
				\label{fig:ipickup}
			}
			
		\section{Ion Profile Monitors}
\begin{figure}[!htb]
				\centering
				\includegraphics[width=.7\linewidth]{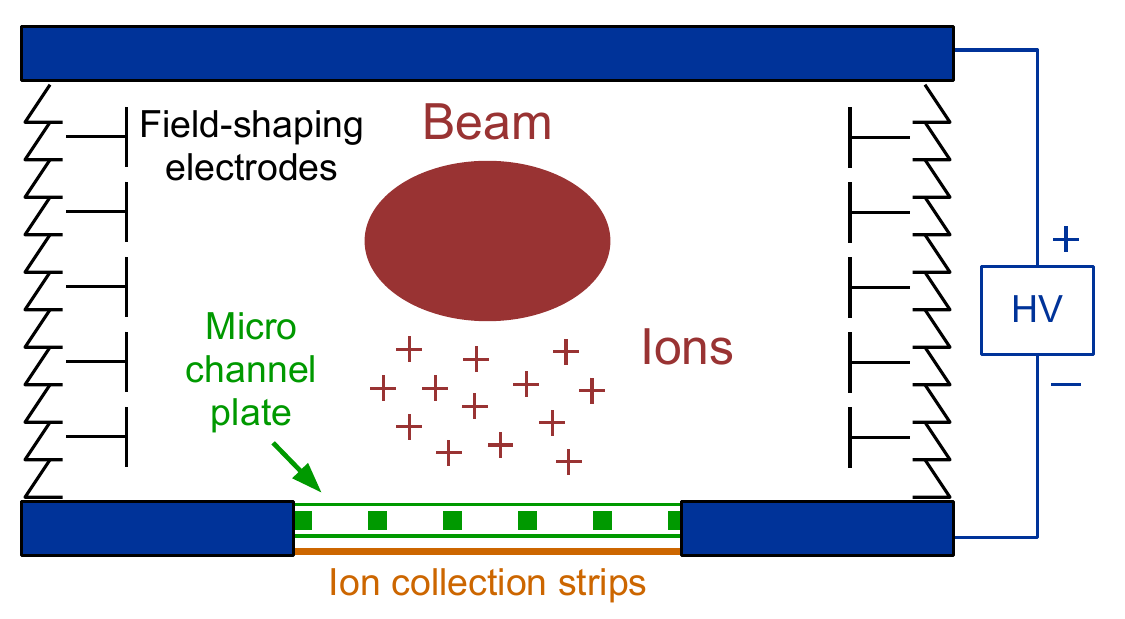}
				\caption{Model of an ion profile monitor}
				\label{fig:IPM}
			\end{figure}

			The \Gls{ipm} is a nondestructive\footnote{By ``nondestructive,'' we mean that the IPM produces no noticeable effect on the beam.} diagnostic tool that measures transverse beam profile or emittance. Charged beam ionizes the air particles left over due to imperfect vacuum. The IPM detector looks at the pattern of these ions to determine transverse beam profile information. A high-voltage potential across the detector generates an electric clearing field, drawing the ions toward a mircochannel plate. A model of the IPM is pictured in \textit{Figure~\ref{fig:IPM}}.
		
\begin{figure}[!htb]
				\centering
				\includegraphics[width=.7\linewidth]{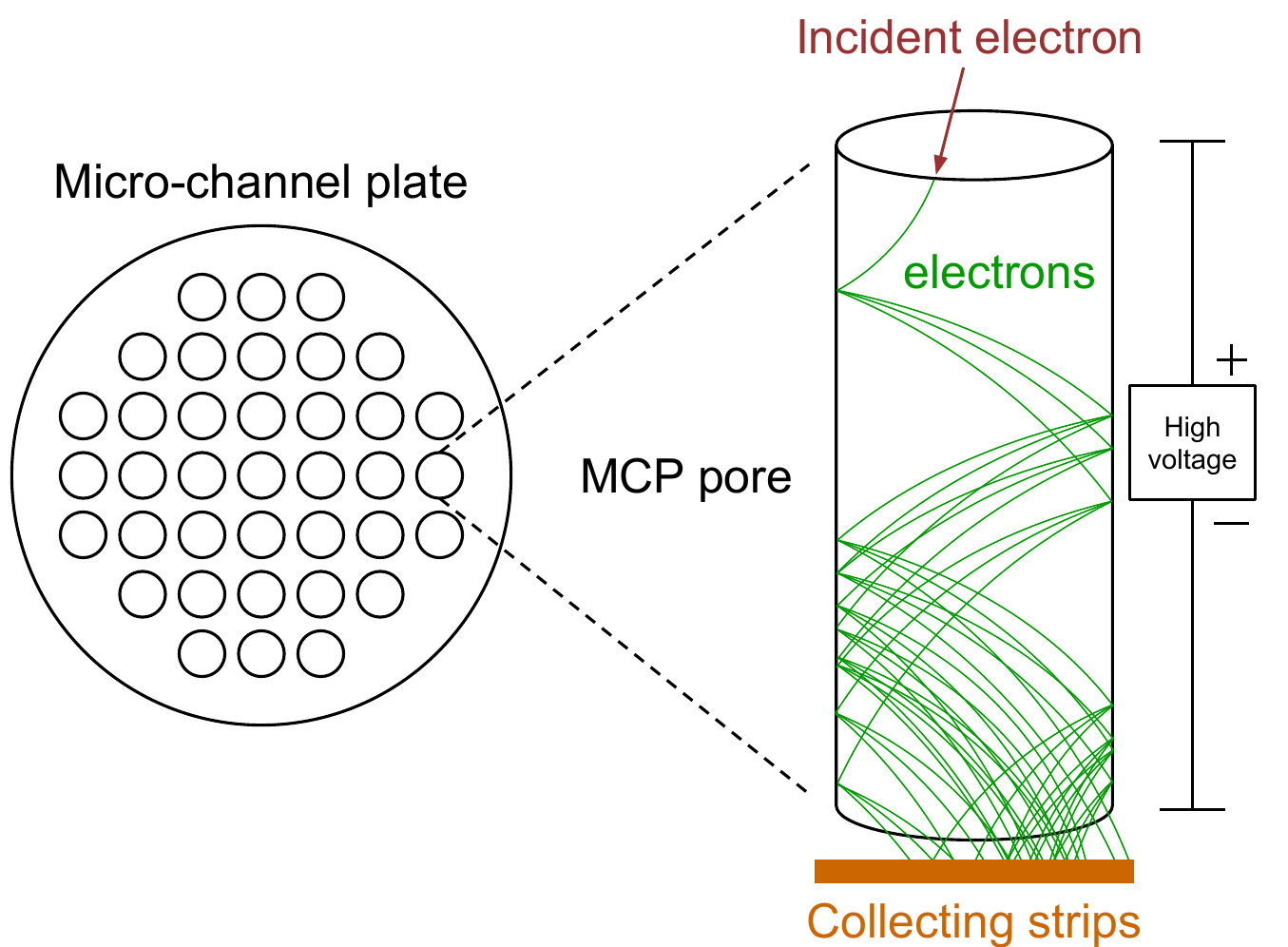}
				\caption{The micro-channel plate}
				\label{fig:MCP}
			\end{figure}

			The microchannel plate, or ``MCP,'' amplifies ions by cascading the secondary emission of electrons down the the walls of each small channel; \textit{Figure~\ref{fig:MCP}} shows the MCP and how it amplifies electrons. Electrons accelerate into collecting strips that preserve the ion pattern from the beam, thus generating a signal that represents the transverse beam profile. 
			
		\chapter{Utilities}\label{chap:con_utils}
	\section{Power Distribution}
	In this section, we describe the system of substations and underground power cables that distribute electrical power around the lab. We also explain some basic theory that extends to almost any power supply found at the lab. This chapter also introduces the utility systems that support the accelerators; these include cooling water, power distribution, and beam pipe vacuum.
	
		\subsection{Substations and Feeders}
			Commonwealth Edison provides electrical power to Fermilab via high-voltage transmission lines that connect from off-site to one of our two substations. The \textit{Master Substation} at the north and \textit{Kautz Road Substation} on the south end of the site use large transformers to step down the voltage on the transmission lines from 345 kV to 13.8 kV. Underground cables known as ``feeders'' carry the power out to each individual service building, where transformers step the 13.8 kV down for local use. Each feeder cable has a unique number that we use to refer to it.
			
			The electrical power from Com-Ed all the way to the service building transformer oscillates sinusoidally at 60 Hz; this is known as AC or ``alternating-current'' power. Electrical power at a constant voltage is called DC or ``direct-current.'' Transformers only work at AC, and power distribution is more efficient at AC, so voltages are sinusoidal until converted to DC by local power supplies. This AC power is often referred to as ``line voltage,'' which can refer to 120 V, 240 V, or 480 V after step-down at a particular building's house transformer.
			
			When we prepare enclosures for access, we must cut electrical power to the major supplies for safety reasons. In addition to turning off the power supplies themselves, we also open breakers at the appropriate substation to prevent power flow through relevant feeder cables; this process is called ``switching off.''\footnote{You may also hear this referred to as ``racking out.'' This is an old term related to the way the breakers used to be opened, by physically pulling out power racks to break the circuit.}
			
		\subsection{Basic Power Supply}
\begin{figure}[!htb]
				\centering
				\includegraphics[width=.8\linewidth]{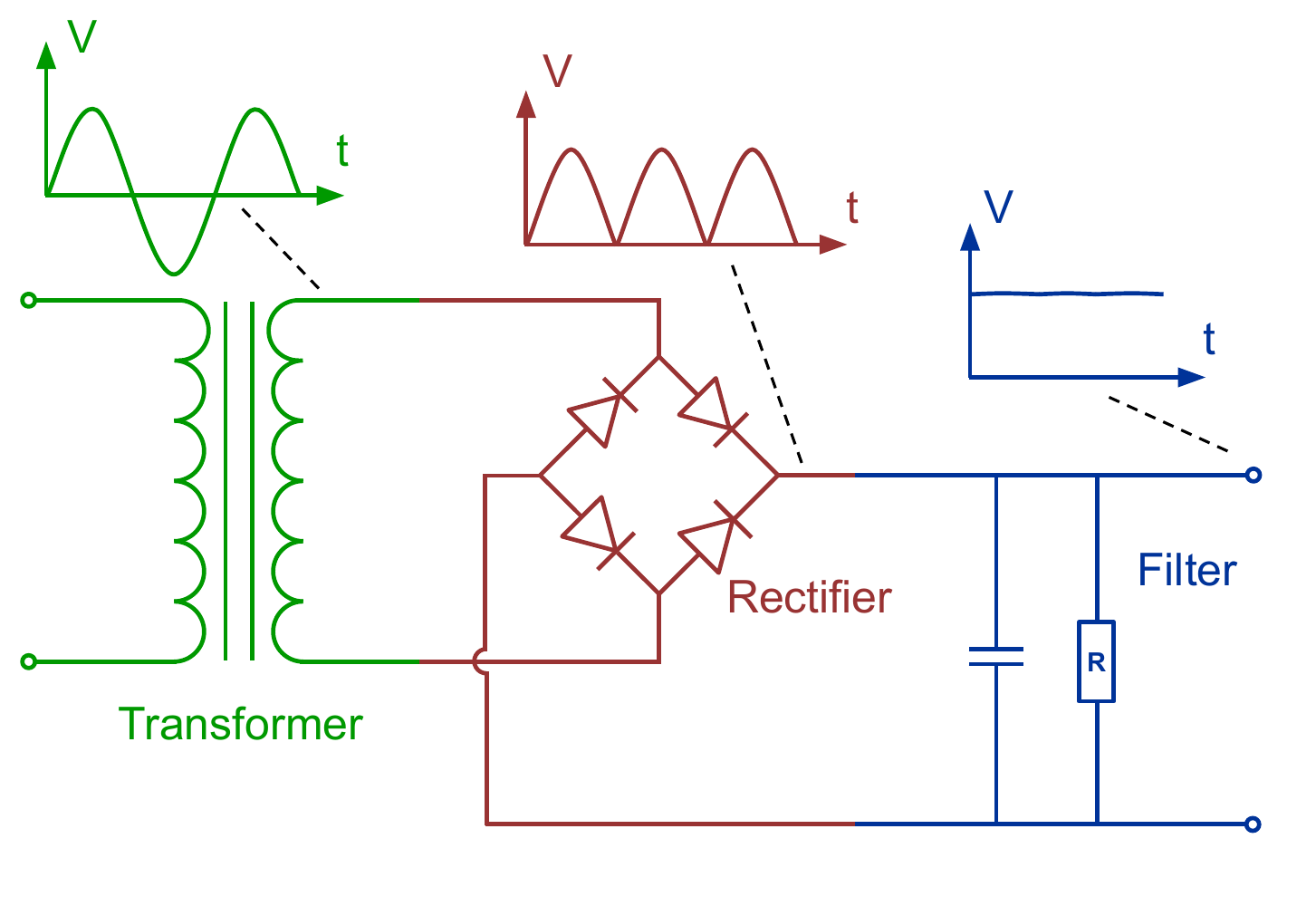}
				\caption{Model of a basic DC power supply}
				\label{fig:powersupply}
			\end{figure}

			We now explain the basics of electrical power supply theory. This knowledge will provide a foundation to understand almost any power supply at the lab, because they all function on the same fundamental principles. \textit{Figure~\ref{fig:powersupply}} breaks down a basic AC/DC (alternating-current to direct-current) power supply into three main parts. This power supply converts a high-voltage sinusoidal input into a lower-voltage flat output.
			
\begin{figure}[!htb]
				\centering
				\includegraphics[width=.9\linewidth]{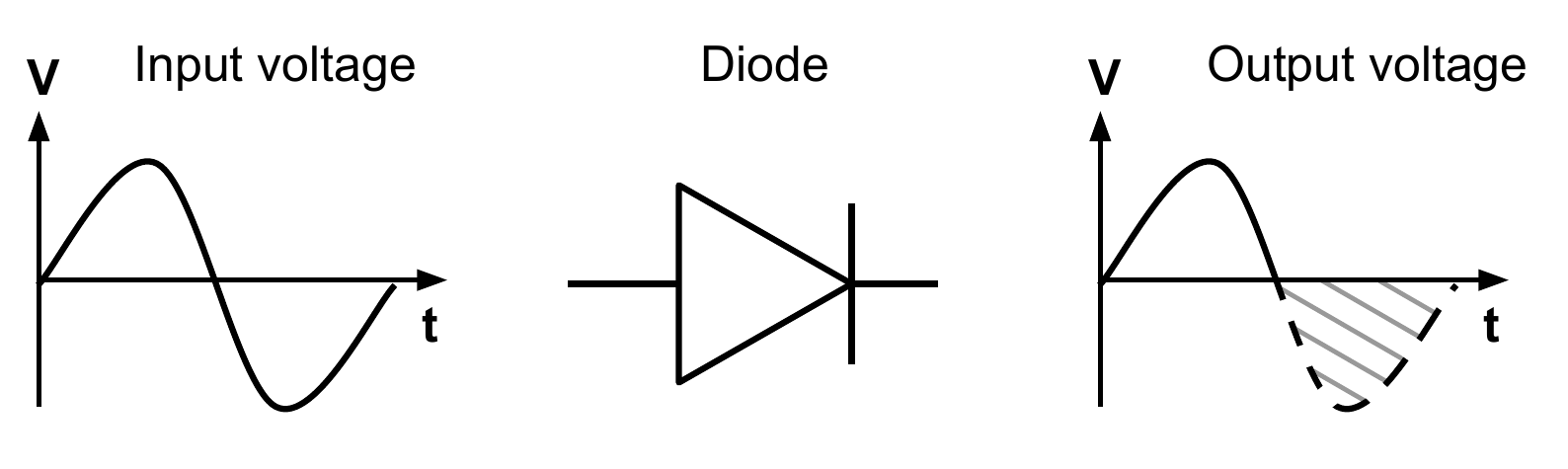}
				\caption{The diode's effect on a sinusoidal input voltage}
				\label{fig:diodeclip}
			\end{figure}

			The \textit{transformer} steps down the AC line voltage to whatever is required for the power supply, outputting a \textcolor{nal-green}{sinusoidal voltage}. The \textit{rectifier} uses diodes, which only conduct current in one direction, to change the shape of the voltage waveform; this flips the negative portions of the voltage to output a series of \textcolor{nal-red}{positive voltage pulses}. \textit{Figure~\ref{fig:diodeclip}} shows how a diode removes the negative portion of an oscillating input voltage.\footnote{This common configuration of diodes is called a ``full-wave bridge rectifier.''} Finally, the filter smooths the pulses from the rectifier into a \textcolor{nal-blue}{relatively flat ``DC'' voltage}. This filter is called ``low-pass'' because it conducts any quick voltage variations to ground, only passing a smooth voltage with minimal variations to the output.\footnote{The variations in output voltage of a power supply are called ``ripple''. Extra filtering and protection is often needed to remove or protetct against adverse effects of power supply ripple.}
			
			To provide even more control over the output waveform shape, power supplies may use multiple rectifiers whose outputs are summed together. This is known as ``multi-phase'' rectification, because there is a slight phase difference between each input signal. \textit{Figure~\ref{fig:multiphase}} shows how the combination of multiple input phases can provide a \textcolor{nal-green}{smooth DC output} when combined.
			
			\marginpar{
				\centering
				\includegraphics[width=1.0\linewidth]{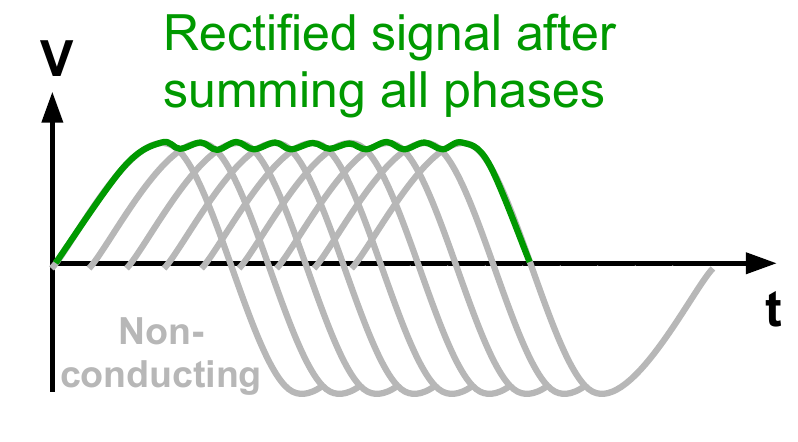}
				\captionof{figure}{Green DC output from multi-phase rectification}
				\label{fig:multiphase}
			}
			
			Sometimes we need a power supply to provide an AC voltage; for example, the \index{Booster}Booster GMPS supply converts the 60 Hz feeder voltage to a 15 Hz voltage to power the magnets. We need our rectifier and filter combination to provide any arbitrary frequency output, including DC if necessary.\footnote{DC is equivalent to a frequency of 0 Hz. Think of it this way: plug a frequency $f=0$ into a simple oscillating voltage expression $V(t)=A\cos(2\pi ft)$. Since $cos(0)=1$, a DC frequency means that $V(t)=A$, which is a constant voltage.}
			
			We can make use of a component known as a \textit{\gls{scr}}\footnote{Also known as a ``thyristor.''}, to take the place of the diodes in the rectifier part of the power supply. SCRs function like diodes, but they can be ``gated''; in other words, a control signal applied to the \gls{scr} tells it when to start acting like a diode. \textit{Figure~\ref{fig:SCRclip}} shows that an \gls{scr} behaves like a diode beginning at the rising edge of its gate pulse. The \gls{scr} continues to conduct until the input voltage becomes negative, where it ceases to conduct and must be gated again. Now we can choose which portion of the input sinusoid to rectify, which provides more control over the shape of the output waveform. Instead of a low-pass filter, we can use a filter tuned to allow a specific frequency range through to the output; this is called a ``band-pass'' filter, and allows us to smooth the \gls{scr}-rectified output into a waveform of arbitrary frequency.
			
			\marginpar{
				\centering
				\includegraphics[width=.7\linewidth]{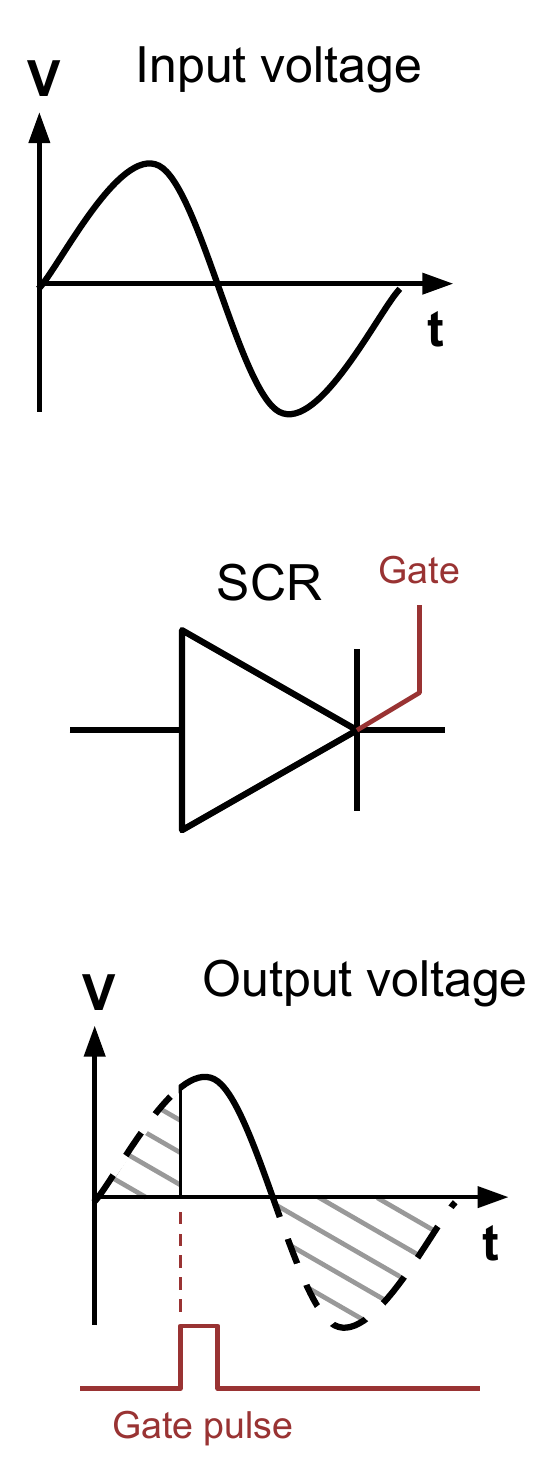}
				\captionof{figure}{Gated \gls{scr}'s effect on sinusoidal input voltage}
				\label{fig:SCRclip}
			}
			
			Most power supplies also contain some kind of protection from glitches that cause too much power to flow into the load. A ``fuse'' is a wire that burns up and breaks the circuit when its current limit is exceeded. If something goes wrong in the load that causes it to draw too much current, the fuse will burn out and stop the flow of current as long as it is wired in series with the load. We can also use an \gls{scr} in parallel with the load that will shunt current to ground and away from the load when triggered by a protection circuit. Also known as a ``crowbar circuit''\footnote{This is called a ``crowbar circuit'' because it shunts current so quickly that it is like dropping a steel crowbar across the power supply output leads.}, this circuit senses overvoltages or other problems in the power supply and triggers the \gls{scr} to dump all current to ground if necessary.\footnote{Other fast high-power switching mechanisms are also used as crowbars, such as the mercury-vapor ``ignitron'' or the vacuum tube equivalent of the thyristor known as a ``thyratron.''}
			
\begin{figure}[!htb]
				\centering
				\includegraphics[width=.9\linewidth]{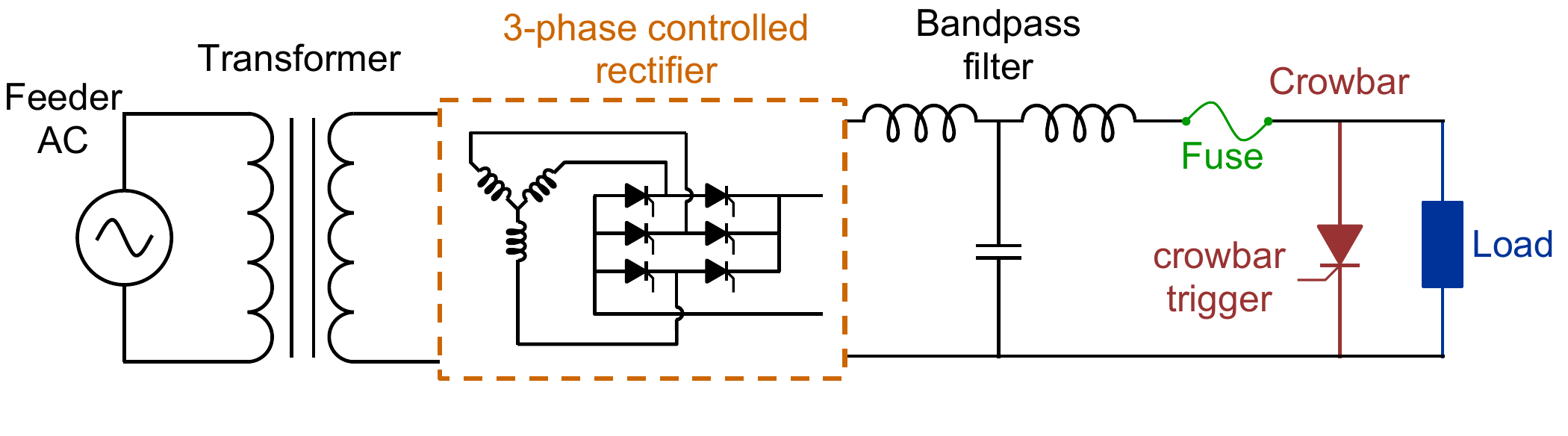}
				\caption{Multi-phase rectified power supply with output protection}
				\label{fig:PScomplete}
			\end{figure}

			\textit{Figure~\ref{fig:PScomplete}} shows a more complete model of a power supply that uses a more complicated rectifier-filter combination to provide an arbitrary AC output, as well as the aforementioned protection circuitry.\cite{allaboutcircuits}
			
	\section{Vacuum}
	There are two basic reasons why the beam pipe in an accelerator is under vacuum. First, collisions with air molecules in the beam pipe cause scattering and lead to emittance growth and beam loss. Second, without an evacuated beam pipe we would be unable to sustain the high electric fields required by devices like RF cavities and septa. These high electric fields would ionize the air, forming a path to ground and causing arcs. Thus we must pump out air molecules wherever the beam travels and in locations with high electric fields.
			
	The basic components of any vacuum system are the airtight beam pipe, some type of vacuum pump, and instrumentation to measure vacuum quality. Each pump type uses a different method to remove air molecules from the beam pipe, and thus works best in specific pressure ranges. This limitation of pump pressure range means that we must evacuate the beam pipe in stages. The following is a description of the basic pump types that we use. We then describe the instrumentation we use to measure vacuum quality. Finally, we describe a behavior known as ``out-gassing'' that can occur during pump-down, and how we can perform a ``bake-out'' to preempt this effect.
	
		\subsection{Roughing Pumps}
		
			\marginpar{
				\centering
				\includegraphics[width=1.0\linewidth]{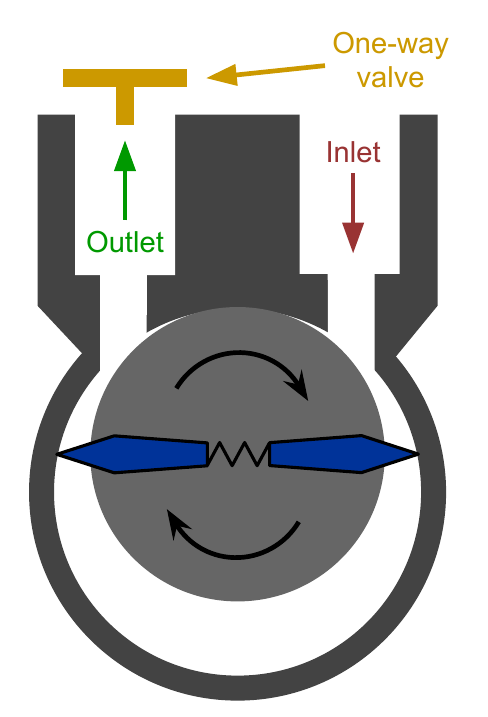}
				\captionof{figure}{A vane-type roughing pump}
				\label{fig:rpump}
                }
		
		Roughing pumps provide the first stage of pumping, effective from 760 to $10^{-3}$ Torr. A specific ``vane-type'' roughing pump is pictured in \textit{Figure~\ref{fig:rpump}}, and consists of an oil-lubricated wheel with vanes along its perimeter that expand to touch the outer wall. As the wheel rotates, the vanes trap air molecules from the vacuum vessel between the wheel and the pump wall and push them to an exhaust port, where the air molecules exit via a one-way valve. 
			
		As the air pressure drops, the number of particles entering the intake port also drops. At an intake pressure of about $10^{-3}$ Torr, the roughing pump is no longer able to effectively remove air molecules from the beam pipe. 
		
		\subsection{Turbomolecular Pumps}
			Turbo pumps work from $10^{-1}$ to $10^{-6}$ Torr, and often provide support as turbochargers to increase the flow of air molecules into the roughing pump intake. Pictured in \textit{Figure~\ref{fig:turbopump}}, the turbo uses several layers of turbine vanes that rotate at very high velocity (20,000-100,000 rpm). The spinning vanes hit air molecules and drive them toward the exhaust port, much like an automobile turbocharger or a jet engine. Due to the delicate nature of the spinning vanes, turbo pumps cannot be turned on until the roughing pump has sufficiently reduced the beam pipe pressure. 
			
			\marginpar{
				\centering
				\includegraphics[width=1.0\linewidth]{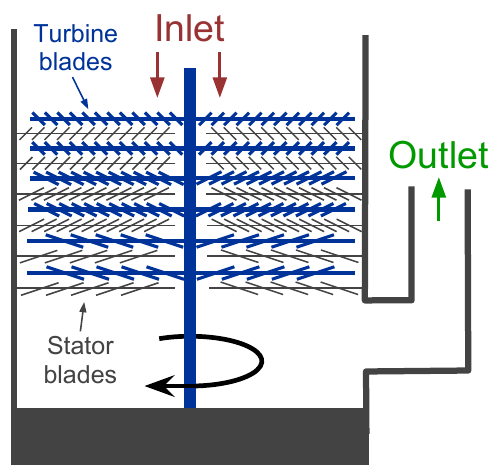}
				\captionof{figure}{Model of a turbo pump}
				\label{fig:turbopump}
			}
			
		\subsection{Ion Pumps}
\begin{figure}[!htb]
				\centering
				\includegraphics[width=.35\linewidth]{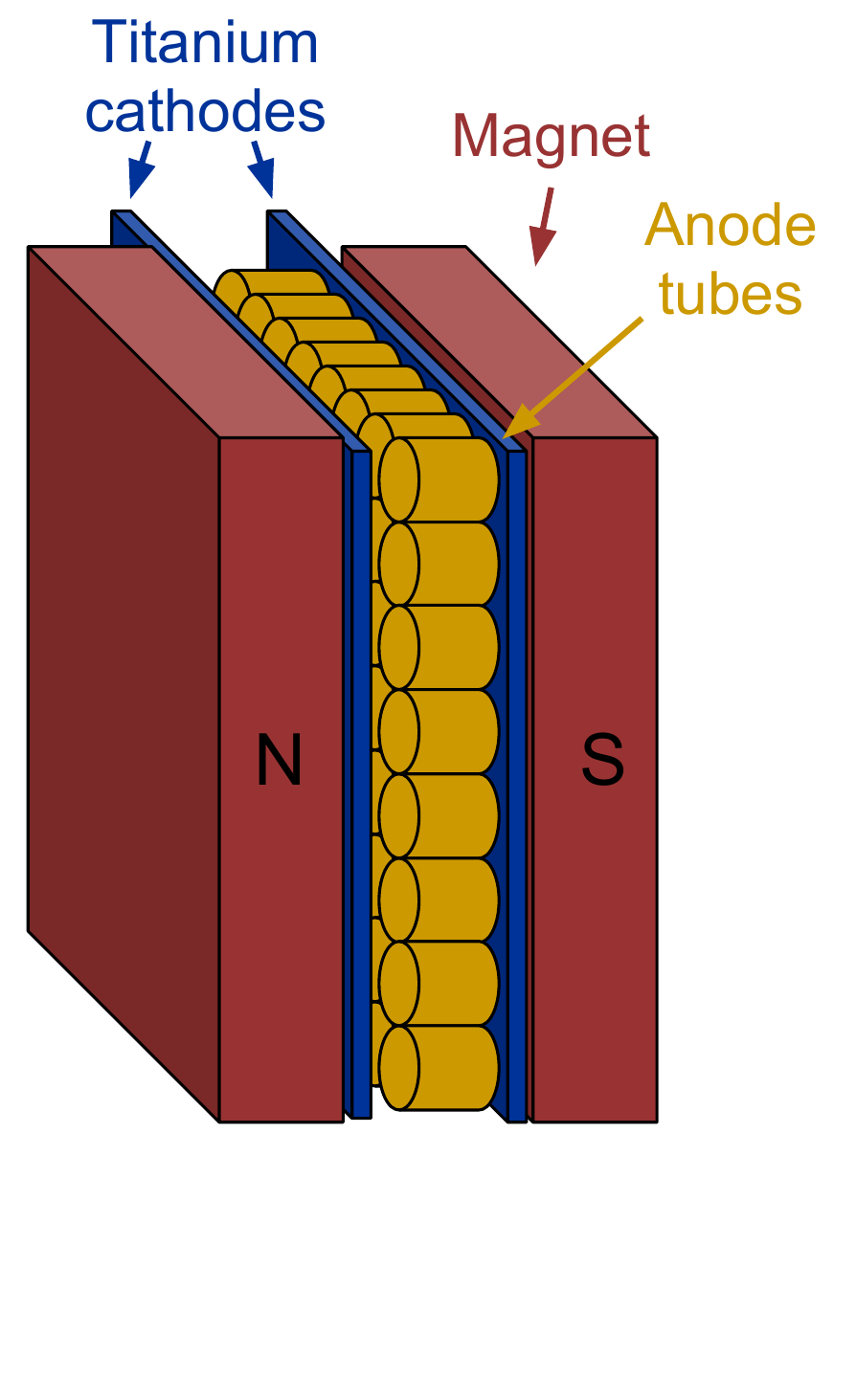}
				\includegraphics[width=.6\linewidth]{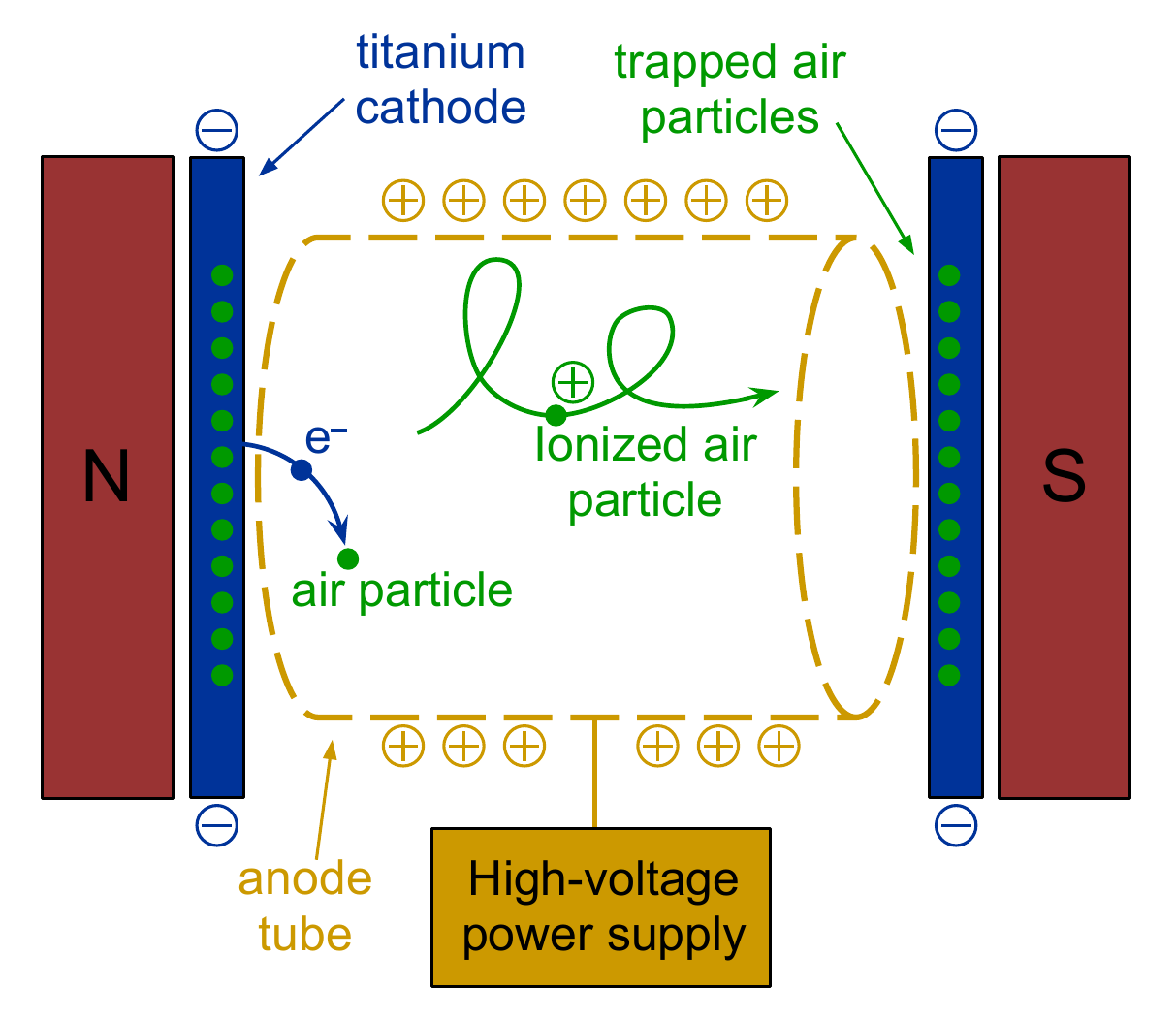}
				\caption{Model of an ion pump.}
				\label{fig:ionpump}
			\end{figure}

			Ion pumps are effective from $10^{-5}$ to $10^{-12}$ Torr, and do not mechanically remove air like the roughing and turbo pumps. Instead, ion pumps capture and trap air molecules by ionizing them and causing them to bind with Titanium atoms\footnote{Titanium readily interacts with ionized air particles, which is why it was chosen.}. Pictured on the left of \textit{Figure~\ref{fig:ionpump}}, the ion pump consists of two parallel Titanium cathode plates with a high-voltage anode in the middle. A magnet surrounds the entire structure to create a uniform field that is perpendicular to the plates.
			
			The high voltage on the anode structure creates an electric field that frees electrons from the anode's surface and accelerates them toward the cathode plates. The freed electrons collide with air particles and knock electrons out of the molecules, creating positively-charged air ions. All of the electrons are attracted to the anode tube, while the positive air ions are attracted to the cathode plates. 
			
			The magnetic field causes the charged particles to travel in a spiral path known as ``cyclotron motion,'' which increases the chance of further ionizing other air molecules. The ionized air particles become embedded in the Titanium cathode plates, which effectively lowers the air pressure in the entire chamber. By measuring the electrical current caused by the ions hitting the cathode plates, we can measure the quality of the vacuum. If this current becomes too high (i.e. poor vacuum), the ion pump turns off to prevent damage. Due to the high voltage of the anode and the potential of sparking, ion pumps cannot work in poor vacuum. Ion pumps have a finite lifetime, because the Titanium cathodes become contaminated with trapped air molecules over time.
			
		\subsection{Titanium Sublimation Pumps}
			\Glspl{tsp} work on a principle similar to the ion pump. The sublimation pump consists of titanium rods connected to an external power supply. When this supply turns on, small amounts of the titanium rods sublimate; in other words, they go directly from solid to vapor without boiling. The sublimated titanium vapor condenses the inside of a chamber internal to the beam pipe, where it readily interacts with any nearby air molecules. The air molecules chemically interact with the Titanium and become trapped, thus lowering the air pressure.
			
		\subsection{Vacuum Valves}
			Vacuum valves create an air-tight seal that lets us isolate areas of the beam pipe, so a leak in one area does not have to affect another. These valves obstruct the entire beam pipe aperture with a strong plate that also impedes the flow of beam. Any issues in the vacuum system that degrade vacuum quality cause the valves to close very quickly. Vacuum valves typically require pneumatic pressure to remain open, so they can drop quickly if necessary. However, closed vacuum valves are detrimental to beam quality, so they must be open when running beam to prevent equipment damage. Some locations also have manual valves that must be actuated by hand from the tunnel, and these provide vacuum isolation during long shutdown periods.
			
			Valves specifically designed to obstruct the beam itself are called ``beam valves.'' For example, storage machines like MI and the Recycler have ``coasting beam valves'' that close when we access the tunnel. This deliberately kills any beam that may still be circulating due to the relatively long lifetimes in these machines.
			
		\subsection{Measuring Vacuum Quality}
			The three main instruments used at Fermilab to measure vacuum quality are thermocouples, cold-cathodes, and ion gauges.
			
			\textit{Thermocouple gauges} work on the principle of heat convection. A heated wire is
			exposed to an evacuated chamber. A temperature sensing circuit called a
			thermocouple in contact with the heated wire constantly
			measures the wire temperature. Because the heated wire will naturally
			transfer some of its energy to the air molecules around it as the pressure
			around the wire is reduced, the thermocouple will detect a proportionally
			higher wire temperature. Thermocouples are usable from atmospheric
			pressure to about $10^{-3}$ Torr.
			
			\textit{Cold-cathode gauges} function like miniature ion pumps. A high voltage wire is
			placed into the vacuum chamber. The amount of current flowing from this
			wire to the outer casing of the gauge is proportional to the amount of
			ionized gas in the system, which is a measure of the vacuum quality. Cold
			cathodes operate in the range from $10^{-3}$ to $10^{-8}$ Torr.
			
			\textit{Ion gauges} are similar to cold-cathode gauges. Instead of a high voltage wire, ion
			gauges have a filament and a collector. The filament is heated, liberating
			electrons, which accelerates toward the collector. As they travel this
			short path, the electrons from the filament ionize some of the neutral
			air molecules in the area. The collector picks up the electrons from the
			filament and the newly liberated electrons, and measures the total
			electrical current. The magnitude of this current is related to the pressure.
			Ion gauges are operable from $10^{-2}$ to $10^{-10}$ Torr.
			
		\subsection{Out-gassing and Bake-outs}
		Molecules have a tendency to attach to the surface of the metal, and slowly break away
		when the surface is under reduced pressure. The surfaces of some metals are porous, providing
		small cavities where gases can be trapped at standard pressure only to slowly bleed into the
		vacuum once the pressure is reduced. The process of releasing gases from the surface of a
		material under reduced pressure is known as ``out-gassing.'' Out-gassing can be a source of
		seemingly spurious vacuum bursts, and is a problem to overcome when re-evacuating a
		component that has been ``let up'' to atmosphere (brought from vacuum up to atmospheric pressure).
		
		In some cases, the outgassing is not caused by trapped gases being liberated from inside
		the metal, but rather by some contaminant on the surface evaporating. Oils and certain rubber
		components can out gas in this way; even the natural body oils from a person touching the inside
		of a beam pipe can cause problems during pump down.
		
		Sometimes, out-gassing is intentionally induced in high vacuum components to remove
		trapped gases or to boil off any oils collected on the inner surfaces. This procedure is known as a
		``bake out.'' Special electric heating blankets, commonly called ``bake-out blankets,'' are wrapped
		around the outside of the beam pipe or device to be baked. Raising the temperature of the vacuum vessel frees any gases
		trapped on the inner surface and forces the molecules back into the system where they can be
		pumped out. Due to the expected increase in the number of air molecules in the vacuum vessel
		during a bake out, extra portable vacuum pumps are usually added to the system. Once the
		system has been sufficiently baked, the blankets are turned off, the vacuum valves
		connecting the extra pumping stations to the vessel are shut, and the ion pumps and sublimation
		pumps are turned on.
		
	\section{Low Conductivity Water}
	Magnets, RF cavities, and other high-power devices dissipate significant amounts of heat. Removing this heat efficiently is crucial for machine operation. Fermilab uses multiple water cooling systems to deal with these heat loads.
	
	Tap water and ground water are usually full of ions that make water electrically conductive. The \Gls{cub} manages deionizing and cooling tap water into \Gls{lcw}. \Gls{cub} water may be added to a system if pressures or water levels are low. This deionization process makes water resistive enough to flow through high voltage systems without creating a path to ground.

	\marginpar{
		\centering
		\includegraphics[width=1.0\linewidth]{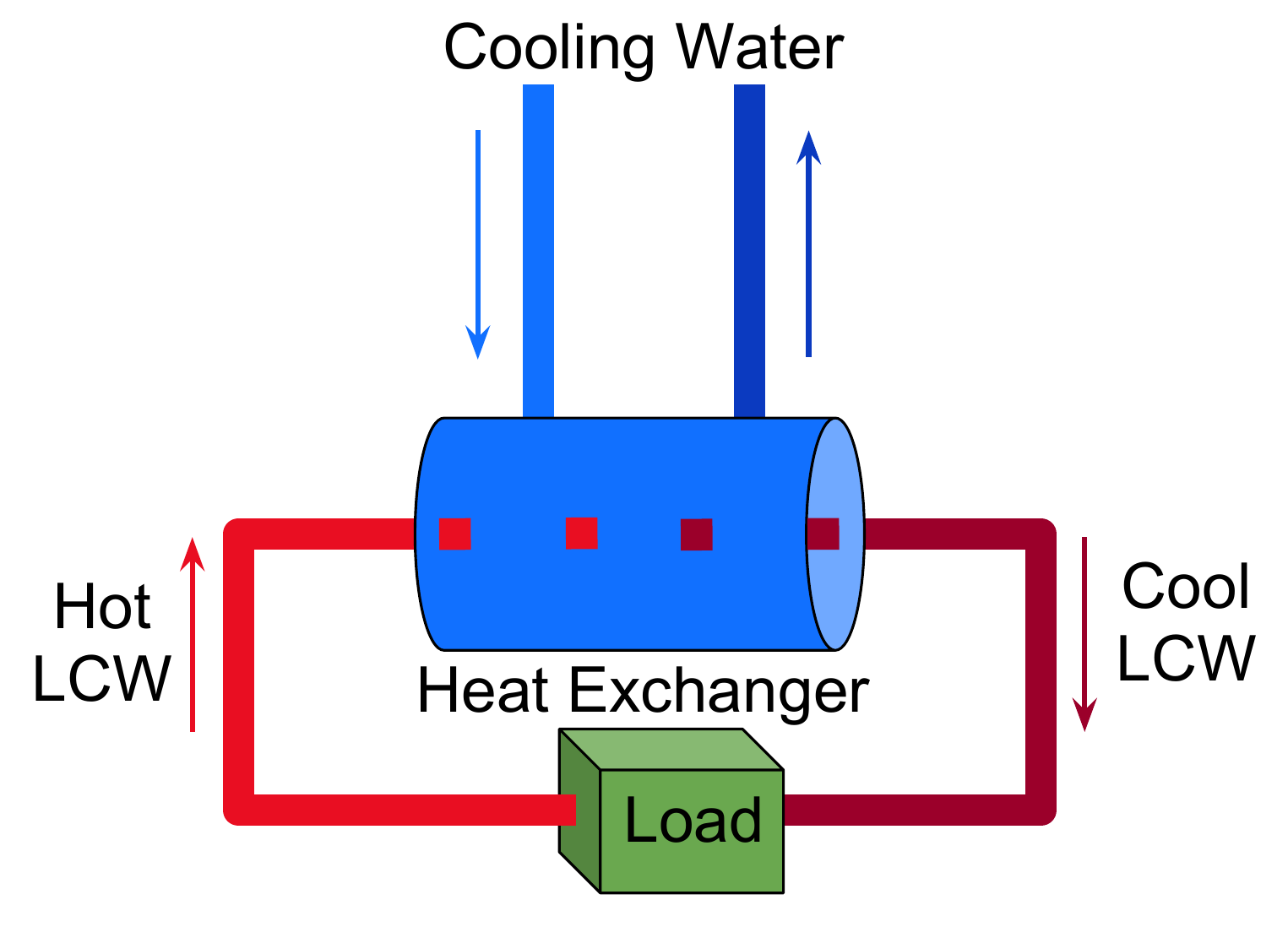}
		\captionof{figure}{Diagram of a simple heat exchanger}
		\label{fig:hex}
	}
			
	We use \gls{lcw} to cool devices due to its efficiency and low cost. Each local \Gls{lcw} system cools its water via a heat exchanger as shown in \textit{Figure~\ref{fig:hex}}. The cooling water comes from an outside source (ie. pond water) to cool the system's \Gls{lcw} as it passes through the heat exchanger. As the colors suggest in \textit{Figure~\ref{fig:hex}}, the local system outlet water is hot because it has heat exchanged with the load.

		\chapter{Safety Systems}\label{chap:con_safe}

The Operations department has a responsibility to maintain the safety of all those interacting with the accelerators. There are many systems that work together to make this possible: \Glspl{bbm}, \Glspl{cdc}, a Radiation Monitoring System, \Glspl{ess}, Enclosure Key Systems, and Configuration Control Lockout all ensure that the accelerators operate while minimizing hazards for lab personnel and surrounding communities.

\marginpar{A \keyterm{\gls{failsafe}} device is one that, in the event of failure, responds in a way that will cause a minimum amount of harm.}

These systems are \gls{failsafe} to prevent electrical or radiological harm to personnel. A \keyterm{\gls{failsafe}} device is one that, in the event of failure, responds in a way that will cause a minimum amount of harm. A system being \gls{failsafe} does not mean that failure is impossible, but rather that the system's design mitigates unsafe consequences of the system's failure; that is, if one of these systems fails, it is no less safe than when it is operating correctly.

	\section{Beam Budget Monitors}
	The \Gls{bbm} is an application that monitors beam delivery rates to each enclosure. These hourly limits, stated in the running conditions, are imposed by the \Gls{doe} and Fermilab and are based on shielding assessments for each enclosure.
	
	The \gls{bbm} conveniently displays a graph of the integrated rate of beam delivered to the operational areas of the lab. It is Operation's job to monitor this rate and make sure that the limits are not exceeded. If a limit is exceeded, a beam switch may be taken until a solution can be found that will reduce the beam rate to that area. Operations has a ``soft'' limit as a warning before we reach the \Gls{doe} limit.
	
		\subsection{Beam Switches}
		The Beam Switch Boxes in the \gls{mcr} provide a quick and easy way for an operator to disable beam to an area.

	\section{Critical Devices and Controllers}
	\Glspl{cdc} permit or deny beam delivery to enclosures by controlling the state of particular beamline devices. These ``critical devices'' are capable of preventing beam from entering each machine's enclosure when turned off by the \gls{cdc}. The \gls{cdc} turns a critical device off if any of its inputs (enclosure ESS, RadLoop, etc.) are bad. The \gls{cdc} also monitors whether the critical device turns off correctly and quickly; in the case of an indeterminate critical device state, the \gls{cdc} issues a ``failure mode'' signal to the upstream \gls{cdc} that turns off its critical device. For the sake of redundancy, each enclosure has two ways to inhibit beam: either an enclosure's \gls{cdc} controls two separate critical devices, or it has two ways of turning off a single device.
	
		\subsection{Personnel Protection}
		The \gls{cdc} (via the critical devices) is one piece of the safety system preventing beam from entering an enclosure while it is being accessed by personnel.
	
		\subsection{Inputs}
		A \gls{cdc} has many inputs that must all provide a constant voltage signal in order for the critical devices to stay on. Common inputs include:
		
			\begin{itemize}
				\item Enclosure \gls{ess} \\*
					*See \textit{Section~\ref{sec:con_safe_ess}} for more information
				\item RadLoop \\*
					*See \textit{Section~\ref{sec:con_safe_radmon}} for more information
				\item Downstream \glspl{cdc} Failure Mode
				\item Digital Control Bit (ACNET On)
				\item Critical Device Status
			\end{itemize}
			
		*Most of these inputs are redundant in some fashion. \\
		
		If any one of these fail to provide the proper signal to the \gls{cdc}, it turns off the critical device. If the critical device status returns an unknown status the \gls{cdc} goes into its failure mode. When in failure mode the upstream machine's \gls{cdc} will trip.
		
	\section{Radiation Monitoring}\label{sec:con_safe_radmon}
	The Radiation Monitoring system protects people from excessive levels of radiation exposure. This system simply watches radiation detectors in the machine areas and trips the \gls{cdc} if the levels get too high. The radiation detectors are summed together to form the Rad Loop. There are two types of detectors throughout the complex, ``chipmunks'' and ``scarecrows.'' The only difference between the two types of detectors is the radiation level for which they are calibrated. Chipmunks are most common and can be found in service buildings and on top of the berms. Scarecrows are built for high radiation areas such as beam dumps and extraction areas. 

	There are two outputs from the Rad Loop that feed into the \gls{cdc} logic. The Real Time Rad Loop gives a good signal as soon as levels have dropped. Operators must acknowledge the Latched Rad Loop and reset the trip from the rad mux display in the \Gls{mcr}, often after checking with the \Gls{rso}.
		
	\section{Electrical Safety Systems}\label{sec:con_safe_ess}
	The \gls{ess} is in place to prevent personnel exposure to electrical hazards in the tunnels. Powered devices in the tunnel that have exposed bus and are capable of producing a potential of \SI{50}{\volt} or greater, \SI{50}{\ampere} or greater, or storing \SI{1}{\joule} of energy or greater are a part of the \gls{ess}. Some exposed bus that qualifies to be on the \gls{ess} has been covered and protected so is not included in the the \gls{ess} circuit.

		\subsection{Personnel Protection}
		The \gls{ess} provides personnel protection from electrical hazards. If an enclosure door is opened or a key is unturned, it will disable the \gls{ess}. This will inhibit energized devices, causing all beam in the machine to ``fall out'' and prevent beam from entering the enclosure via the \gls{cdc}.
		
		When the inputs above have been made up an audio warning is sounded inside the enclosure to warn anyone who might still be in the tunnel that power supplies are going to be turned on. After the audio warning has played for two minutes then the audio permit allows the \gls{ess} to be activated. The emergency loop is a series of buttons or switches throughout the enclosure that may be toggled to prevent the \gls{ess} from coming up if personnel are in the tunnel and they hear the audio warning.
		
			\subsubsection{Search and Secure}
			Operations is solely responsible for searching and securing enclosures, ensuring nobody is left in the tunnel. We train a specific procedure for each enclosure. It is extremely important that search and secures are performed properly and thoroughly! If operators find anyone without a key while searching the tunnel that person is to be escorted to a safe place before the search and secure begins again.
			
			As operators complete the secure they are making up the interlock sequence which must be complete in order to activate the \gls{ess}. This interlock loop not only monitors the status of the secure but the doors as well.
	
		\subsection{Inputs}
		An \gls{ess} has many inputs that must all provide a good signal in order for an enclosure's \gls{ess} devices to remain powered.
		
			\subsubsection{Enclosure Doors}
			Enclosure doors are a part of the interlock system and they must be closed in order to be able to operate the machines. In order to be fail-safe, the local interlock boxes at each enclosure door monitor the door using two switches, a mechanical and a magnetic switch. Operators check the door status while performing a search and secure.
			
			\subsubsection{Enclosure Keys}
			Most enclosure keys can be found in the \Gls{mcr}, although there are a few external key trees. All keys for a certain enclosure, in the \Gls{mcr} and otherwise, must be in the turned position in order to permit the \gls{ess}. Each key cylinder has two switches that determine the key position.
			
			\subsubsection{Interlock Ground Fault Circuit}
			The interlock system has a circuit for ground fault detection. This prevents a ground fault from potentially bypassing interlock boxes or switches and will disable the \gls{ess} if there is potential for the interlock system to be in an unknown state.
			
			\subsubsection{Secure Interlock Sequence}
			The secure interlock sequence is made up by operators as they search and secure the tunnel. Operators reset the secure sequence by inserting and turning the reset key on each interlock box. The sequence will not make up if the doors for a box are not closed or if the previous box in the sequence has not been made up. 
			
			\subsubsection{Emergency Loop}
			The emergency loop is made up of either crash buttons or a crash loop on the wall of the enclosure. These loops, if tripped, prevent the \gls{ess} from being activated. This is a safety measure in case someone is in the tunnel when the audio warning sounds. Operators often double-check these while securing as the \gls{ess} will not make up without all the switches in their appropriate position.
			
	\section{Enclosure Key Systems}
	The enclosure keys held in the key trees are the only keys that allow access the tunnels. Possessing a key for an enclosure ensures those individuals are safe from electrocution while in the tunnel. All of the keys for a particular enclosure must be in their turned position to make up the permit to the \gls{ess}.
	
		\subsection{Access Procedures}
		Because we control when and where beam is sent, Operations also controls when and what enclosures may be accessed. There are a few different types of access that require different levels of preparation.
		
			\subsubsection{Controlled Access}
			Controlled Access is the most basic form of access. In order to make an enclosure ready for this access operators must do the following:
			
				\begin{itemize}
					\item Disable beam to the area
					\item Disable the \gls{cdc} for the enclosure
					\item Perform any turn off procedures/sequencers
					\item Perform any switch-off procedures for the enclosure
					\item Place important keys in a \gls{mcr} group \gls{loto} box
					\item Issue keys to access the tunnel
				\end{itemize}
			
			\subsubsection{Supervised Access}
			Supervised Access extends Controlled Access by reducing the requirements for entry. In addition to Controlled Access requirements, Supervised Access requires:
			
				\begin{itemize}
					\item Radiation survey performed by Rad Safety Department
					\item Enclosure Interlocks dropped
					\item Configuration Control Lockout
				\end{itemize}
			
			\subsubsection{Open Access}
			Open Access is a special type of access commonly used in specific PPD areas (ie. MTest). Open Access is a relaxed form of controlled access that allows most anyone to enter the enclosure after Operations has disabled beam and turned off the \gls{cdc}.
			
			\subsubsection{Power-On Access}
			Power-On is an access done while there is still main power delivered to the tunnel. This is rarely allowed and requires approval from the Directorate.
			
	\section{Configuration Control Lockout}
	Operators must perform configuration control for supervised access by locking off power supplies for beamline components that have exposed electrical connections. The configuration control list is provided by the \Gls{ad} \Gls{esh} and kept in the \Gls{mcr}. The configuration control performed by the Operations Department does not satisfy \gls{loto} requirements because those locks are not applied by the people exposed to the hazard. They are merely an extra layer of safety. Individuals who must operate circuit breakers in order to work on components before beginning work must also have Electrical Safety in the Workplace (\Gls{nfpa}) training and wear appropriate clothing and protective equipment.  

	\appendix

		\loadgeometry{base}

		\chapter{Derivations} \label{chap:con_apndx_a}
This Appendix contains several interesting derivations for some of the concepts mentioned in the main book. A note about this Appendix: a math review is beyond the scope of this book, so the Appendix assumes prior knowledge of vector calculus, differential equations, and some electrodynamics. This material is entirely supplementary to the actual Rookie Book, and \textit{should not be considered required for operator knowledge}.

	\section{Limitation of quadrupole focusing}
	In this section, we show that a quadrupole magnetic field cannot provide focusing in both transverse planes at the same time. Inside the magnet's aperture, any focusing force in one dimension will necessarily defocus in the other. However, by alternating quadrupole magnets of opposite polarity, we can achieve net transverse focusing.
	
	We begin with the \textit{Ampere-Maxwell Law} in differential form:
	
		\begin{equation} \label{eq:ampmax}
			\vec{\nabla} \times \vec{B} = \mu_0\vec{J}+\mu_0\epsilon_0\frac{\partial \vec{E}}{\partial t}
		\end{equation}
	
	Inside the aperture of a magnet, there is no flowing current density $\vec{J}$ or electric field $\vec{E}$. Thus inside a magnet, the \textit{Ampere-Maxwell Law} shown in \textit{Equation~\ref{eq:ampmax}} becomes:
	
		\begin{equation} \label{eq:bcurl0}
			\vec{\nabla} \times \vec{B} = 0
		\end{equation}
	
	If we consider a general magnetic field $\vec{B} = B_x\hat{x} + B_y\hat{y} + B_z\hat{z}$, we can rewrite $\vec{\nabla} \times \vec{B}$ in terms of field components:
	
		\begin{equation}
			\vec{\nabla} \times \vec{B} =
			\hat{x}[(\frac{\partial B_z}{\partial y})-(\frac{\partial B_y}{\partial z})]
			- \hat{y}[(\frac{\partial B_z}{\partial x})-(\frac{\partial B_x}{\partial z})]
			+ \hat{z}[(\frac{\partial B_y}{\partial x})-(\frac{\partial B_x}{\partial y})]
		\end{equation}
	
	Since $\hat{x}$, $\hat{y}$, and $\hat{z}$ are linearly-independent, the only way to make \textit{Equation~\ref{eq:bcurl0}} true is to set each coefficient of the basis vectors equal to zero. In other words, \textit{Equation~\ref{eq:bcurl0}} gives us three simple equations: $\frac{\partial B_z}{\partial y} - \frac{\partial B_y}{\partial z}=0$, $\frac{\partial B_z}{\partial x} - \frac{\partial B_x}{\partial z}=0$, and $\frac{\partial B_y}{\partial x} - \frac{\partial B_x}{\partial y}=0$. 
	
	The last of these equations shows us that the transverse magnetic gradients are equal. For the rest of this derivation, we will refer to the transverse magnetic gradient as $B'$, where $B' = \frac{\partial B_y}{\partial x} = \frac{\partial B_x}{\partial y}$.
	
	For small displacements from the ideal orbit, the particle sees the following magnetic field\footnote{This is a first-order approximation, under the assumption that the square, cube, etc. terms are negligibly small for small particle displacements. This means that we're only considering dipole and quadrupole field components of the multi-pole expansion in this derivation.}\cite{syphers}:

		\begin{equation} \label{eq:bfieldsyphers}
			\vec{B}=\hat{x}[B_x(0,0)+B'y+\frac{\partial B_x}{\partial x}x]+\hat{y}[B_y(0,0)+B'x+\frac{\partial B_y}{\partial y}y]
		\end{equation}
		
	We are interested in \textit{restoring} forces; that is, only forces that are proportional to displacement can focus or defocus the beam. The dipole terms $B_x(0,0)$ and $B_y(0,0)$ are constant and do not depend on displacement, so they cannot provide a focusing force. Ignoring these terms from \textit{Equation~\ref{eq:bfieldsyphers}}, we now consider the force that this field exerts on a particle of charge $q$. For simplicity, the particle travels in the $\hat{z}$ direction with a speed of $\beta c$, where we make use of the relativistic beta $\beta = \frac{v}{c}$, and $c$ is the speed of light in vacuum. The particle's velocity vector is then $\vec{v}=\beta c \hat{z}$. Plugging this velocity and the magnetic field of \textit{Equation~\ref{eq:bfieldsyphers}} without the constant terms into the \textit{Lorentz equation}, we get:
	
		\begin{equation} \label{eq:force}
			\vec{F} = q(\vec{v}\times\vec{B}) = -\beta cB'x\hat{x} + \beta cB'y\hat{y} -\beta c\frac{\partial B_y}{\partial y}y\hat{x} + \beta c\frac{\partial B_x}{\partial x}x\hat{y}
		\end{equation}
	
	The last two terms in \textit{Equation~\ref{eq:force}} provide a force that is perpendicular to the displacement, as indicated by $y\hat{x}$ and $x\hat{y}$, so these terms cannot be restoring (focusing or defocusing) forces. Thus our restoring force ends up being:
	
		\begin{equation} \label{eq:force2}
			\vec{F}_{restoring} = \beta c B'(-x\hat{x}+y\hat{y})
		\end{equation}
	
	Notice that for a positive magnetic gradient $B'$, the horizontal force is $F_x = -\beta c B'x\hat{x}$; the minus sign means that the field pushes in the opposite direction of displacement. However, the vertical force for a positive $B'$ is $F_y = +\beta c B'y\hat{y}$, which pushes with the displacement. Thus a magnet with positive $B'$ focuses particles back to the beam pipe center in the x-direction, and defocuses in the y-direction; this is a ``focusing quadrupole''.
	
	If we flip the sign of $B'$, we get a vertically focusing and horizontally defocusing force. The transverse forces $F_x$ and $F_y$ have opposite signs regardless of our choice of B'. Thus we cannot use a single magnet (a single choice of $B'$) to focus in both planes.
	
	This illustrates the necessity for alternating-gradient focusing: by using a quadrupole of gradient $+B'$, followed by one with gradient $-B'$, we can build a repeating pattern (``lattice'') that provides net focusing in both planes.
	
	\section{Beta and phase function relationship}
	In this section, we derive the relationship between the lattice beta function $\beta(s)$ and the phase function $\psi(s)$. We will focus on the horizontal motion for this derivation, but the vertical derivation is the same. Recall that the differential equation describing transverse particle motion is given by \textit{Hill`s Equation}:
	
		\begin{equation} \label{eq:Hill}
			\frac{d^2x}{ds^2}+K(s)x=0
		\end{equation}
	
	The general solution to \textit{Hill`s Equation}, which describes the transverse particle displacement $x$ as a function of the longitudinal parameter $s$, is as follows:
	
		\begin{equation} \label{eq:Hillsoln}
			x(s)=A\sqrt{\beta_x(s)}\cos[\psi_x(s)+\phi]
		\end{equation}
		
	If we plug this solution back into Hill's Equation, we get the following expression in \textit{Equation~\ref{eq:Hillplug}}. Note that we now suppress the $s$ dependence and $x$ subscript of $\beta$ and $\psi$ for simplicity of notation.
	
		\begin{equation} \label{eq:Hillplug}
			\Omega\cos[\psi]-\frac{1}{\sqrt{\beta}}\frac{d\beta}{ds}\frac{d\psi}{ds}\sqrt{\beta}\frac{d^2\psi}{ds^2}\sin[\psi]=0
		\end{equation}
		
	Note that we have used the arbitrary variable $\Omega$ to represent the cosine coefficient. Since the cosine and sine functions are linearly-independent, the only way to make \textit{Equation~\ref{eq:Hillsoln}} true for all values of $\psi$ is to set the sine and cosine coefficients each equal to zero. Thus \textit{Equation~\ref{eq:Hillsoln}} actually gives us two equations, $\Omega=0$ and the following:
	
		\begin{equation} \label{eq:sincoeff}
			\frac{d\beta}{ds}\frac{d\psi}{ds}+\beta\frac{d^2\psi}{ds^2}=0
		\end{equation}
	
	Consider the product rule of differentiation, where $\frac{d}{ds}[A(s)B(s)]=\frac{dA}{ds}B(s)+A(s)\frac{dB}{ds}$. We can apply this rule in reverse to rewrite \textit{Equation~\ref{eq:sincoeff}} as:
	
		\begin{equation} \label{eq:sinprodrule}
			\frac{d}{ds}(\beta\frac{d\psi}{ds})=0
		\end{equation}
		
	If the rate of change of $\beta\frac{d\psi}{ds}$ is zero, that means that it must be a constant. In fact, since can choose any arbitrary constant, we say that $\beta\frac{d\psi}{ds}=1$. Or in other words, 
	
		\begin{equation} \label{eq:betadpsi}
			\frac{d\psi}{ds}=\frac{1}{\beta}
		\end{equation}
		
	Integrating both sides of \textit{Equation~\ref{eq:betadpsi}} over $s$, we arrive at the conclusion that the phase function $\psi(s)$ is calculated by integrating the inverse of the beta function $\beta(s)$ over the beam trajectory $s$:
	
		\begin{equation} \label{eq:psibeta}
			\psi(s)=\int{\frac{1}{\beta(s)}ds}
		\end{equation}
		
	\section{Transition slip factor}
		The following is a derivation of the relationship between a fractional change in transit time and momentum $\frac{\Delta v}{v}=\eta\frac{\Delta p}{p}$. The ``slip factor'' $\eta=\frac{1}{\gamma_t^2}-\frac{1}{\gamma^2}$ changes sign on either side of transition, which changes the effect that a particle's momentum offset has on its transit time.
	
		Consider the time interval $\tau$ between successive passes through the accelerating gap in the RF cavity. \textit{Equation~\ref{eq:tau}} shows that fractional changes in transit time $\frac{\Delta\tau}{\tau}$ can come from either a fractional change in path length $\frac{\Delta L}{L}$ or in velocity $\frac{\Delta v}{v}$ \cite{syphers}.
		
		\begin{equation} \label{eq:tau}
			\frac{\Delta\tau}{\tau}=\frac{\Delta L}{L}-\frac{\Delta v}{v}
		\end{equation}
		
		We will now express $\frac{\Delta L}{L}$ and $\frac{\Delta v}{v}$ each in terms of a fractional momentum change $\frac{\Delta p}{p}$. The momentum for a relativistic particle is $p = \gamma mv$, where $\gamma^2 = \frac{1}{1-(\frac{v}{c})^2}$ and $c$ is the speed of light in vacuum. Differentiating the momentum with respect to the velocity, we get $\frac{\Delta p}{\Delta v} = \gamma^3 m$. Using the relationship $m=\frac{p}{\gamma v}$, we arrive at \textit{Equation~\ref{eq:vmomentum}}, which shows the fractional velocity change a function of fractional momentum change. 
		
		\begin{equation} \label{eq:vmomentum}
			\frac{\Delta v}{v} = \frac{1}{\gamma^2}\frac{\Delta p}{p} 
		\end{equation}
		
		Since the path length deviation $\frac{\Delta L}{L}$ also depends on the momentum, we can define the variable $\gamma_t$ as a proportionality between fractional path length and momentum changes:
		 
		\begin{equation} \label{eq:pmomentum}
			\frac{\Delta L}{L} = \frac{1}{\gamma_t^2}\frac{\Delta p}{p}
		\end{equation}
		
		If we now substitute \textit{Equation~\ref{eq:vmomentum}} and \textit{Equation~\ref{eq:pmomentum}} back into \textit{Equation~\ref{eq:tau}}, we get the following:
		
		\begin{equation} \label{eq:tautransition2}
			\frac{\Delta\tau}{\tau} = (\frac{1}{\gamma_t^2} - \frac{1}{\gamma^2}) \frac{\Delta p}{p}
		\end{equation}
		
		Defining the slip factor as, $\eta=\frac{1}{\gamma_t^2}-\frac{1}{\gamma^2}$ we arrive at the following compact expression:
		
		\begin{equation} \label{eq:tautran}
			\frac{\Delta\tau}{\tau}=\eta\frac{\Delta p}{p}
		\end{equation}
		
		The slip factor $\eta$ is negative below transition, so a larger particle momentum corresponds to a shorter transit time. Above transition, $\eta$ is positive, and thus a larger particle momentum corresponds to a longer transit time.
		
	\section{Normalized emittance}
		Liouville's theorem states that in the absence of external forces or the emission of energy, particle beam emittance is conserved. However, the definition of emittance in \textit{Chapter~\ref{chap:con_physics}} is incomplete, and does not actually qualify as conserved under Liouville's theorem. The proceeding is a relativistic correction to the emittance that satisfies the conservation criteria.\cite{lee}

		Liouville's theorem only applies to a phase space as defined by the system's Hamiltonian conjugate parameters. For a general system parameter $q$, its Hamiltonian conjugate momentum $p$ must satisfy $p=\frac{\partial L}{\partial q'}$ where $q'=\frac{dq}{dt}$ and $L$ is the system's Lagrangian function $L=T-U$ for kinetic energy function $T$ and potential energy function $U$.\cite{amech} The area in phase space defined by $(q,p)$ is the conserved quantity under Liouville's theorem. 
		
		This means that we need to adjust our phase space coordinates, $x$ and $\frac{dx}{ds}$ in the horizontal plane, so that they qualify as Hamiltonian conjugates. We accomplish this by considering the relativistic momentum $p=\gamma m_0\frac{dx}{dt}$ and how it relates to $\frac{dx}{ds}$. Rewriting $\frac{dx}{dt}$ as the equivalent expression $\frac{dx}{dt}=\frac{dx}{ds}\frac{ds}{dt}$, and realizing that $\frac{ds}{dt}=\beta$, we can say that:
		
		\begin{equation} \label{eq:relmom}
			p=m_0c(\beta\gamma)\frac{dx}{ds}
		\end{equation}
		
		where we make use of the familiar relativistic variables $\beta=\frac{v}{c}$ for the speed of light in vacuum $c$, and $\gamma=\sqrt{\frac{1}{1-\beta^2}}$.
		
		So for the horizontal plane, our Hamiltonian conjugate phase space that satisfies Liouville's theorem uses the variables $x$ and $p=m_0c(\beta\gamma)\frac{dx}{ds}$. Defining our new emittance in this Hamiltonian conjugate space as the ``normalized emittance'' $\epsilon^*$, we see that it is simply proportional to the emittance $\epsilon$ of \textit{Chapter~\ref{chap:con_physics}} by a factor of $\beta\gamma$: in other words, $\epsilon^*=(\beta\gamma)\epsilon$.
		
		We call the emittance defined in \textit{Chapter~\ref{chap:con_physics}} the ``physical emittance'' or unnormalized emittance. Solving for the physical emittance, we see that:
		
		\begin{equation} \label{eq:physicalemitt}
			\epsilon=\frac{\epsilon^*}{\beta\gamma}
		\end{equation}
		
		To reiterate, the normalized emittance $\epsilon^*$ is \textit{conserved} under Liouville's theorem; that means that it is a constant regardless of beam energy. However, \textit{Equation~\ref{eq:physicalemitt}} shows that the physical emittance is inversely proportional to $\beta\gamma$. In other words, the subjective beam size as given by $\epsilon$ \textit{decreases} as the beam energy increases. This effect is known as ``adiabatic damping'', and explains why transverse beam size is much smaller at higher energies. Adiabatic damping has an effect on accelerator design, since lower-energy accelerators must usually have a larger aperture than those at higher-energy.
		
		Since the normalized emittance $\epsilon^*$ remains the same regardless of beam energy, any measured decrease tells us about beam degradation due to external forces or energy loss. This means that $\epsilon^*$ is a very useful parameter to determine beam quality.
		
		It should be noted that the space defined by $(x,p)$ is the actual definition of transverse horizontal phase space. The space defined in \textit{Chapter~\ref{chap:con_physics}} using $(x,\frac{dx}{ds})$ is actually called the ``trace space'', but colloquially you may here these terms used interchangeably.
		
	\section{Magnetic curvature radius}
		The magnetic field imparts a change in beam velocity direction known as ``centripetal acceleration''. Since the magnetic force is always perpendicular to the particle velocity\footnote{Recall that this is a consequence of the vector cross-product term in the Lorentz force equation $\vec{F}=q(\vec{E}+\vec{v}\times\vec{B})$}, only the direction of the beam velocity can change, not the magnitude. This is called ``centripetal acceleration'', defined in \textit{Equation~\ref{eq:centrip}} where $r$ is the radius of the resulting path and $v$ is the magnitude of the particle velocity..
		
		\begin{equation} \label{eq:centrip}
			a_c = \frac{v^2}{r}
		\end{equation}
		
		Consider the case where the magnetic field is perpendicular to the velocity: the Lorentz force equation simplifies to \textit{Equation~\ref{eq:lorentzperp}}.

		\begin{equation} \label{eq:lorentzperp}
			F = qvB
		\end{equation}
		
		If we plug \textit{Equation~\ref{eq:centrip}} into Newton's Second Law $F=ma$, and set the resulting force equal to \textit{Equation~\ref{eq:lorentzperp}}, we get:
		
		\begin{equation} \label{eq:force_equating}
			F = m(\frac{v^2}{r}) = qvB
		\end{equation}

		Now we can solve for the resulting path radius $r$ as shown in \textit{Equation~\ref{eq:radiusappendix}}, where $p$ is the beam momentum, $q$ is the particle charge, and $B$ is the magnetic field strength.
	
		\begin{equation} \label{eq:radiusappendix}
			r=\frac{p}{qB}
		\end{equation}

	\backmatter

		\cleardoublepage
		\addcontentsline{toc}{chapter}{Glossary}
		\printglossary
		
		\cleardoublepage
		\addcontentsline{toc}{chapter}{Index}
		\printindex
		
		\bibliographystyle{plain}
		\bibliography{citations}
			
\end{document}